\documentclass[9pt,journal,twocolumn]{IEEEtran}

\usepackage[utf8]{inputenc}
\usepackage{graphicx}
\usepackage{epstopdf}
\usepackage{amsfonts}
\usepackage[cmex10]{amsmath}
\usepackage{mathtools}
\usepackage{framed,xcolor}
\usepackage{cite}
\usepackage{subfigure}
\usepackage{bm}
\definecolor{shadecolor}{gray}{0.90}

\DeclareMathAlphabet{\mathantt}{OT1}{antt}{li}{it}
\DeclareMathAlphabet{\mathantt}{OT1}{pzc}{m}{it}

\graphicspath{{./figs/}}

\begin{document}
\title{The effect of hardware-computed travel-time on localization accuracy in the inversion of experimental (acoustic) waveform data}
\author{Mika~Takala,  Timo~D.~H\"{a}m\"{a}l\"{a}inen, Sampsa~Pursiainen
\thanks{M.\ Takala and S.\ Pursiainen (corresponding author) are  with the Laboratory of Mathematics, Tampere University of Technology, Finland.}
\thanks{M.\ Takala and T.\ D.\ H\"{a}m\"{a}l\"{a}inen are  with Laboratory of Pervasive Computing, Tampere University of Technology, Finland.} 
\thanks{Copyright  \textcircled{c} 2017 IEEE. Personal use of this material is permitted. However, permission to use this material for any other purposes must be obtained from the IEEE by sending a request to pubs-permissions@ieee.org.}
\thanks{This paper has supplementary downloadable material available at http://ieeexplore.ieee.org., provided by the author. The material includes a set of audio wave measurement data files, a readme file and a Matlab plot scrip. Contact sampsa.pursiainen@tut.fi for further questions about this work.} } 
\maketitle
\begin{abstract}
This study aims to advance hardware-level computations for travel-time tomography applications in which the wavelength is close to the diameter of the information that has to be recovered. Such can be the case, for example, in the imaging applications of  ({\bf 1}) biomedical  physics,  ({\bf 2}) astro-geophysics and  ({\bf 3}) civil engineering. Our aim is to shed light on the effect of that preprocessing the digital waveform signal has on the inversion results and to find computational solutions that guarantee robust inversion when there are incomplete and/or noisy measurements. We describe a hardware-level implementation for integrated and thresholded travel-time computation (ITT and TTT). We compare the ITT and TTT approaches in inversion analysis with experimental acoustic travel-time data recorded using a ring geometry for the transmission and measurement points. The results obtained suggest that ITT is essential for maintaining the robustness of the inversion with imperfect signal digitization and sparsity. In order to ensure the relevance of the results,  the specifications of the test setup were related to those of applications  ({\bf 1})--({\bf 3}). 
\end{abstract}

\IEEEpeerreviewmaketitle

\section{Introduction}

This paper concerns waveform tomography in which either an acoustic or electromagnetic wave travels through a target object and, based on the external measurements of the wave,  the task is to estimate the distribution(s) of a given parameter within the target \cite{kaipio2004} such as  the velocity of the wave or the absorbtion parameter. Tomographic imaging based on wave propagation requires computationally heavy mathematical inversion of the data in order to retrieve the relevant result set, such as an image or three-dimensional model of the test subject. 

In this paper, we explore the problem of determining the travel-time of the signal  \cite{tarantola2005,hooi2014} and its relation to accuracy of the the inverse localization. In particular, we focus on the effect of hardware-level computations in applications where the signal wavelength may be expected to be close to the diameter of the details that are to be recovered. Such can be the case, for example, in biomedical microwave or ultrasonic tomography \cite{grzegorczyk2012,meaney2010,fear2002,ruiter2012, opielinski2013, ranger2012,bond2000,kepler2000}, in astro-geophysical subsurface imaging    \cite{kofman2015,kofman2007,su2016,pursiainen2014,pursiainen2014b} and in non-destructive testing of civil engineering materials and structures \cite{yoo2003,chai2010,chai2011,acciani2008}.  Our aim is determine what effect the preprocessing of digital waveform signal has on the inversion results and in that context to find computational solutions that guarantee robust inversion even with incomplete and noisy measurements. We use travel-time data, as it is known to yield robust information of the unknown parameter \cite{barriot1999}, and also because it requires minimal data transfer between the hardware and the inversion routines. Data preprocessing performed on embedded hardware has its limitations, and therefore, we aim to find out how the hardware-level evaluation of the travel-time is related to the tomography results. To perform the  preprocessing operations, a field programmable gate array (FPGA) \cite{wolf2004} chip on a development board is used as a  platform for the design. In order to make the implementation fast and flexible, the hardware on the FPGA was implemented by adapting Matlab scripts \cite{hunt2006} to C code \cite{kochan2004} and then utilizing Catapult C hardware synthesis \cite{coussy2008}  to generate the hardware from the adapted C code.  

\begin{figure*}[!] \begin{footnotesize} \begin{center}
\begin{minipage}{2.5cm}({\bf 1})\begin{center}
  	\includegraphics[height=2.4cm]{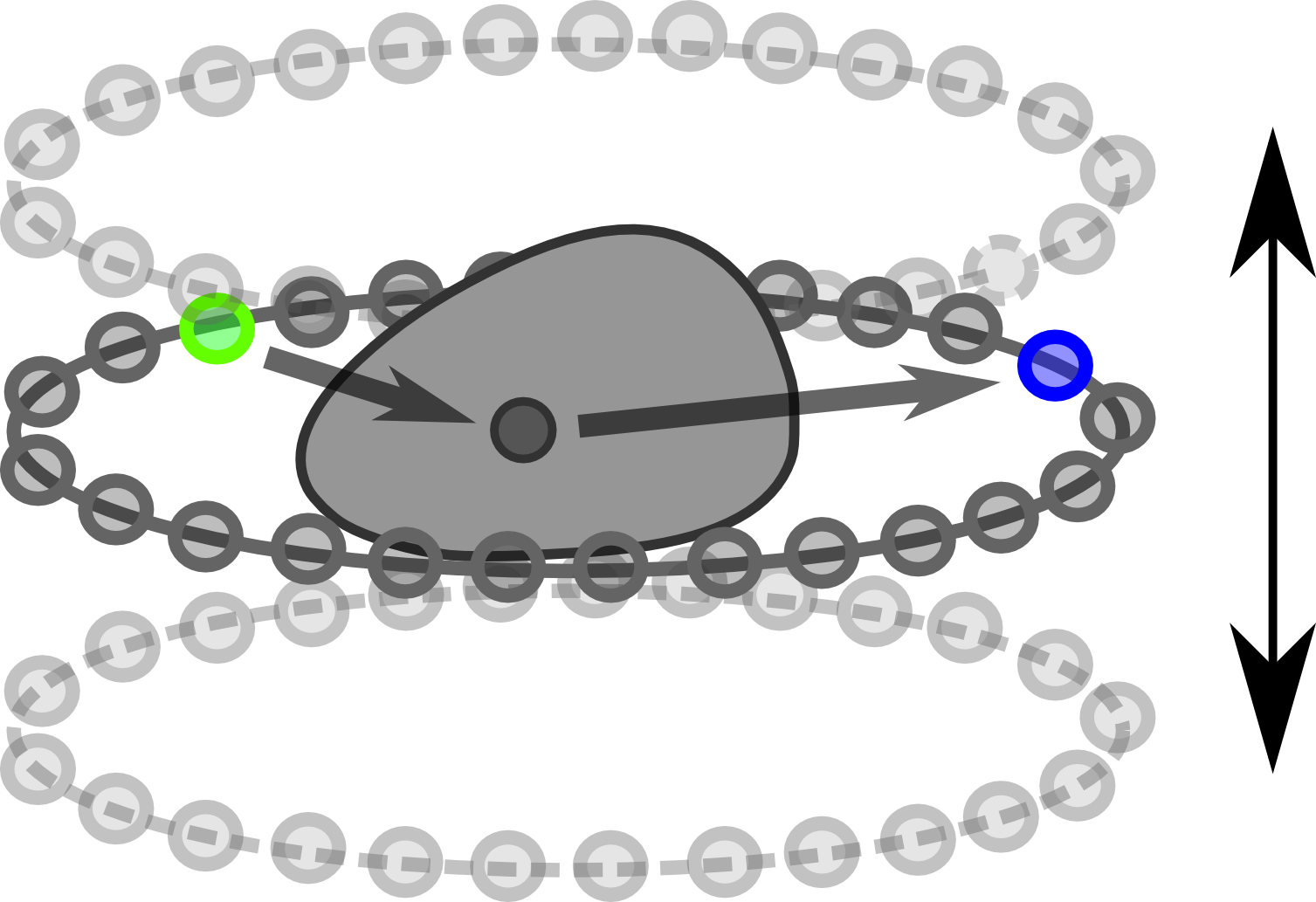}    
  	\end{center}
\end{minipage}
\hskip2.0cm
\begin{minipage}{2.5cm}({\bf 2})\begin{center}
  	\includegraphics[height=2.8cm]{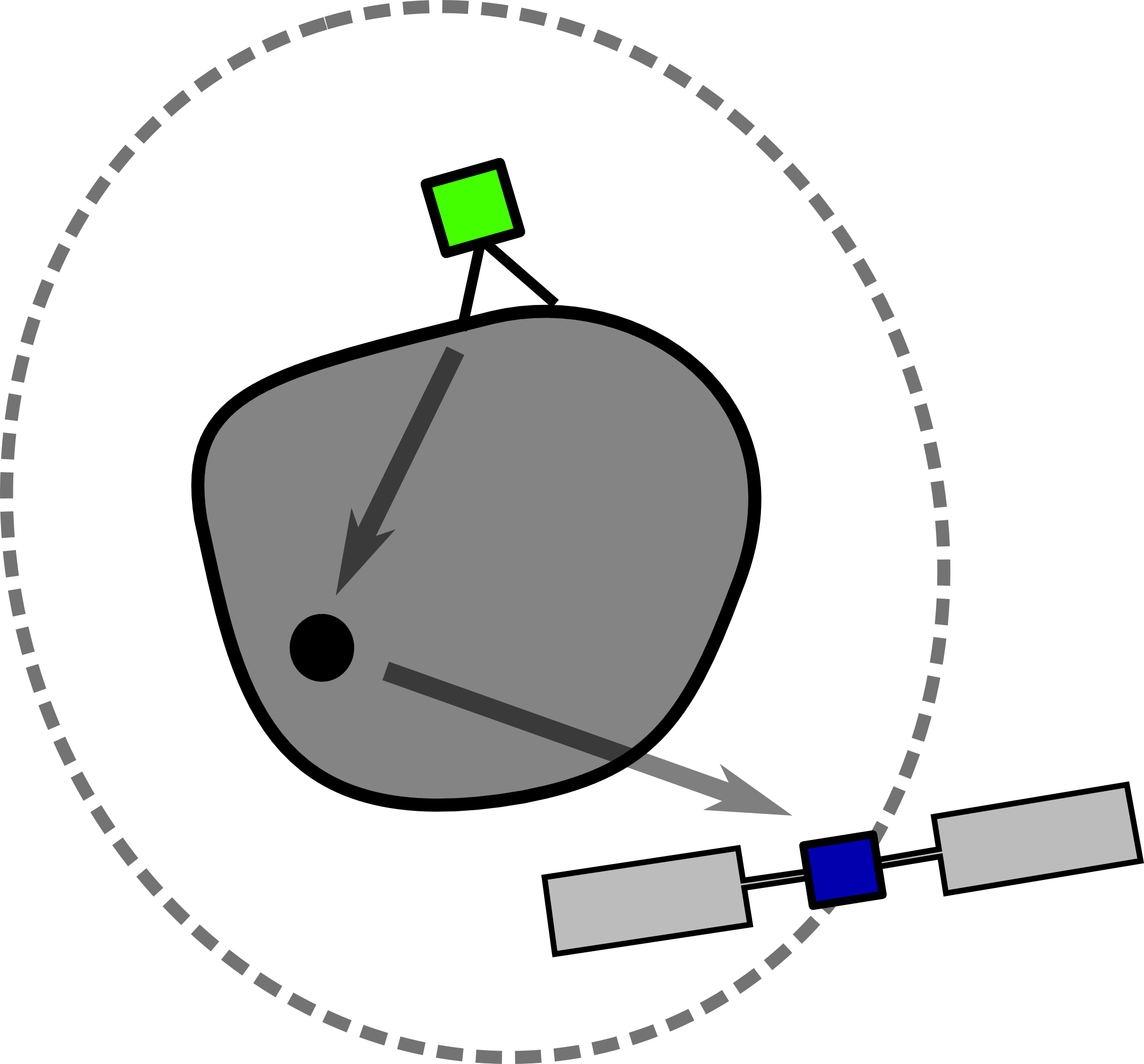} 
  	\end{center}
\end{minipage}
\hskip2.0cm
\begin{minipage}{2.5cm}({\bf 3})\begin{center} \mbox{} \vskip0.3cm
  	\includegraphics[height=2.4cm]{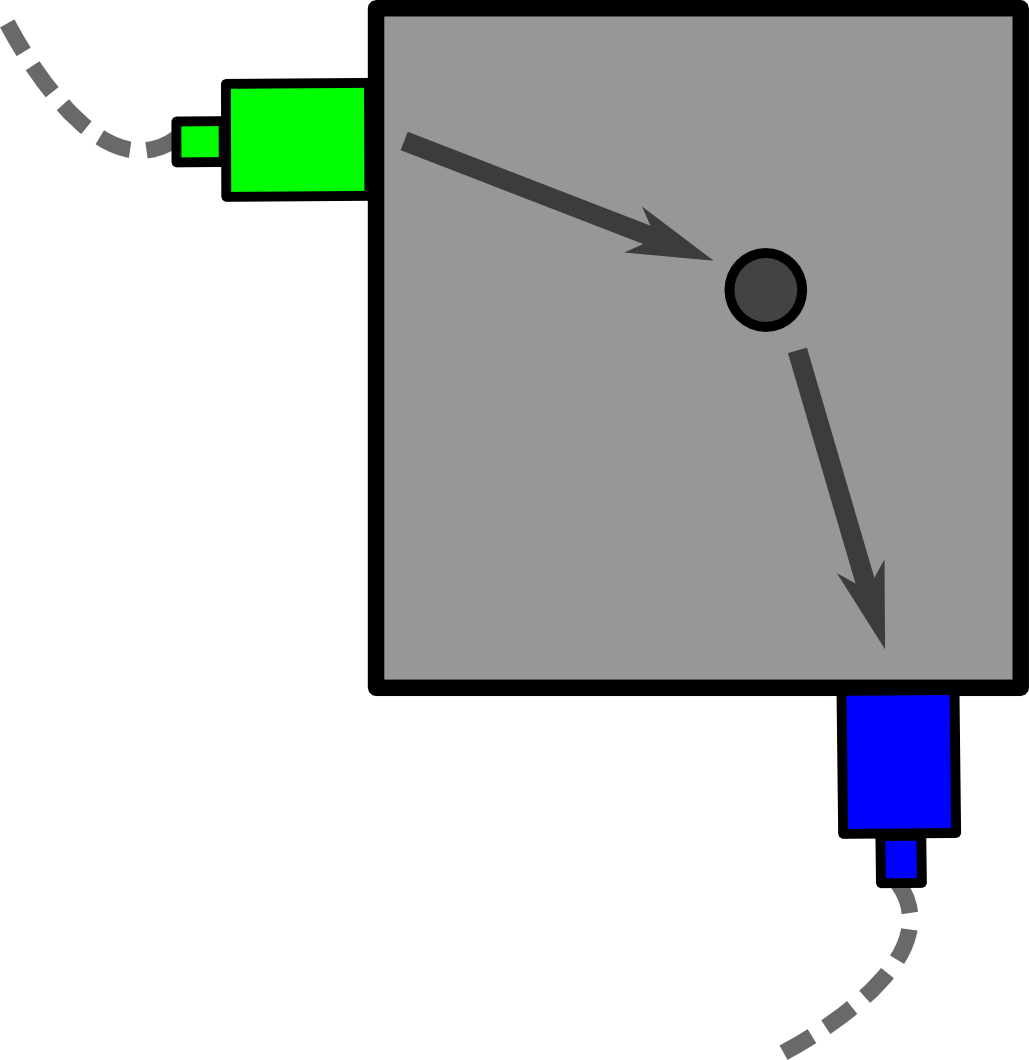} 
\end{center}
\end{minipage} \end{center}
 \end{footnotesize}
  \caption[Orbiter-to-orbiter and lander-to-orbiter scenario]{A schematic picture of the travel-time measurement in the applications ({\bf 1})--({\bf 3}). The signal path between the  transmitter (green) and receiver (blue) is refracted by a perturbation (circle) in the wave speed distribution.  {\bf Left:} A movable sensor ring  that can be used to scan a 3D target object of M/UCT in slices \cite{opielinski2013,grzegorczyk2012,son2015}.  {\bf Center:}  The CONSERT experiment in which the objective was to recover the internal structure of the nucleus of the comet 67P/Churyumov-Gerasimenko based on sparse lander-to-orbiter signal transfer  between the Rosetta spacecraft and a single comet lander Philae \cite{kofman2007}. {\bf Right:}  Ultrasonic detection of defects within a concrete block via two transducers and a semi-direct transmission  \cite{bond2000}.}
  \label{omegadrawing}
\end{figure*}

FPGA-based processing of tomographic travel-time data is utilized in all above-mentioned application fields. As specific examples we consider ({\bf 1}) the microwave and ultrasonic computed  tomography (MCT and UCT) of the breast \cite{grzegorczyk2012,meaney2010,fear2002, ruiter2012, opielinski2013, ranger2012}, ({\bf 2})  tomography of small solar system bodies (SSSBs), e.g., the COmet Nucleus Sounding Experiment by Radiowave Transmission (CONSERT)  \cite{kofman2015, kofman2007,pursiainen2014,pursiainen2014b}, and ({\bf 3}) the acoustic/electromagnetic imaging and testing of concrete structures   \cite{yoo2003,chai2010,chai2011,acciani2008,bond2000,kepler2000}. An M/UCT breast scan (Figure \ref{omegadrawing}) can be performed by utilizing  a 2D sensor ring  \cite{opielinski2013,grzegorczyk2012,son2015} which records data slices sharing the direction of  the plane normal with the ring. MCT and UCT have recently been shown to have the potential to detect and classify breast lesionsat least as reliably as other imaging methods, such as computed X-ray tomography (CT) and magnetic resonance imaging (MRI)  \cite{opielinski2013, opielinski2015, meaney2007}. Other recently studied methods for breast cancer diagnosis include, e.g., optical tomography \cite{choe2005,corlu2007}. The CONSERT experiment took place as a part of the European Space Agency's {\em Rosetta} mission. The objective of CONSERT was to recover the internal structure of the nucleus of the comet 67P/Churyumov-Gerasimenko based on sparse lander-to-orbiter signal transfer  between the Rosetta spacecraft and a single comet lander {\em Philae} (Figure \ref{omegadrawing}). Space technology also involves also the challenging space environment as a limitation \cite{fortescue2003, agrawal2014}, leading, e.g.,  to sparse measurements. In concrete testing \cite{yoo2003,chai2010,chai2011,acciani2008}, the task can be for example to detect interior defects within a concrete element (Figure \ref{omegadrawing}) in a similar way to, e.g. the way voids are localized in geoscientific applications. 

In the numerical experiments, we compared integrated and thresholded travel-time (ITT and TTT)  approaches via inversion analysis utilizing experimental acoustic waveform data. A 16-bit and 8-bit analog-to-digital (A/D) conversions were tested together with two different threshold criteria and normalization levels. The results obtained suggest that ITT is essential for maintaining the robustness of the inversion if the A/D conversion is incomplete. In order to ensure the relevance of the results,  the specifications of the test setup were related to those of applications  ({\bf 1})--({\bf 3}). 

This paper is organized as follows. Section 2 describes the materials and methods, including the mathematical  forward modeling and inversion techniques, the test setup, the equipment and the FPGA implementation.  Section 3 presents the inversion results. Section 4 sums up the findings of this work and discusses the results and the potential direction of future work.

\section{Materials and methods}

\subsection{Forward modeling}
\label{forwardmodel}

\subsubsection{Signal wave}

In this study, a  waveform signal is modeled as a scalar field $u$ representing an electromagnetic or acoustic wave. The computational domain $\Omega$ is assumed to contain the target object of the tomography together with its immediate surroundings. During the measurements, that is, when $t \in [t_1, t_2]$, the transmitters and receivers can be either fixed or moving or touching the surface or remote from it. In $(t, {\vec x}) \, \in \, [t_1, t_2] \times \Omega$, the scalar field $u$ is assumed to obey the following second-order wave equation system:
{\setlength\arraycolsep{2pt} \begin{eqnarray} \label{kaava1}
\frac{1}{\mathtt{c}^2} \frac{\partial^2 u}{\partial t^2} - \Delta_{\vec x} u  & = & f(t) \,   \delta(\vec{x} - \vec{x}^{({0})}), \\  u(0, {\vec x}) &   = & \frac{\partial u}{\partial t}(0, {\vec x }) =  0,
\end{eqnarray}}
where $\mathtt{c}$ denotes the signal velocity  in $\Omega$, $f(t)$ is the time dependence of the signal transmission,  $\vec{x}^{({0})}$ is the point (spatial location) of the transmission, and $\delta(\vec{x} - \vec{x}^{({0})})$ is a Dirichlet delta function, i.e., it is zero everywhere except at $\vec{x} = \vec{x}^{({0})}$ and satisfies  the integral identity $\int_{\Omega} \delta(\vec{x} - \vec{x}^{({0})}) \, d \Omega = 1$. The left-hand side of Equation (\ref{kaava1}) is the standard hyperbolic operator of the wave equation \cite{evans1998}, and the right-hand side represents a point source transmitting an isotropic signal pulse.  The isotropic radiation pattern is altered, if the source is placed in front of a reflector (with $\mathtt{c} = 0$). In this study, a loudspeaker profile is used as a reflector.  Additionally, it is assumed that only the set of points belonging to $\Omega$ can  transmit a signal and that there are no other signal sources present. Defining a new variable  $ \vec g(t, \vec x) = \int_0^t \nabla u(\tau, \vec x) \, d \tau$ and $h(t) = \int_0^t f(\tau) \, d \tau$, the resulting system is of the following first-order form:
{\setlength\arraycolsep{2 pt} \begin{eqnarray}
\frac{1}{\mathtt{c}^2}  \frac{\partial u}{\partial t} -\nabla\cdot\vec g  & =&   {h}(t) \, \delta(\vec{x} - \vec{x}^{{{(0)}}}),  \\       \frac{\partial \vec g}{\partial t} - \nabla u  & =&   0.
\end{eqnarray}}
This system can be discretized spatially using the finite element method (FEM) \cite{braess2007} and temporally via the finite-difference time-domain (FDTD) method \cite{schneider2016} leading to so-called leap-frog formulae which enable the simulation of the complete wave  \cite{pursiainen2014}.

\subsubsection{Measurement model} 

In this study, we aim at localizing perturbations in the inverse of the velocity distribution $\mathtt{n} = {1}/{\mathtt{c}}$ based on the travel-time $T$ of the signal, as this is known to provide robust information about the velocity  \cite{barriot1999,tarantola2005}. The wave $u_m(t, \vec{x})$ measured at $\vec{x}$ is a sum of the propagating wave $u(t, \vec{x})$ and a random noise term $\varepsilon(t, \vec{x})$, i.e., \begin{equation} u_m(t, \vec{x}) = u(t, \vec{x}) + \varepsilon(t, \vec{x}). \end{equation} The noise can include both modeling and measurement errors. In order to minimize the effect of the noise,  {\em a priori} information should be used, for example, to determine the most relevant time interval for each measurement point. However, deriving a complete statistical model for $\varepsilon(t, \vec{x})$ can be difficult due to potential but unknown error sources, such as reflections, refractions and absorption. There is also no unique way to obtain $T$ based on the waveform measurement $u_m$ \cite{tarantola2005}.  


\subsubsection{Integrated and thresholded travel-time estimates}

\begin{figure}
\begin{center}
\subfigure[Measured travel-time calculation.]{
\label{th_vs_integ_mes_fig}
\includegraphics[width=6cm]{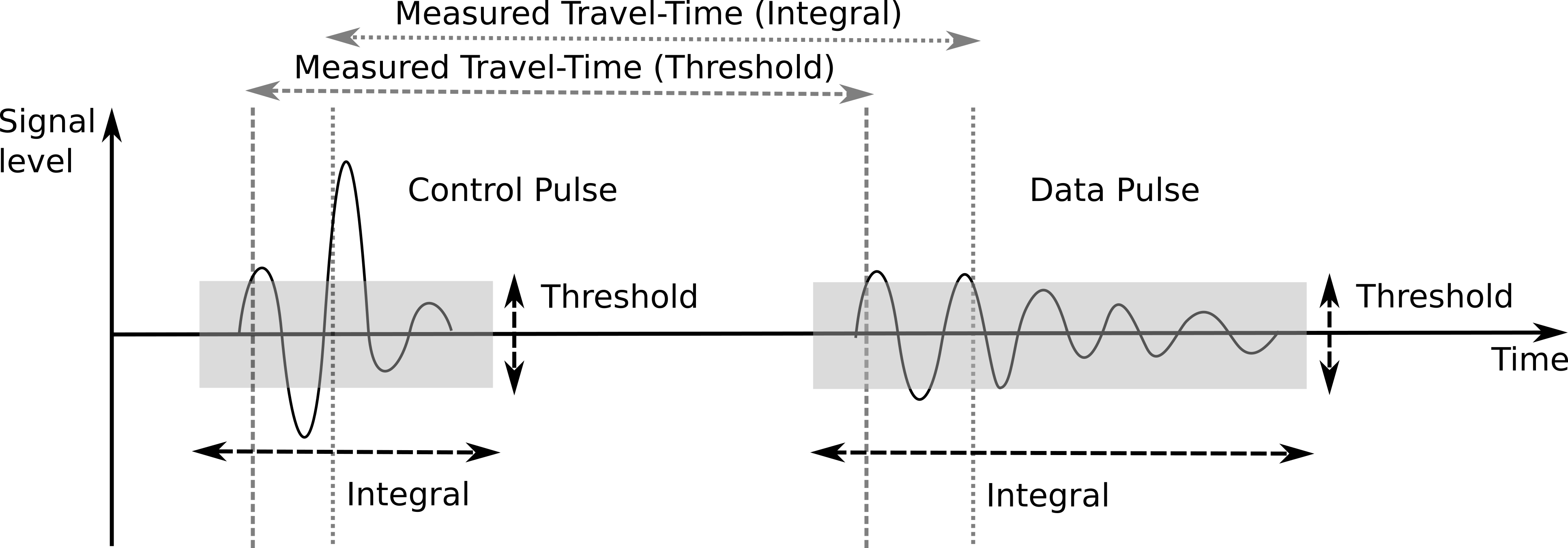}
}
\subfigure[Simulated travel-time calculation.]{
\label{th_vs_integ_sim_fig}
\includegraphics[width=6cm]{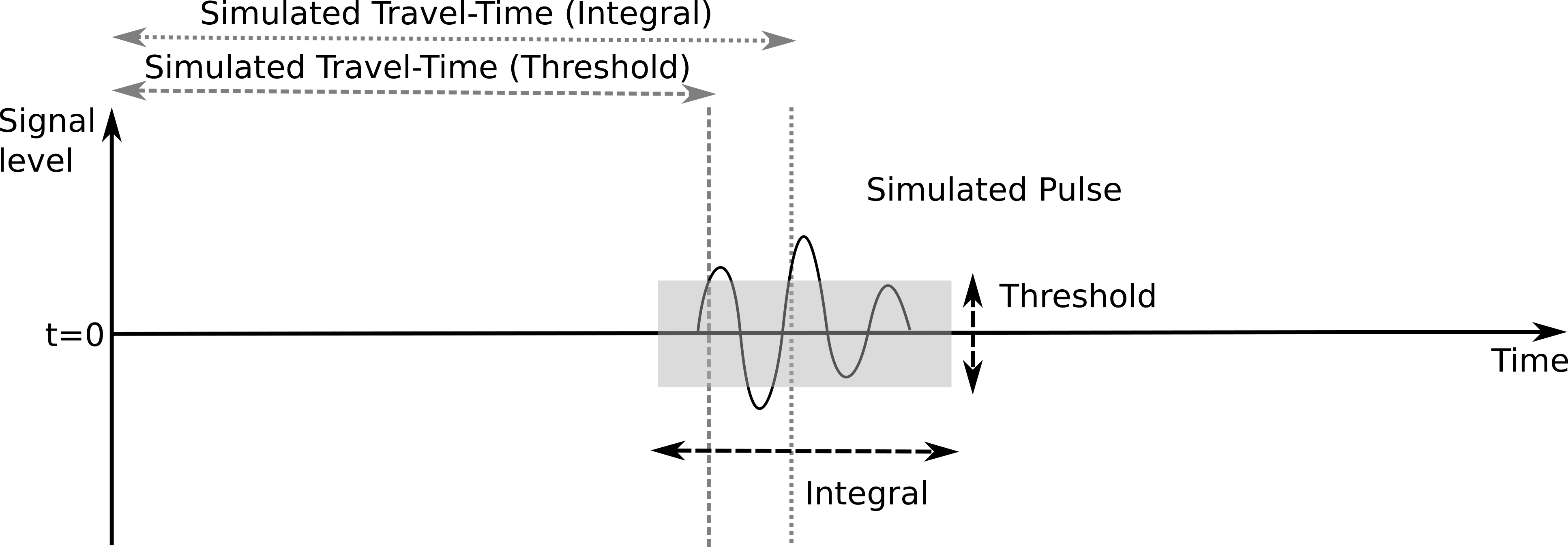}
}
\caption[Illustration of measured and simulated travel-time calculation]{Visualisation of travel time calculation types. The greyed out areas represent the integrated signal. The (b) image highlights how the methods must also be used in a simulation.}
\label{signaldiagram}
\end{center}
\end{figure}
In this paper, we study integration and thresholding as two alternative techniques for estimating $T$. Given the measured wave $u_m$, the integrated travel-time (ITT) estimate for a signal received at  $\vec{x}$ within the time interval $[t_1, t_2]$ can be defined via the formula  \cite{pursiainen2015electromagnetic}
\begin{equation}
\label{traveltimeintegration}
T(\vec{x}) = \frac{\int_{t_1}^{t_2} t \,u_m(t, \vec{x})^2 dt}{\int_{t_1}^{t_2} u_m(t, \vec{x})^2 dt}.
\end{equation}
To interface the mathematic model and real signals, some decisions have to be made. These include how to decide the time interval [$t_1, t_2]$ and whether to use the ITT or an alternative strategy, e.g., the thresholded travel-time (TTT) estimate. ITT and TTT were calculated as follows: 
\begin{enumerate}
\item Detect the time value $t_0$ (TTT) where the amplitude reaches a pre-selected threshold value. 
\item Set $t_1 = t_0 - \tau_1$ and $t_2 = t_0 + \tau_2$, where $\tau_1$ and $\tau_2$ are auxiliary {\em a priori}  parameters ensuring that the essential part of the signal pulse will be contained in $[t_1, t_2]$. Then, obtain ITT through Equation (\ref{traveltimeintegration}).
\end{enumerate}

Here,  $\tau_1$ extends the inspected time interval in the reverse direction, that is, (before) the signal detection point $t_0$. Parameter $\tau_2$  cuts the signal based on the {\em a priori} information of the pulse length in order to prevent noise due to reflections. Figure \ref{signaldiagram} visualizes the calculation of ITT and TTT for measurement and simulated data. Figure \label{th_vs_integ_mes_fig} shows that in the measurements, the travel-time can only be calculated if the starting point (control pulse) is also measured. The measured travel-time can be compared to the simulated case (Figure \ref{th_vs_integ_sim_fig}) where the starting point (zero-point) is defined exactly.

Based on the measurements, exactly one travel-time value $T$ is calculated for each transmitter-receiver position pair within the measurement configuration. The resulting set of values is referred to as the measurement data. The simulated signal refers to the wave that can be obtained via the FDTD method \cite{pursiainen2014} using  an initial (constant) estimate  for the unknown parameter.

\subsubsection{Path integrals and ray-tracing}

In addition to FEM/FDTD computations, a forward simulation for the travel-time can be obtained via an  integral of the form \begin{equation}T = \int_\mathcal{C} \mathtt{n} \,  d s, \label{path_integral} \end{equation}  where $\mathcal{C}$ is the signal path \cite{barriot1999}. In this approach, the prediction for $\mathcal{C}$ has a central role. This can be done, e.g., based on Snell's law of reflection and reftraction \cite{pursiainen2013}. We use the approximation  (\ref{path_integral}) in order to find a reconstruction of $\mathtt{n}$. We assume that the signal paths $\mathcal{C}_1, \mathcal{C}_2, \ldots, \mathcal{C}_M$ are straight line segments and that $\mathtt{n}$ is of the form $\mathtt{n} = \mathtt{n}_p + \mathtt{n}_0$, where  $\mathtt{n}_0$ is an {\em a priori} known constant background distribution and $\mathtt{n}_p$ is an unknown perturbation.  A discretized version of (\ref{path_integral}) can be obtained by subdividing the computational domain into pixels $P_1, P_2, \ldots, P_N$ and estimating $\mathtt{n}_p$ with a pixelwise constant distribution.  Using ${\bf x}$ to denote the vector of pixel values, the equation (\ref{path_integral}) can be written in the  discretized form 
\begin{equation}
\label{inversionformula}
{\bf y} = {\bf L } {\bf x}  + {\bf y}_0, 
\end{equation}
where $L_{i, j} = \int_{\mathcal{C}_i \cap P_j}  \, d s$ and $y_0$ is a simulated vector estimated from the travel-time data corresponding to $\mathtt{n}_0$.

\subsection{Inversion procedure}

To invert the data, we use a classical total variation based regularization technique, which aims at minimizing (Appendix) the regularized objective funcion $\Psi({\bf x}) = \| {\bf  y} - {\bf y}_0 - {\bf L x} \|_2 + \alpha \| {\bf D}  {\bf x} \|_1$ via the iterative recursion procedure \begin{equation} {\bf x}^{(k)} = \left( {\bf L}^T {\bf L} + \alpha {\bf D} {\bm \Gamma}^{(k)} {\bf D} \right)^{-1} {\bf L}^T ({\bf y} - {\bf y}_0), \label{tv_iteration} \end{equation}  where ${\bf \Gamma}^{(k)}$ is a diagonal weighting matrix defined as ${\bf \Gamma}^{(k)} = \hbox{diag}\left(\|{\bf D} {\bf x}^{(k)}\|\right)^{-1}$ and ${\bf D}$ is a regularization matrix whose entries satisfy 
\begin{equation}
D_{i, j} = \beta \delta_{i,j} + \frac{\int_{P_i \cap P_j} \left( 2\delta_{i,j}-1\right) ds}{\max_{i,j}\left(\int_{P_i \cap P_j} ds\right)}
\end{equation}
with $\delta_{i,j} = 1$ if $i=j$, and zero if otherwise. In ${\bf D}$, the first term penalizes the norm of ${\bf x}$ and the second one the total variation, i.e., the total sum $\sum_{i,j} \int_{P_i \cap P_j} | x_i - x_j| ds$ in which each non-zero term equals to the total jump of ${\bf x}$ between two adjacent pixels multiplied by the pixel's side-length \cite{kaipio2004}. This simple inversion approach usually converges sufficiently in a relatively low number of iteration steps, e.g., five. The value of the parameter $\beta$ determines the balance between the regularization matrices. A small value for $\beta$ leads to inverse estimates with low total variation and larger values can be expected to result in well-localized estimates \cite{pursiainen2013, pursiainen2014, pursiainen2015electromagnetic}.

\subsection{Test setup and scenario}
\label{testsetup}

\begin{table}[!]
\caption{Transmitters and receivers used in recording the experimental data.}
\label{technicaldata}
\begin{center}
\begin{tabular}{@{} llp{3.5cm}} \hline
Type& Item&Details\\
\hline
Transmitter & Two-way active speaker & 20 Hz -- 18 kHz frequency response. \\
Receiver &   Dynamic microphone &  Unidirectional 60 Hz -- 13 kHz frequency response, -72 dB sensitivity.\\
\hline
\end{tabular}
\end{center}
\end{table}
\begin{table}[!]
  \small
    \caption{The diameter and center coordinates of the domain $\Omega$ and the foam cylinders {\bf A}--{\bf C}.}
    \label{objdetails} \begin{center}
    \begin{tabular}{@{}l r r r} \hline 
      {Target} & x  (cm)& y  (cm) & {Diameter} (cm)\\
      \hline
      {\bf A} & -11.0 & 12.0 & 15.0\\
      {\bf B} & 12.0 & 10.0 & 10.0\\
      {\bf C} & -2.5 & -13.0 & 10.0\\
      $\Omega$ & 0.0 & 0.0 & 59.0\\
      \hline
    \end{tabular} \end{center}
  \end{table}
\begin{figure*}[!]
  \begin{center}
    \begin{minipage}{4.7cm} \begin{center} \includegraphics[width=3.3cm]{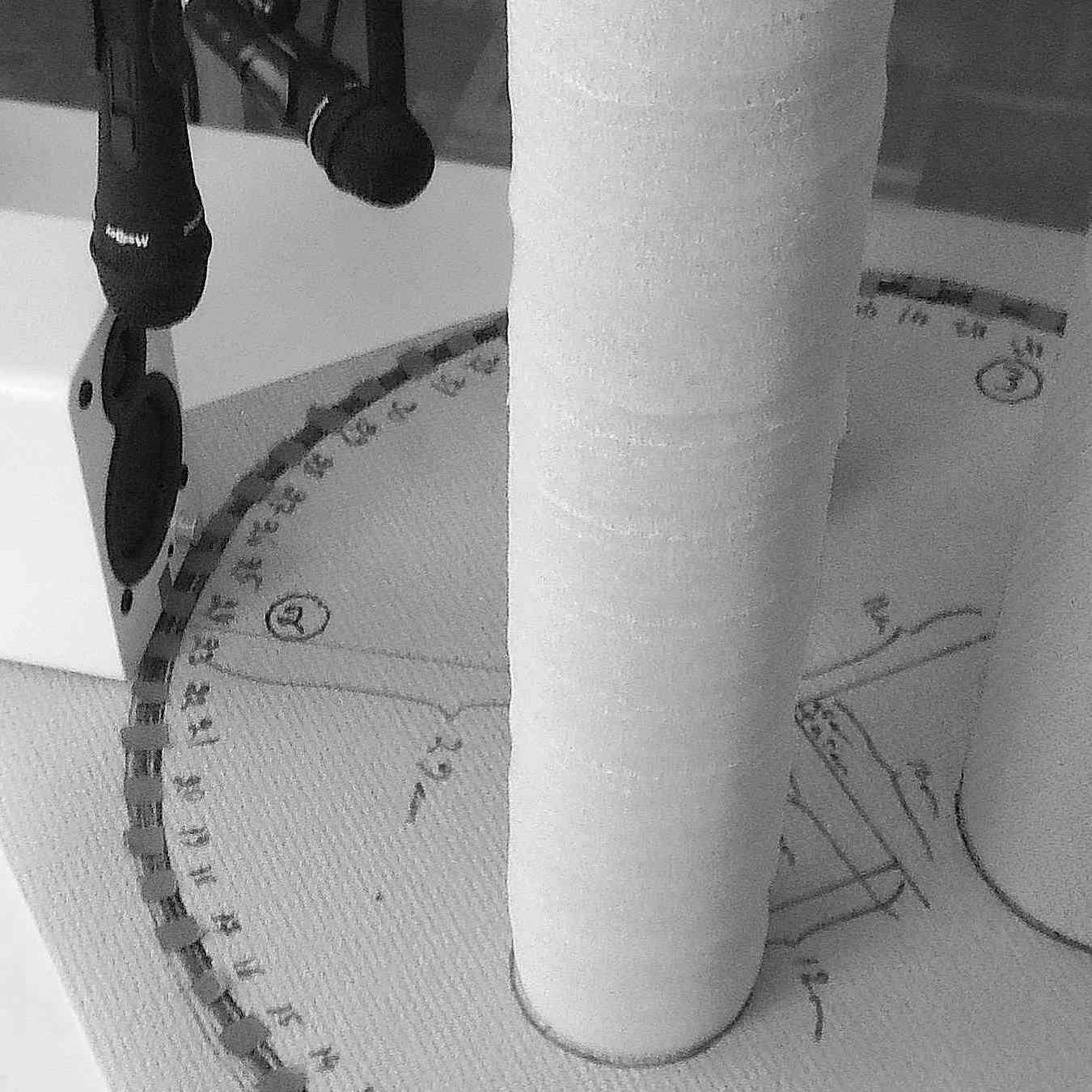} \\ \mbox{} \end{center} \end{minipage}    
\begin{minipage}{4.7cm} \begin{center} \includegraphics[width=3.7cm]{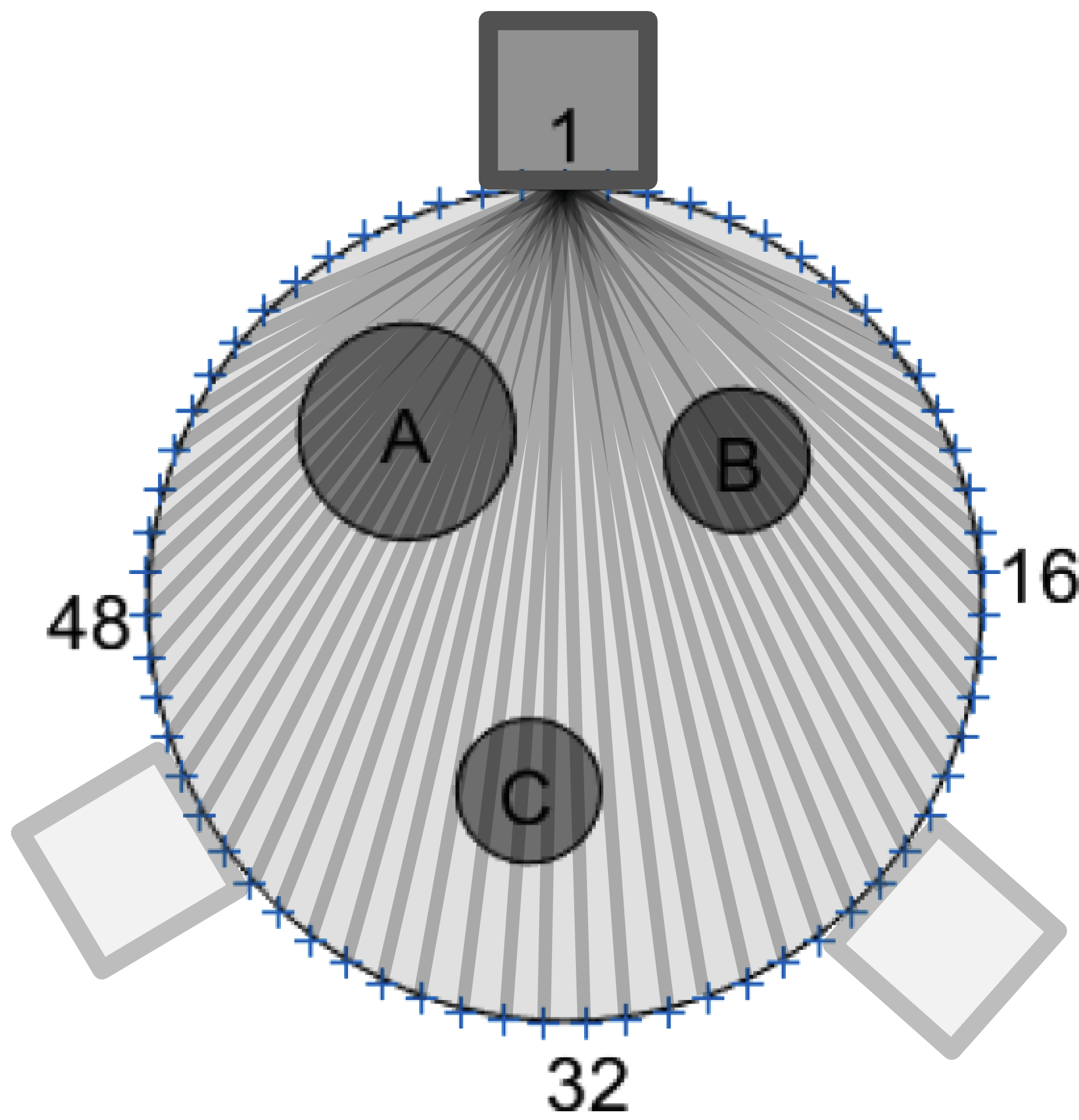} \\ \vskip0.1cm Set ({\bf I}) \end{center} \end{minipage}
\begin{minipage}{4.7cm} \begin{center} \includegraphics[width=3.7cm]{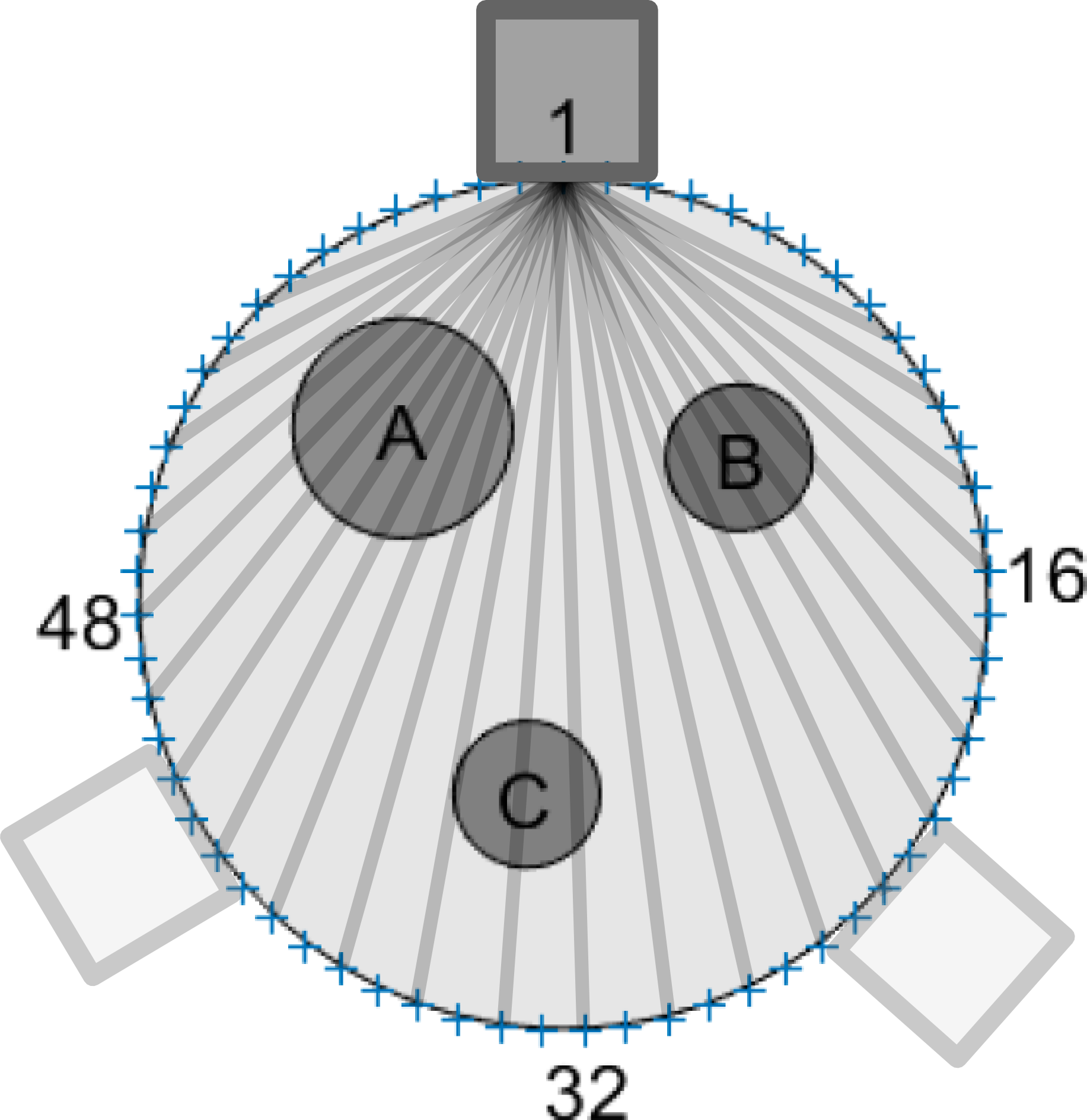}  \\ \vskip0.1cm Set ({\bf II}) \end{center} \end{minipage}
  \end{center}
  \caption{{\bf Left:} The test setup during the measurement procedure. {\bf Center and Right:} A top-down diagram of the setup. The control points {\bf 1}--{\bf 64} are shown as light blue crosses on the perimeter. Four of them are numbered to show the clockwise ordering. Points {\bf 1}, {\bf 24} and {\bf 43} were used as transmitter locations. The dark grey box represents the transmitter (speaker)  when positioned at {\bf 1} and the light grey boxes show it in the case of {\bf 24} and {\bf 43} (clockwise). Based on the measurements two travel-time data sets ({\bf I})  and ({\bf II}) were formed. In ({\bf I}) (dense set, center),  the centermost 47 points were utilized in the final data set consisting of $3 \times 47 = 141$  individual travel-time values, while only every other point of the set ({\bf I}) was included in ({\bf II}) (sparse set, right) resulting in $3 \times 24 = 72$ travel-time values.   }
  \label{equipment_diagram}
\end{figure*}
\begin{figure}[!]
\begin{scriptsize}
\begin{center}
\begin{minipage}{7cm}
\begin{center}
\includegraphics[width=7.0cm]{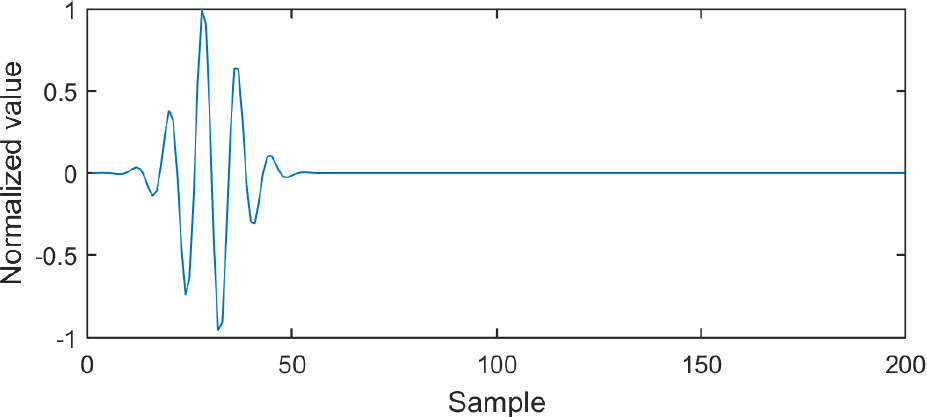} 
\end{center}
\end{minipage} \hskip1cm
\begin{minipage}{7cm}
\begin{center}
\includegraphics[width=7.0cm]{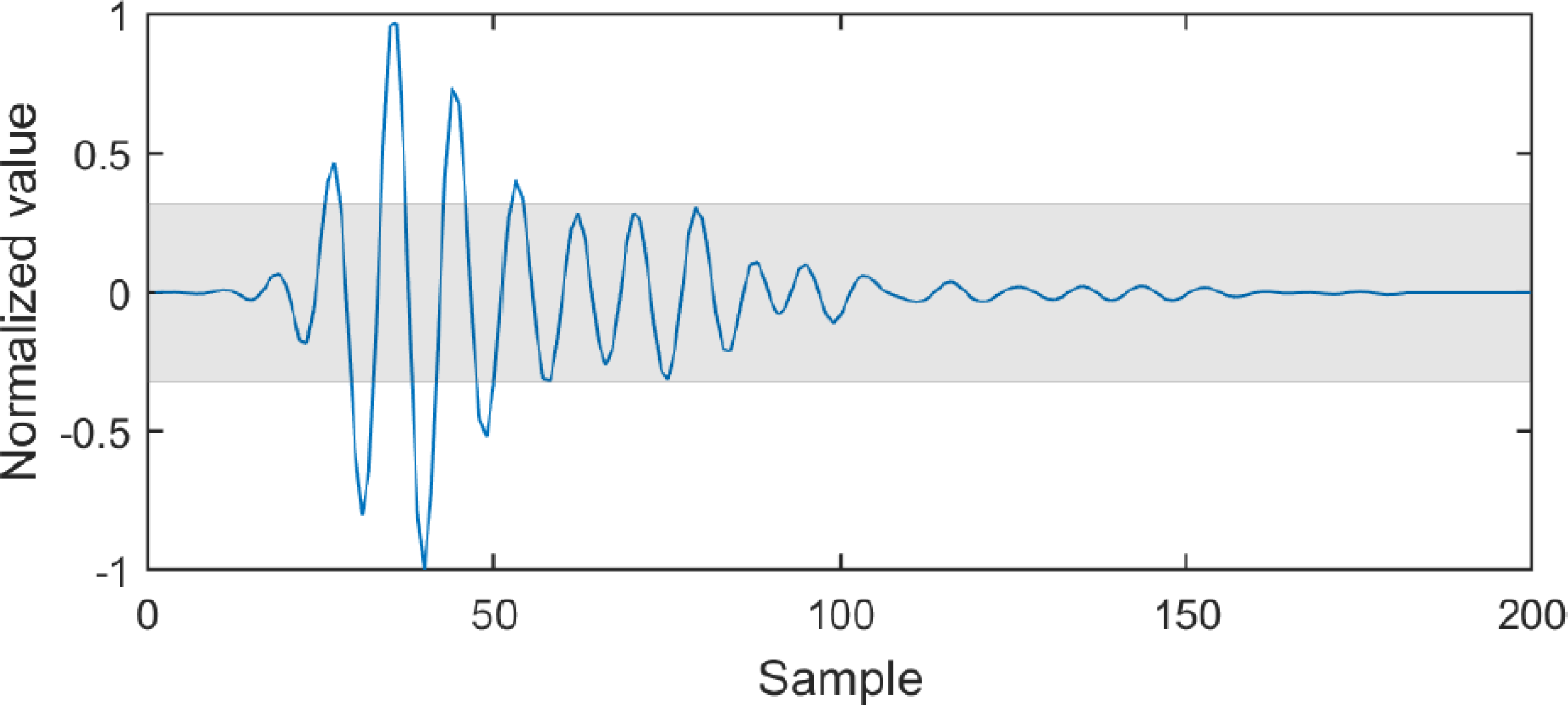} 
\end{center}
\end{minipage} 
\end{center}
\end{scriptsize}
\caption{{\bf Top:} The transmitted signal pulse as given by equation (\ref{signal_pulse}). {\bf Bottom:}  The signal pulse received at the perimeter of the (empty) test domain $\Omega$ without the foam cylinders placed inside. The tail of the received signal  contains noise due to echoes and  inaccuracies in transmission (ringing). Based on a comparison with the original signal the noise peaks were observed to be mainly 10 dB (grey area) below the main peak. }
\label{pulsefigures}
\end{figure}

\subsubsection{Domain}

An acoustic setup with one speaker and two microphones was utilized to gather the experimental data. The computational domain $\Omega$ was a 59 cm diameter disk drawn on a 0.5 cm thick soft foam covering (Figure \ref{equipment_diagram}). In the experiment, the locations of three foam cylinders {\bf A}, {\bf B} and {\bf C} with diameters 15, 10 and 10 cm, respectively, were to be detected. These were placed in the disk $\Omega$ in the upright position and apart from each other. The perimeter of $\Omega$ contained 64 equally spaced control points {\bf 1}--{\bf 64} for localizing the transmitter and the receiver of the signal. A top-down view of the experiment setup is included in Figure  \ref{equipment_diagram}. The diameters and positions of $\Omega$ and {\bf A}--{\bf C} can be found in Table \ref{objdetails}. 

\subsubsection{Signal}

\begin{figure*}[!]
\begin{scriptsize}
\begin{center}
\begin{minipage}{4.7cm}
\begin{center}
\includegraphics[width=4.5cm]{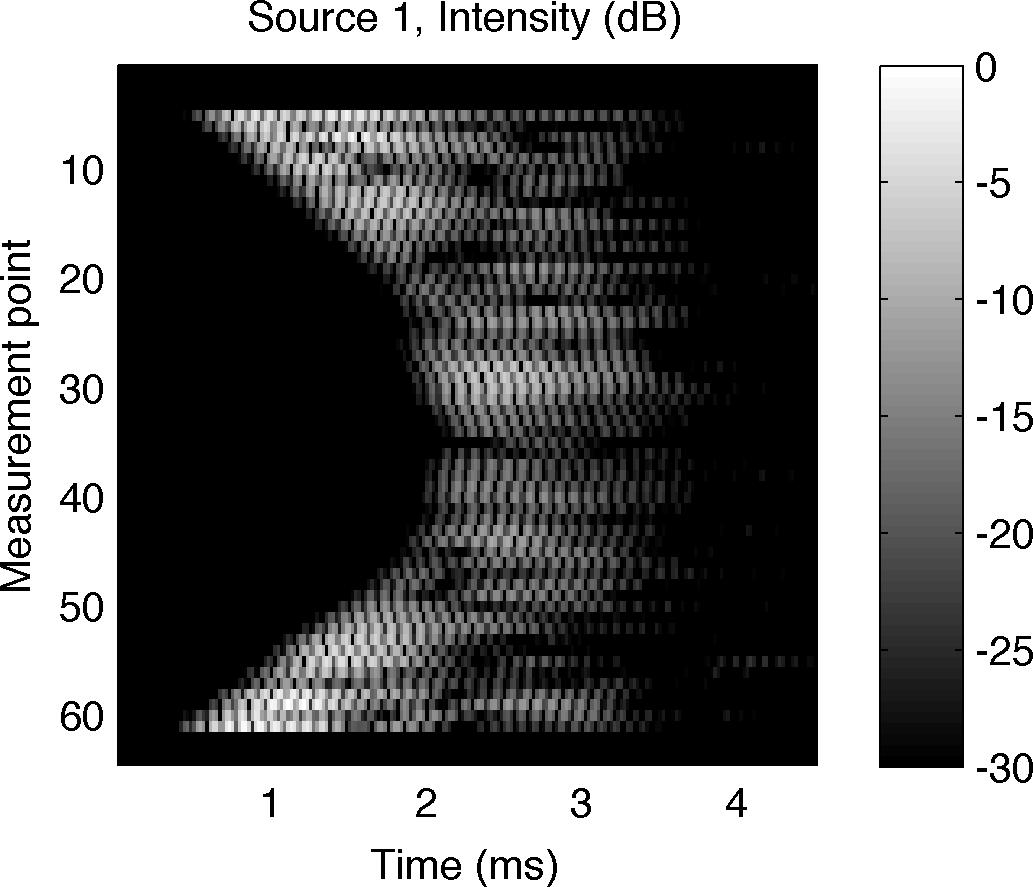} \\   {\bf 1}
\end{center}
\end{minipage} \hskip0.2cm
\begin{minipage}{4.7cm}
\begin{center}
\includegraphics[width=4.5cm]{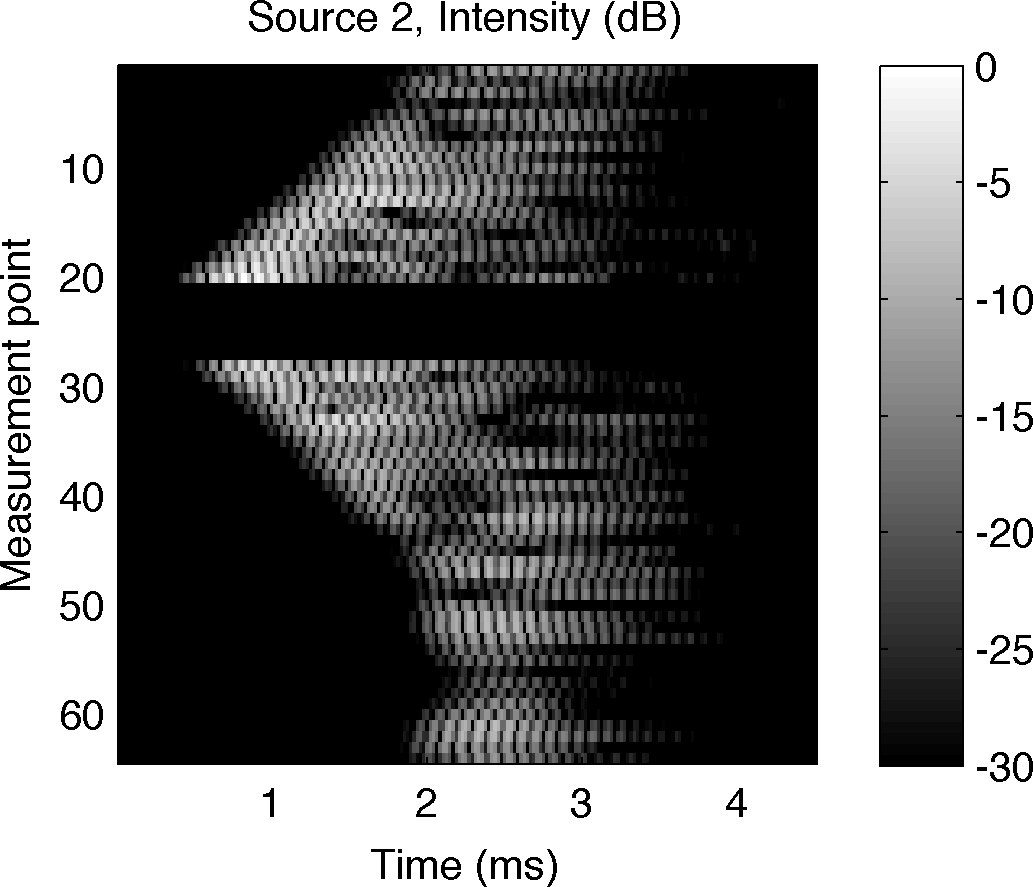} \\   {\bf 24}
\end{center}
\end{minipage} \hskip0.2cm
\begin{minipage}{4.7cm}
\begin{center}
\includegraphics[width=4.5cm]{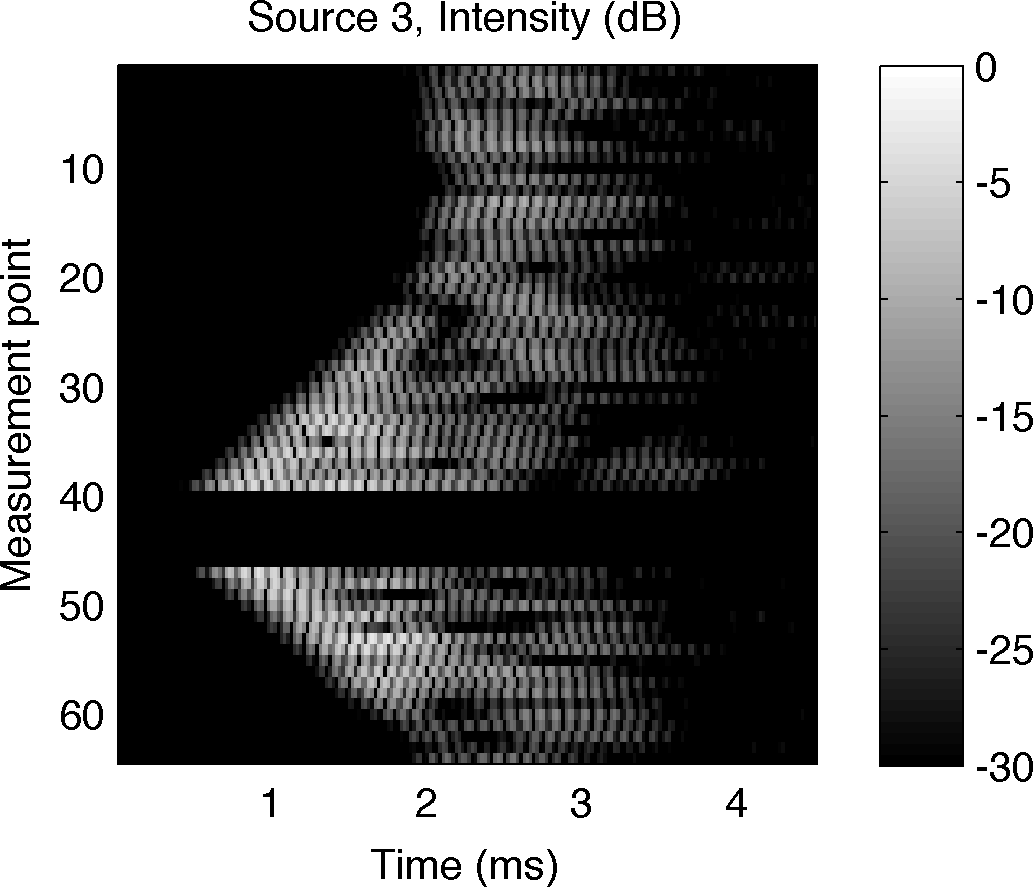} \\   {\bf 43}
\end{center}
\end{minipage} 
\end{center}
\end{scriptsize}
\caption[Measurement data]{Measurement data for transmitter locations {\bf 1}, {\bf 24} and {\bf 43} visualized on a decibel (dB) scale.\label{measurement_data}}
\end{figure*}

The signal pulse  transmitted (Figure \ref{pulsefigures}) was of the form 
\begin{equation}
\label{signal_pulse}
f(t) = \sin(36 t) \exp[-35(t - 0.60)^2] \hbox{ for }  t = [0, 1.2] \hbox{ (milliseconds)},
\end{equation}
and $f(t) = 0$, in all other times. The resulting signal with 5.8 kHz center frequency was transmitted from the points  {\bf 1}, {\bf 24} and {\bf 43} using an active two-way  speaker. Corresponding to each point of transmission, the signal was recorded for the 57 centermost control points opposite the transmitter. Based on the measurements, two travel-time data sets ({\bf I})  and ({\bf II})  (Figure \ref{equipment_diagram}) were formed. In ({\bf I}) (dense set),  the 47 centermost points were utilized in the final data set consisting of $3 \times 47 = 141$  individual travel-time values, while only every other point of the set ({\bf I}) was included in ({\bf II}) (sparse set) resulting in $3 \times 24 = 72$ travel-time values. The measured waveform data can be found in Figure \ref{measurement_data}.  The simulated data were obtained using the FDTD method using the constant $334$ m/s as the signal velocity (the speed of sound in air at 20 $\mbox{}^\circ$C). Both the measured and simulated data have been included as supplementary material.

\subsubsection{Equipment}

The data recording device was an ordinary laptop computer equipped with an external sound card interface. Two-channel audio data was recorded using the waveform audio file format (WAV), 24 bit  resolution and a 48 kHz sample rate. To eliminate any possible hardware or software based delays, a control pulse was recorded by one microphone placed 3.5 cm directly above the speaker. The actual data pulse was recorded with the other microphone positioned on the perimeter of the target area in the radial direction. Figure \ref{equipment_diagram} shows the test setup with both microphones near each other. The technical data of the hardware used in the setup can be found in Table \ref{technicaldata}.

\subsubsection{Noise}

Figure \ref{pulsefigures} includes a comparison between the transmitted signal pulse and the one  received at the perimeter of the (empty) test domain $\Omega$ without the foam cylinders placed inside. The tail of the signal received was observed to contain noise due to echoes and  inaccuracies in transmission (ringing). Based on a comparison with the original signal the noise peaks were estimated to be mainly 10 dB (greyed area) below the main peak.

\subsubsection{Relevance}

The relevance of the test setup with respect to a real travel-time tomography application is the following. The transmitter sends a waveform signal pulse at a known position which is then recorded by the receiver in a known position, and the resulting signal recordings are then sent via a communication link. Further processing will consist of compression or some other processing technique, such as calculating travel-time values. 

\subsection{Experimental FPGA hardware}

\begin{figure}[!]
  \begin{center}
    \includegraphics[width=8cm]{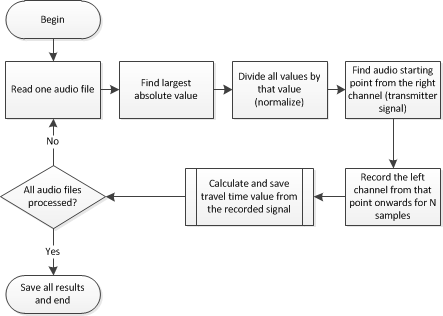}
  \end{center}
  \caption[Matlab script flow chart]{A simplified high-level flow chart of the travel-time calculation algorithm.}
  \label{matlab_high_level}
\end{figure}

The travel-time calculation was implemented using a high-level synthesis of hardware on an FPGA development board (Altera DE2) with the typical performance of an embedded signal acquisition system. FPGA  enables fast  processing of data which is essential in waveform imaging because of the massive data input needed to record a complete wave. A high-level data flow chart of the implementation has been included in Figure \ref{matlab_high_level}. 

In this study, we investigated different bit resolution, threshold and normalization levels. These were set in a separate parameters file. A functioning Matlab script was first transformed directly into C code and the result was then further modified to accept parameters. The hardware was generated by using the Mentor Graphics \textit{Catapult} High-Level Synthesis (HLS) tool \cite{calyptocatapult}. HLS is a method where digital hardware is generated from a high level programming language, such as C \cite{coussy2009}. The tool takes the modified, algorithmic C code and generates register transfer level (RTL) code, which is synthesized as digital logic on an Field Programmable Gate Array (FPGA) chip \cite{grout2011digital}.

The basic hardware for signed 16-bit input (audio) data was implemented first for ITT and TTT, and then adapted separately for the signed 8-bit input data. The software on computer uses common datatypes \cite{matlabdatatypes}, but efficient hardware requires accurately defined bit resolutions for all inputs, outputs and intermediate variables \cite{hlsbluebook}. The HLS tool does not offer a way to calculate the maximum bit resolution for an arbitary integer number. An example of this is the divisor in Equation (\ref{traveltimeintegration}), which has the sum of the squares of the numbers for a part of the signal. Knowing the bit resolution of the element and thus the maximum values, the bit resolution for the square could be calculated.  The 8-bit and 16-bit versions of the hardware required individual optimizations.

\subsection{Numerical experiments}

In the numerical experiments, the initial guess  for the signal velocity distribution was set to be $\mathtt{c} = 334$ m/s (the speed of sound in air at 20 $\mbox{}^\circ$ C).  The number of inverse iteration steps was set at three and the regularization parameters $\alpha$ and $\beta$ were both given the value 0.01 which, based on our preliminary tests, is a reasonable approximaitoion of the midpoint of  the interval of the workable values.  To evaluate the data processing artifacts, bit resolutions of the signed 16 and the signed 8 bits were used as in the A/D conversion units of practical applications \cite{leucci2008}. 

The maximum amplitude in the measurement data set was normalized to a given level $\nu$ dB FS (full of the bit scale) and the other signals proportional thereto.  The control signal was normalized to 0 dB FS for each individual measurement point. 16-bit preprocessing was evaluated at 100 \% ($\nu = 0$ dB FS) normalization and 8-bit at 100 \% ($\nu = 0$ dB FS) and 6 \% ($\nu = -24$ dB FS) normalization.  The 6 \% level represents an extreme case where the signal noise is large, thus simulating either a weak reception of transmitted signals or an insensitive receiver. Two different threshold levels 90 \% and 70 \% for initial signal detection were tested corresponding to around -1 dB FS and -3 dB FS of the maximum value of a normalized signal. 

The data vector ${\bf y}$ was obtained as the difference between the measurement and simulation based travel-time both of which were computed using either the ITT or TTT approach. The interval of the ITT was determined by $\tau_1 = 5$ and $\tau_2 = 250$ (samples at 48 kHz) resulting in a total length of 256 samples (5.3 ms).  TTT was evaluated using two alternative strategies TTT 1 and TTT 2. In the former, ITT was applied to the control pulse and TTT to the  data pulse. In the latter, the travel-time of both the control and data pulse was computed via TTT. The motivation for investigating TTT 1, was the potential situation in which the simulated and measured travel-time are obtained via different techniques, e.g., due to different suppliers of computer software and measurement equipment. 

The reconstructions were analyzed by measuring the Relative Overlapping Area (ROA) and the minimum relative overlap $\hbox{ROA}_{\hbox{\scriptsize min}}$ between the foam cylinders $\mathcal{S}_{\hbox{\footnotesize \bf A}}$, $\mathcal{S}_{\hbox{\footnotesize \bf B}}$, $\mathcal{S}_{\hbox{\footnotesize \bf C}}$  the set $\mathcal{S}_{\hbox{\footnotesize  rec}}$  in which the value of the reconstruction was  less than a fixed limit such that $\hbox{Area}(\mathcal{S}_{\hbox{\footnotesize rec}}) = \hbox{Area}(\mathcal{S}_{\hbox{\footnotesize \bf A}} \cup \mathcal{S}_{\hbox{\footnotesize \bf B} } \cup \mathcal{S}_{{\hbox{\footnotesize \bf C }}})$. ROA  and $\hbox{ROA}_{\hbox{\scriptsize min}}$ were calculated as given by the equations
{\setlength\arraycolsep{2 pt}\begin{eqnarray}
\hbox{ROA} & = & \frac{\hbox{Area}(\mathcal{S}_{\hbox{\footnotesize rec}} \cap [\mathcal{S}_{\hbox{\footnotesize \bf A}} \cup \mathcal{S}_{\hbox{\footnotesize \bf B} } \cup \mathcal{S}_{{\hbox{\footnotesize \bf C }}}])}{\hbox{Area}(\mathcal{S}_{\hbox{\footnotesize \bf A}} \cup \mathcal{S}_{\hbox{\footnotesize \bf B} } \cup \mathcal{S}_{{\hbox{\footnotesize \bf C }}})} \\
\hbox{ROA}_{\hbox{\scriptsize min}} & = & \min \Big( \frac{\hbox{Area}(\mathcal{S}_{\hbox{\footnotesize rec}} \cap \mathcal{S}_{\hbox{\footnotesize \bf A}})}{\hbox{Area}(\mathcal{S}_{\hbox{\footnotesize \bf A}})}, \frac{\hbox{Area}(\mathcal{S}_{\hbox{\footnotesize rec}} \cap \mathcal{S}_{\hbox{\footnotesize \bf B}})}{\hbox{Area}(\mathcal{S}_{\hbox{\footnotesize \bf B}})}, \nonumber \\ & &   \hphantom{\min \Big( } \frac{\hbox{Area}(\mathcal{S}_{\hbox{\footnotesize rec}} \cap \mathcal{S}_{\hbox{\footnotesize \bf C}})}{\hbox{Area}(\mathcal{S}_{\hbox{\footnotesize \bf C}})} \Big).
\end{eqnarray}}
The reconstructions were computed using a laptop computer equipped with the 2.8 GHz Intel Core i7 processor 2640M and 8 GB of RAM. Computing a single reconstruction took 9 seconds of CPU time. 

\section{Inversion results}

\begin{figure*}[t]
\begin{scriptsize}
\begin{center}
\begin{framed}
Threshold 90 \% (-1 dB FS) \\ \mbox{} \vskip-0.1cm
\begin{minipage}{5.0cm}
\begin{center}
\begin{framed}
ITT, 16-bit, Norm.\ 100 \% (0 dB FS) \\ \mbox{} \vskip-0.1cm
\includegraphics[width=2.0cm]{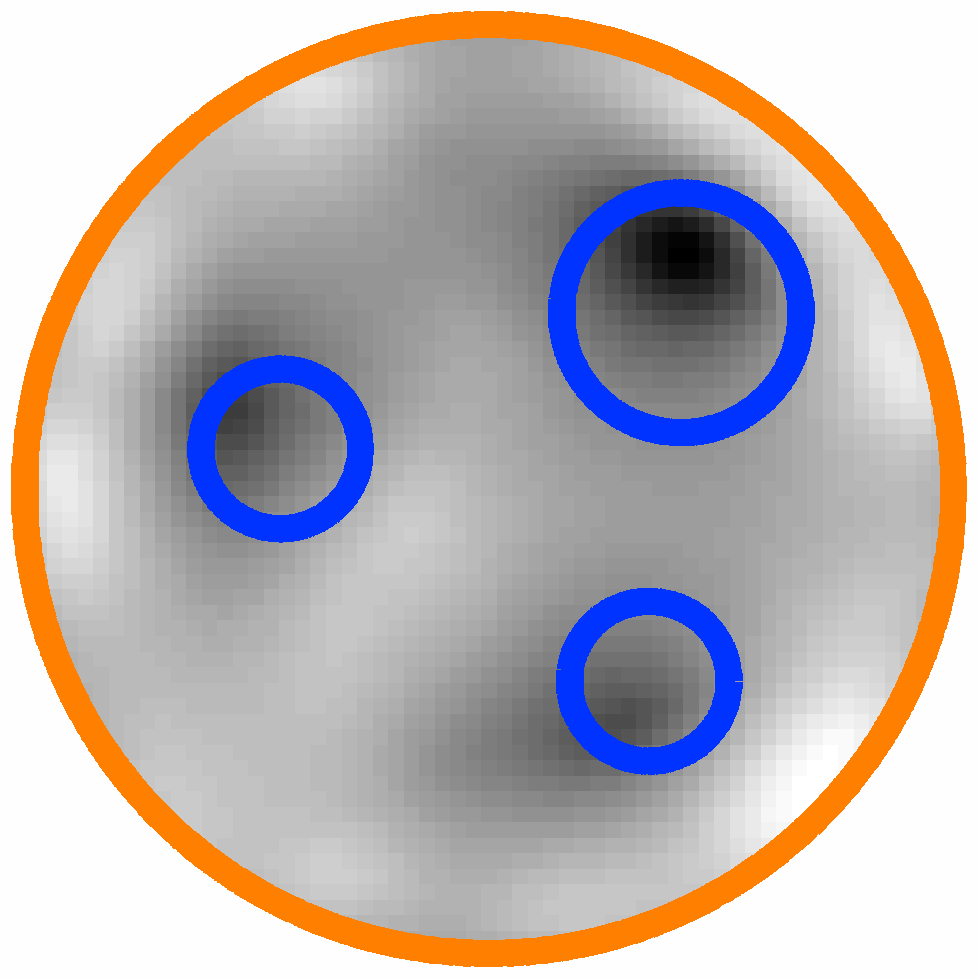} \hskip0.1cm
\includegraphics[width=2.0cm]{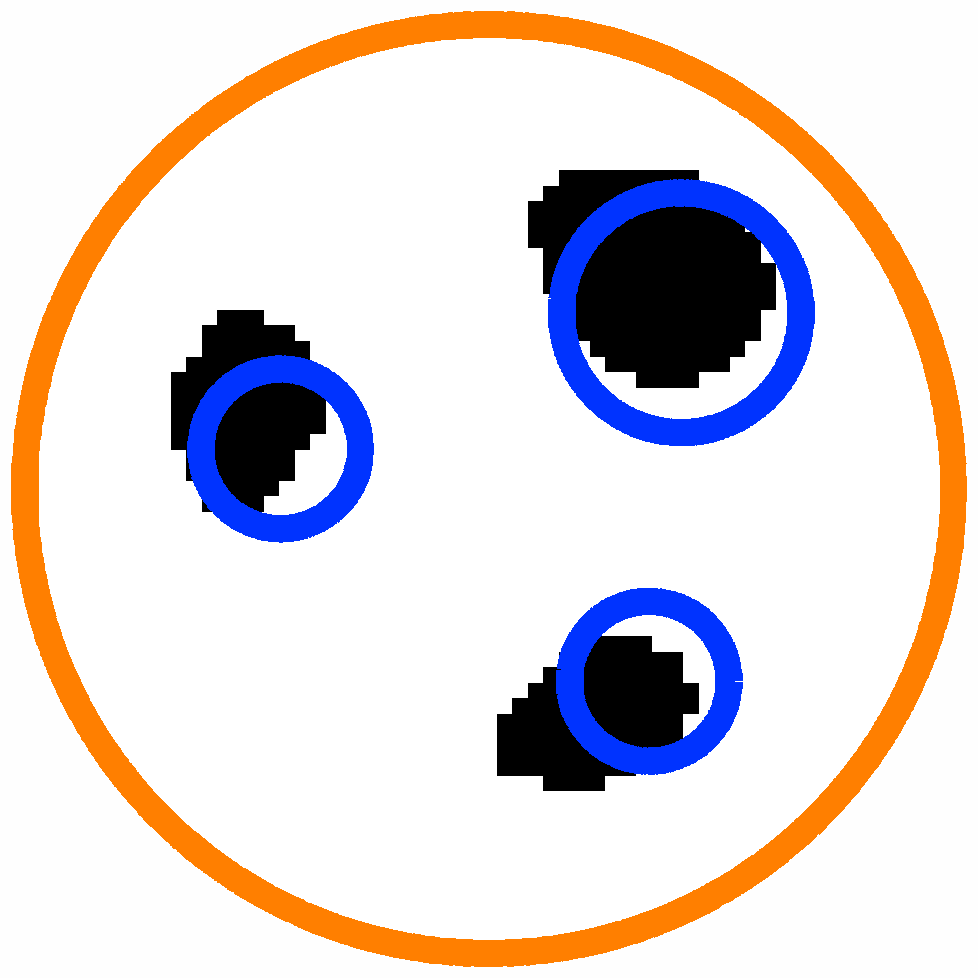} \\
ROA = 63 \%, \\ ROA$\mbox{}_{{\min}}$ = 54 \%
\end{framed}
\end{center}
\end{minipage} \hskip0.2cm
\begin{minipage}{5.0cm}
\begin{center}
\begin{framed}
ITT, 8-bit, Norm.\ 100 \% (0 dB FS) \\ \mbox{} \vskip-0.1cm
\includegraphics[width=2.0cm]{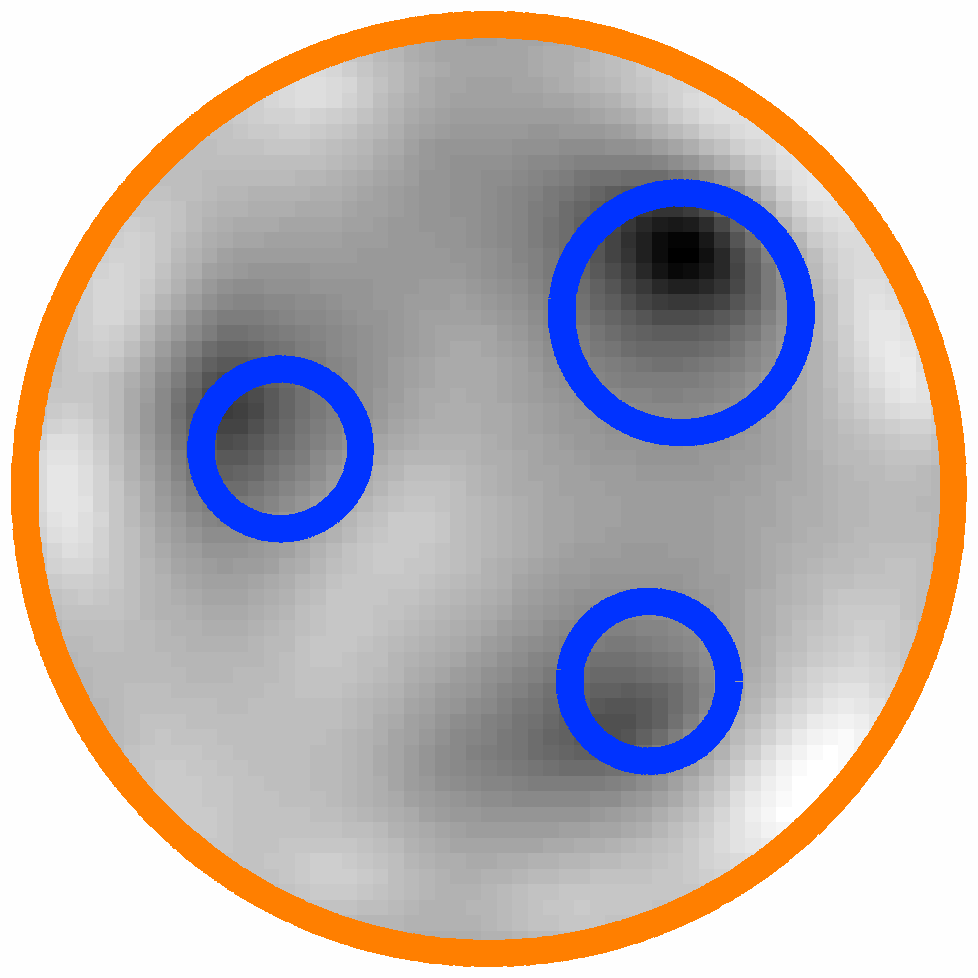} \hskip0.1cm
\includegraphics[width=2.0cm]{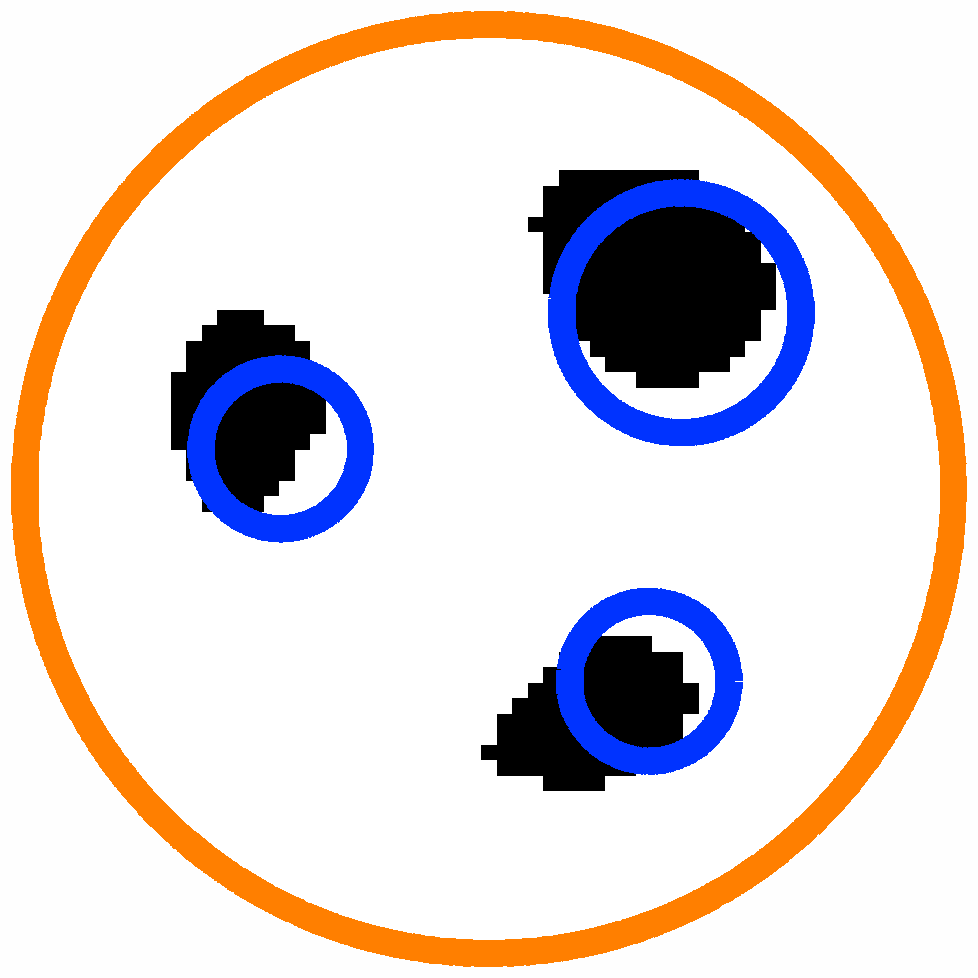} \\
ROA = 63 \%, \\ ROA$\mbox{}_{{\min}}$ = 54 \%
\end{framed}
\end{center}
\end{minipage} \hskip0.2cm
\begin{minipage}{5.0cm}
\begin{center}
\begin{framed}
ITT, 8-bit, Norm.\ 6 \% (-24 dB FS)  \\ \mbox{} \vskip-0.1cm
\includegraphics[width=2.0cm]{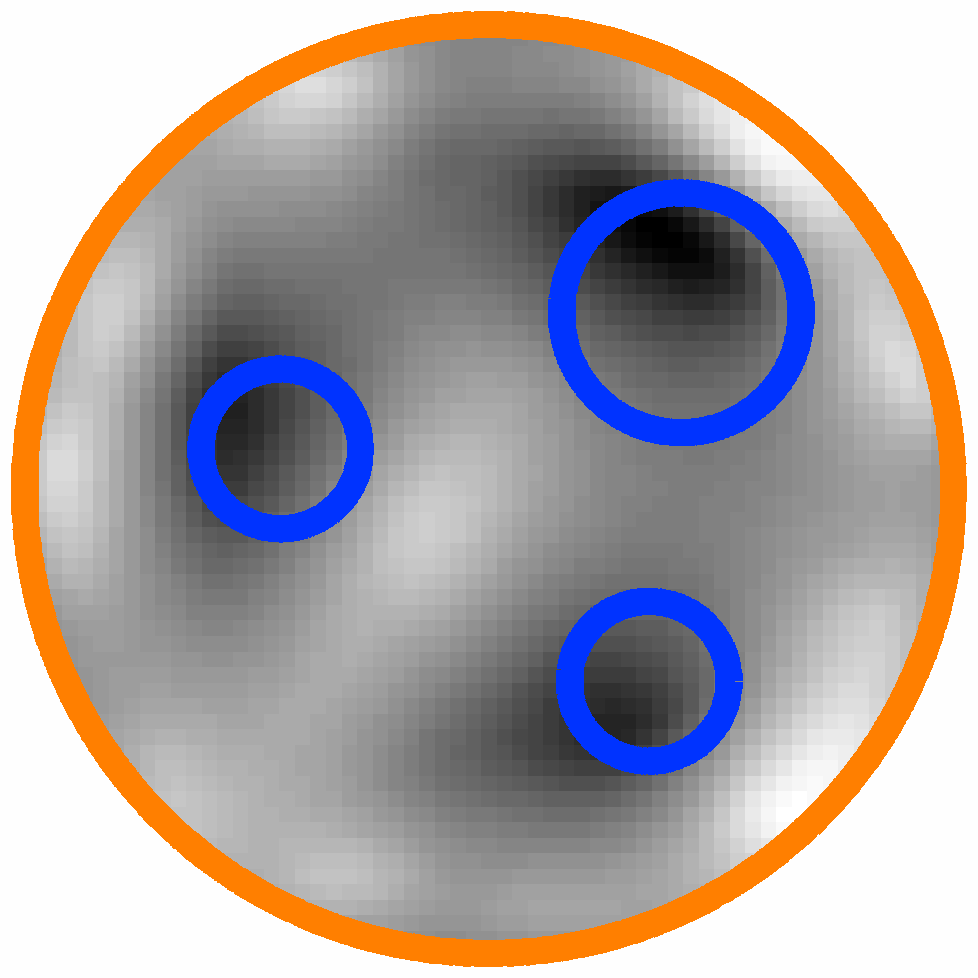} \hskip0.1cm
\includegraphics[width=2.0cm]{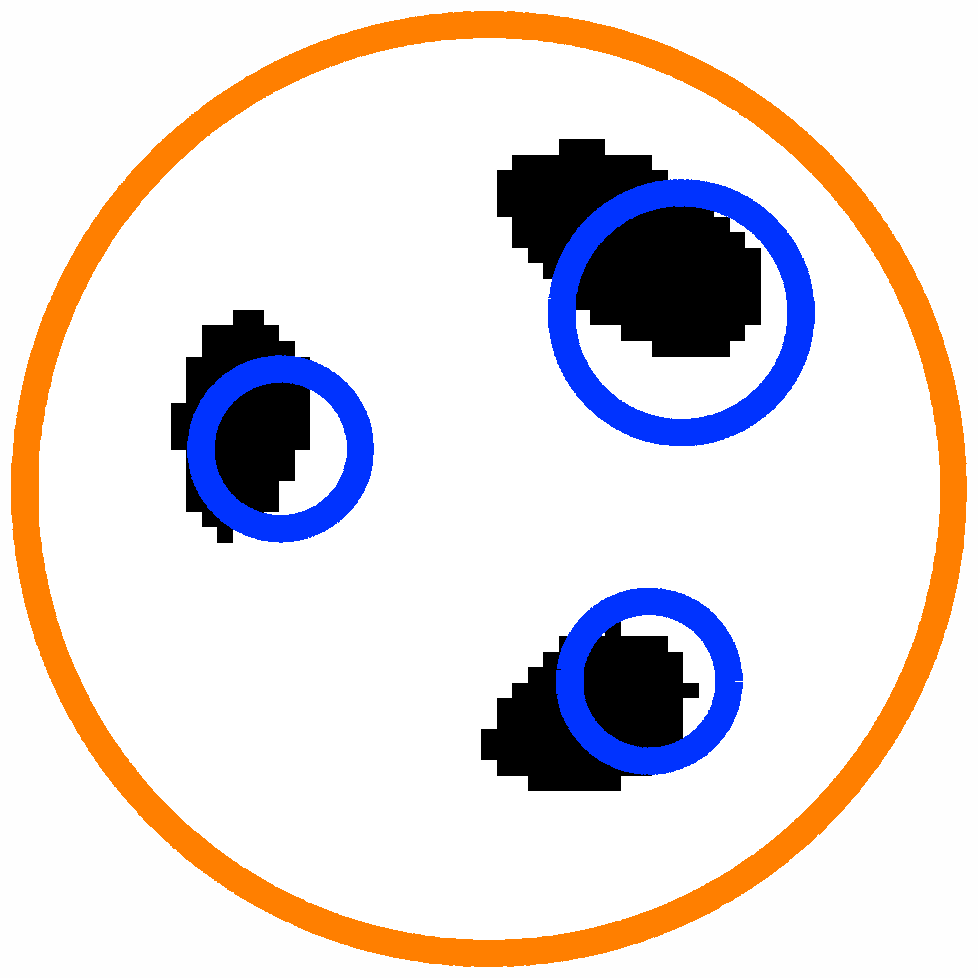} \\
ROA = 54 \%, \\ ROA$\mbox{}_{{\min}}$ = 53 \%
\end{framed}
\end{center}
\end{minipage} \end{framed} \mbox{} \vspace{-0.5cm}
\begin{framed}
Threshold 70 \% (-3 dB FS) \\ \mbox{} \vskip-0.1cm
\begin{minipage}{5.0cm}
\begin{center}
\begin{framed}
ITT, 16-bit, Norm.\ 100 \% (0 dB FS) \\ \mbox{} \vskip-0.1cm
\includegraphics[width=2.0cm]{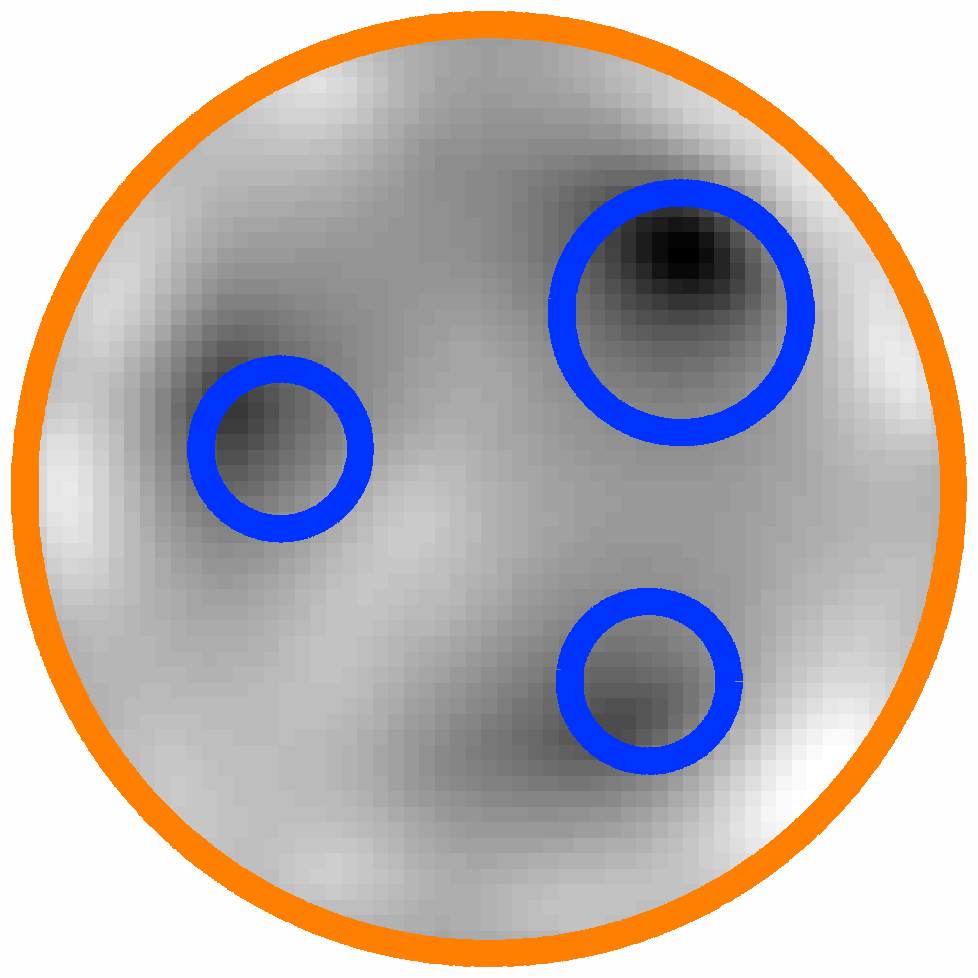} \hskip0.1cm
\includegraphics[width=2.0cm]{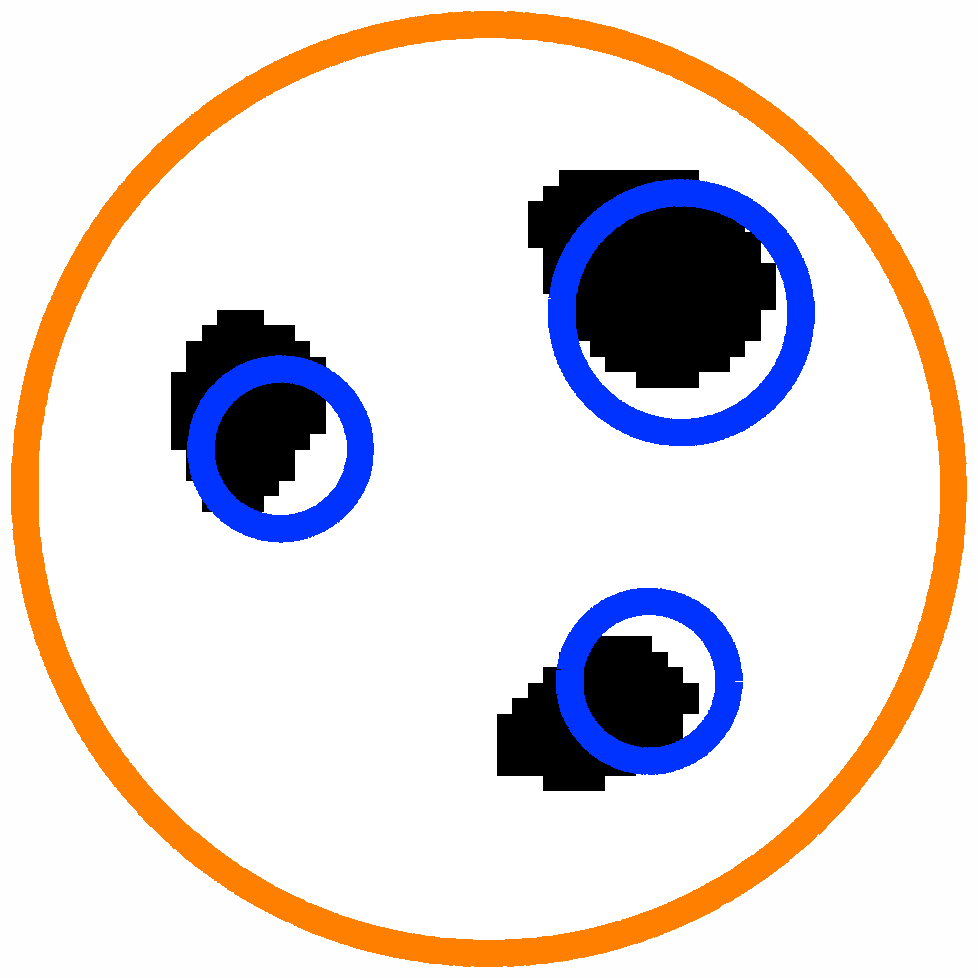} \\
ROA = 63 \%, \\ ROA$\mbox{}_{{\min}}$ = 53 \%
\end{framed}
\end{center}
\end{minipage} \hskip0.2cm
\begin{minipage}{5.0cm}
\begin{center}
\begin{framed}
ITT, 8-bit, Norm.\ 100 \% (0 dB FS) \\ \mbox{} \vskip-0.1cm
\includegraphics[width=2.0cm]{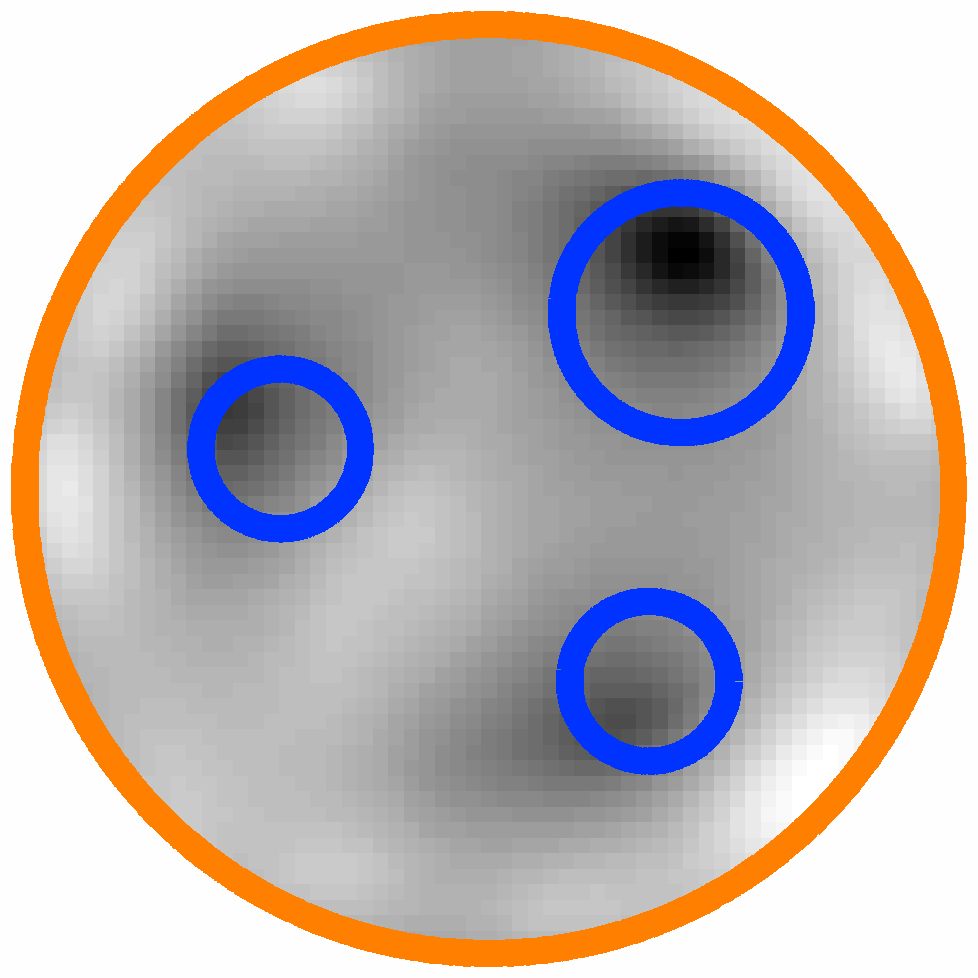} \hskip0.1cm
\includegraphics[width=2.0cm]{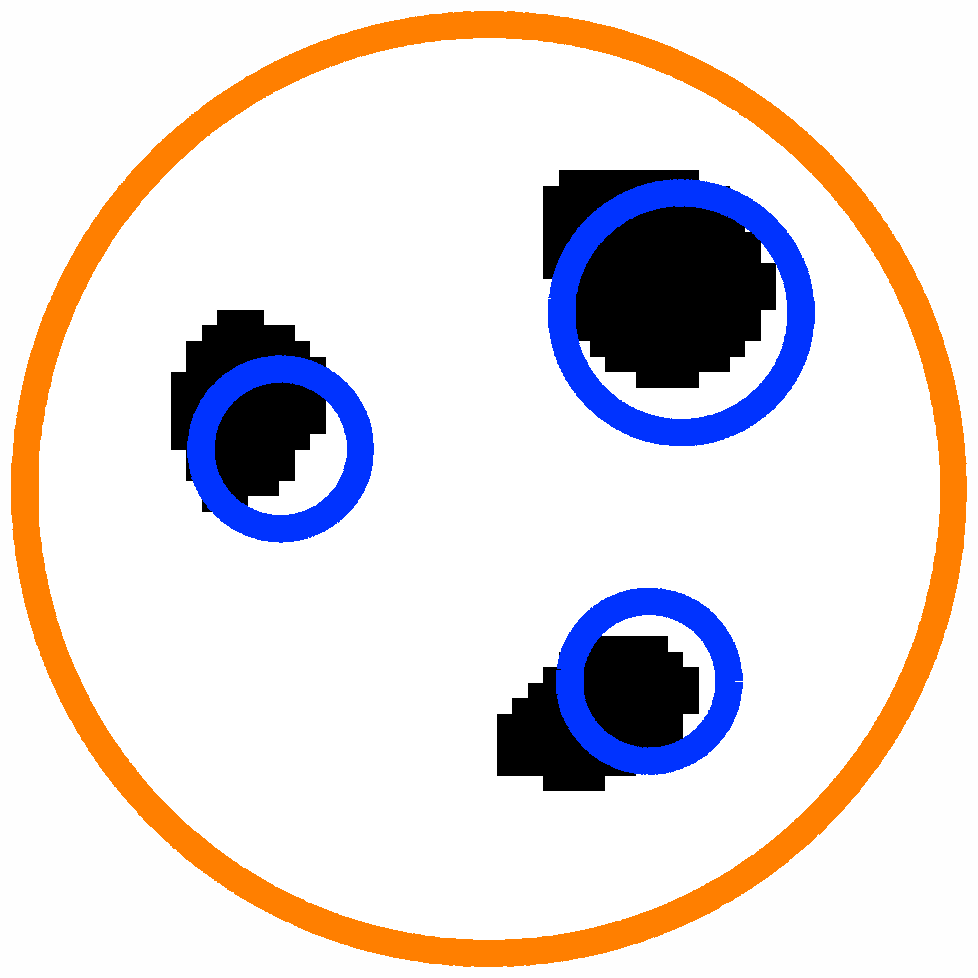} \\
ROA = 64 \%, \\ ROA$\mbox{}_{{\min}}$ = 56 \%
\end{framed}
\end{center}
\end{minipage} \hskip0.2cm
\begin{minipage}{5.0cm}
\begin{center}
\begin{framed}
ITT, 8-bit, Norm.\ 6 \% (-24 dB FS) \\ \mbox{} \vskip-0.1cm
\includegraphics[width=2.0cm]{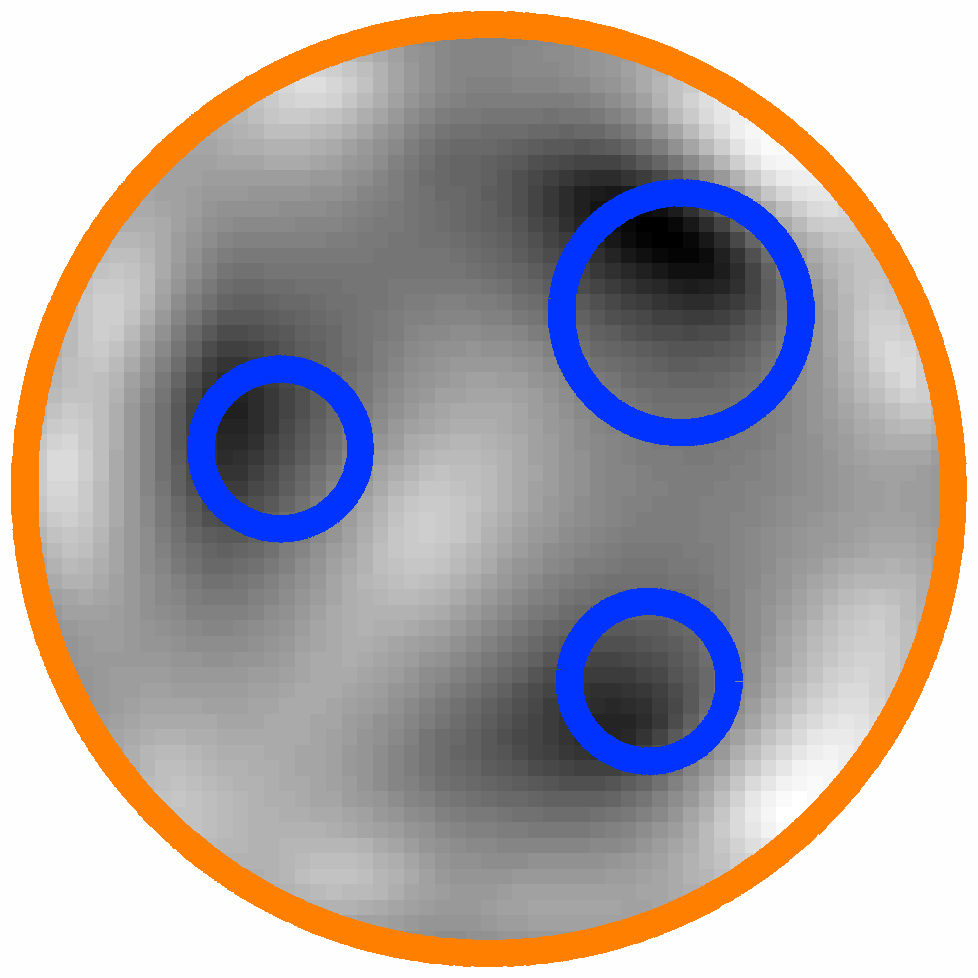} \hskip0.1cm
\includegraphics[width=2.0cm]{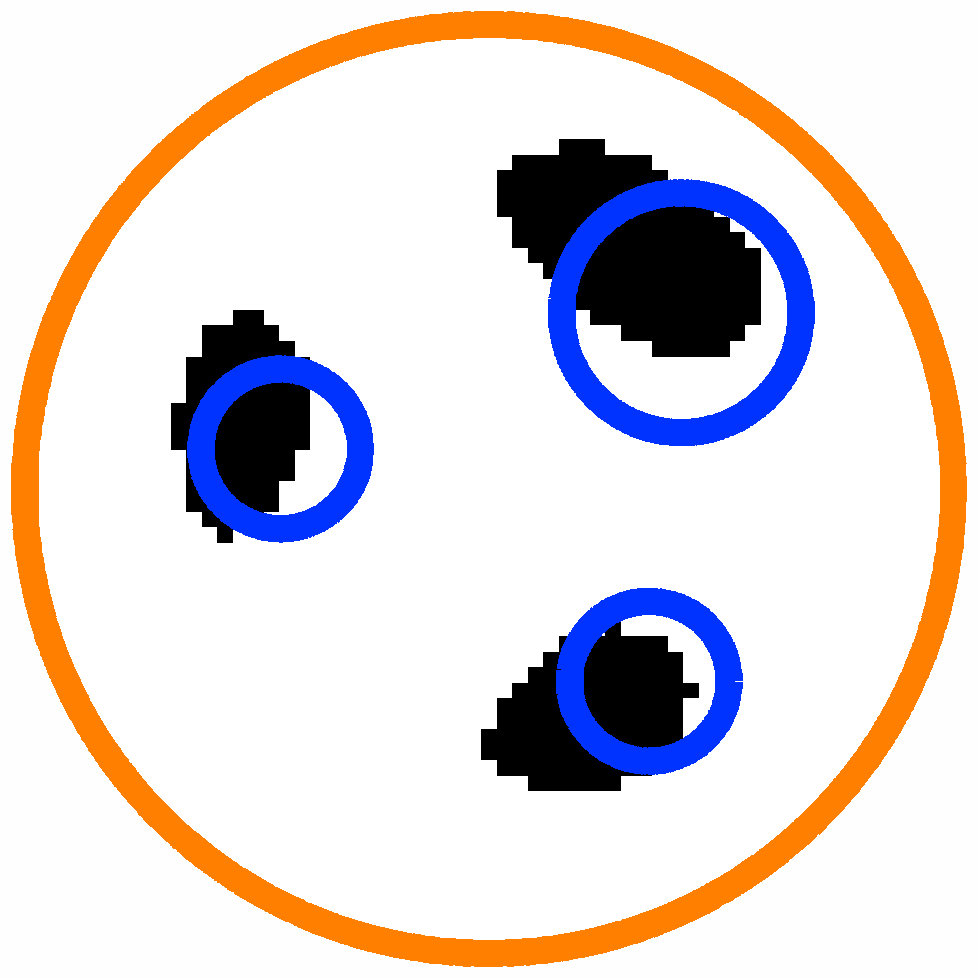} \\
ROA = 54 \%, \\ ROA$\mbox{}_{{\min}}$ = 53 \%
\end{framed}
\end{center}
\end{minipage} \end{framed}
\end{center}
\end{scriptsize}
\caption[Integrated/integrated travel time results]{Results with Integrated travel time (ITT) used for both the control signal and measured signal.}
\label{ittfigure}
\end{figure*}

\begin{figure*}[t]
\begin{scriptsize}
\begin{center}
\begin{framed}
Threshold 90 \% (-1 dB FS) \\ \mbox{} \vskip-0.1cm
\begin{minipage}{5.0cm}
\begin{center}
\begin{framed}
TTT 1, 16-bit, Norm.\ 100 \% (0 dB FS) \\ \mbox{} \vskip-0.1cm
\includegraphics[width=2.0cm]{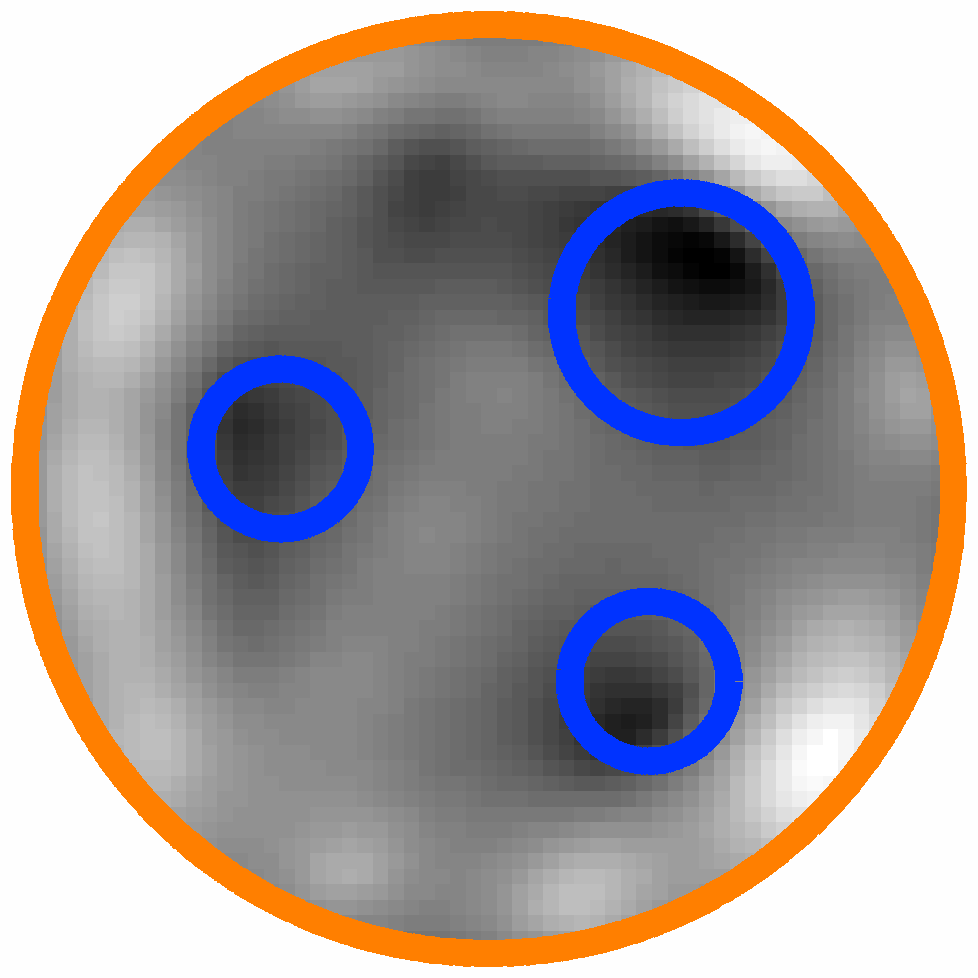} \hskip0.1cm
\includegraphics[width=2.0cm]{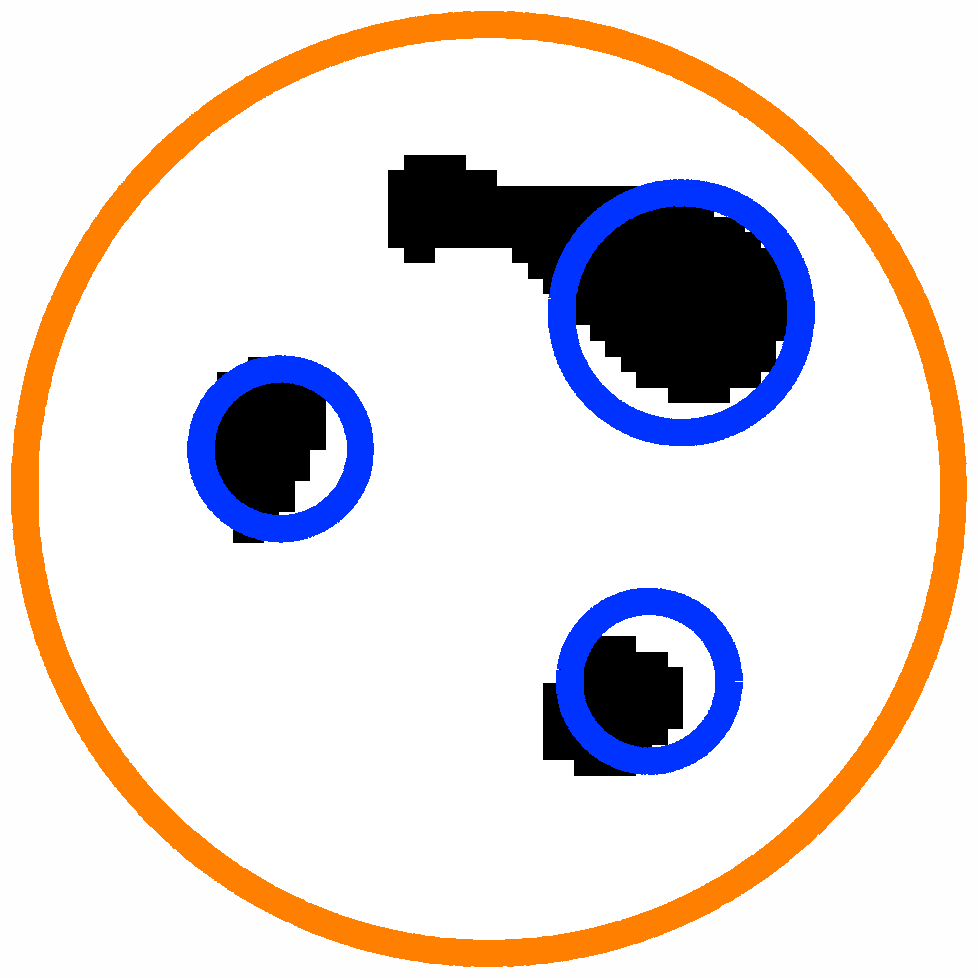} \\
ROA = 68 \%, \\ ROA$\mbox{}_{{\min}}$ = 47 \%
\end{framed} 
\end{center}
\end{minipage} \hskip0.2cm
\begin{minipage}{5.0cm}
\begin{center}
\begin{framed}
TTT 1, 8-bit, Norm.\ 100 \% (0 dB FS) \\ \mbox{} \vskip-0.1cm
\includegraphics[width=2.0cm]{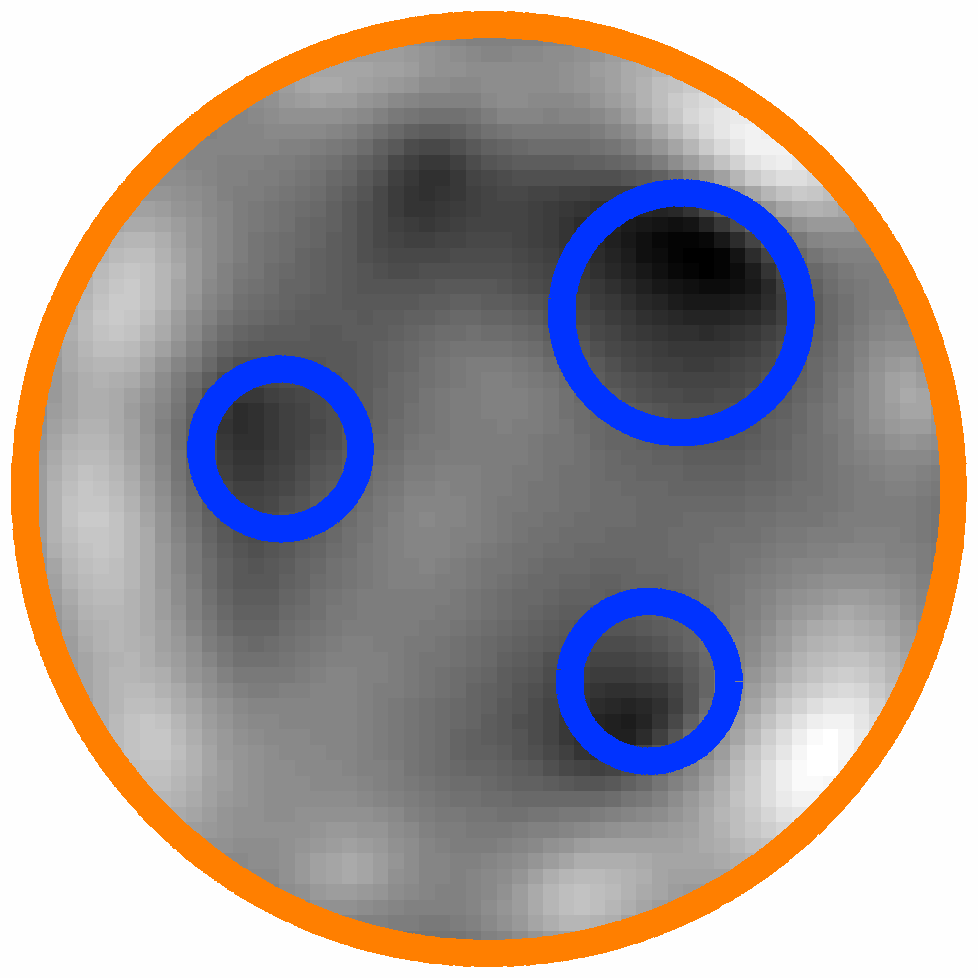} \hskip0.1cm
\includegraphics[width=2.0cm]{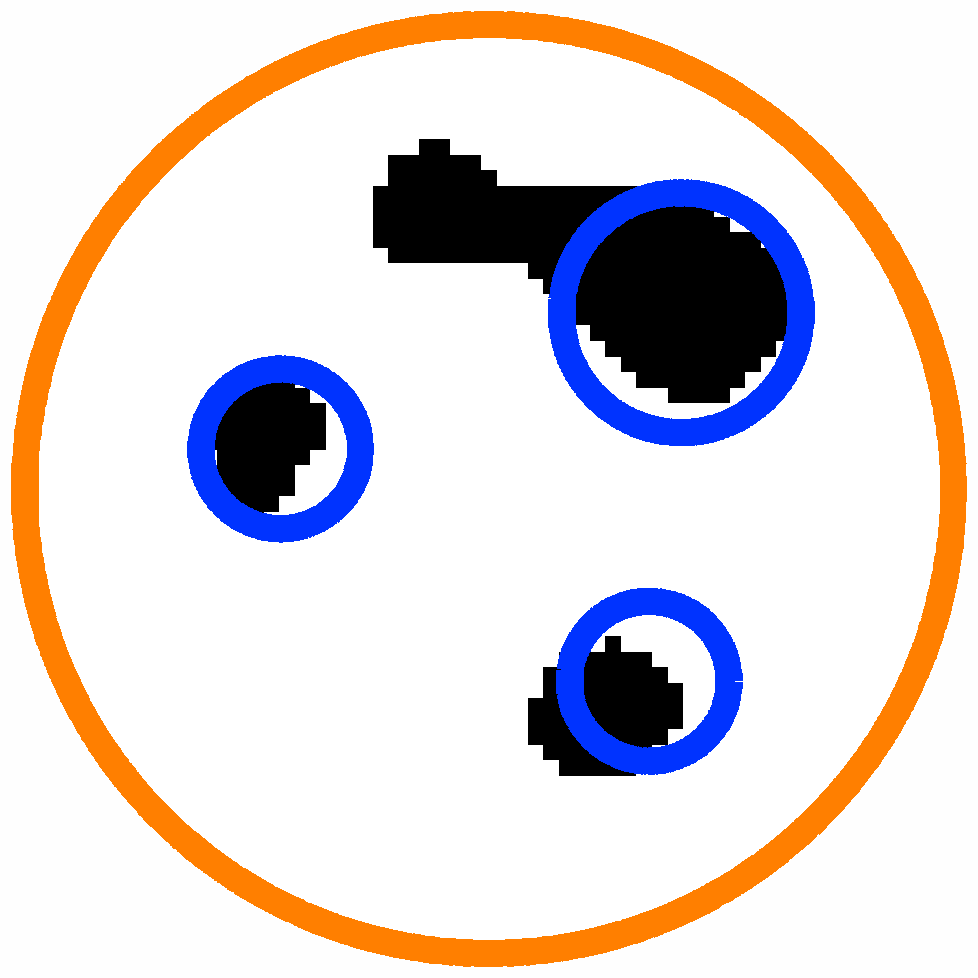} \\
ROA = 64 \%, \\ ROA$\mbox{}_{{\min}}$ = 42 \%
\end{framed}
\end{center}
\end{minipage} \hskip0.2cm
\begin{minipage}{5.0cm}
\begin{center}
\begin{framed}
TTT 1, 8-bit, Norm.\ 6 \% (-24 dB FS) \\ \mbox{} \vskip-0.1cm
\includegraphics[width=2.0cm]{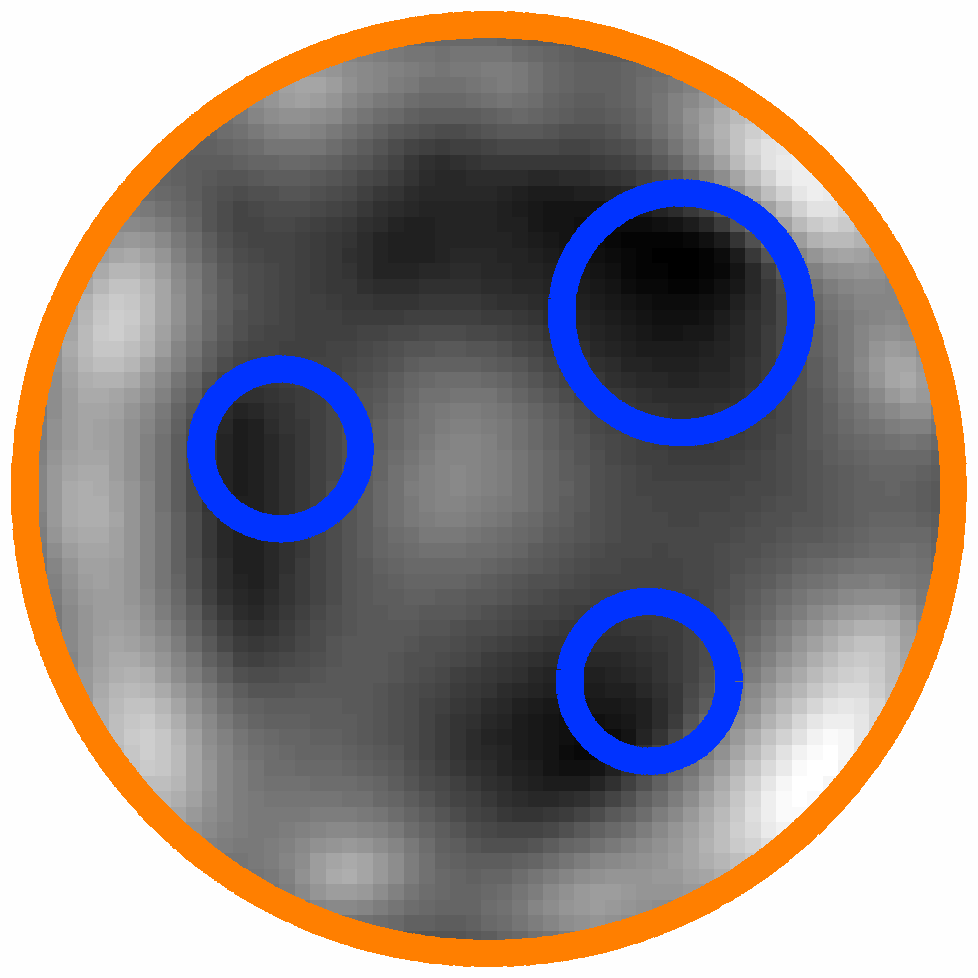} \hskip0.1cm
\includegraphics[width=2.0cm]{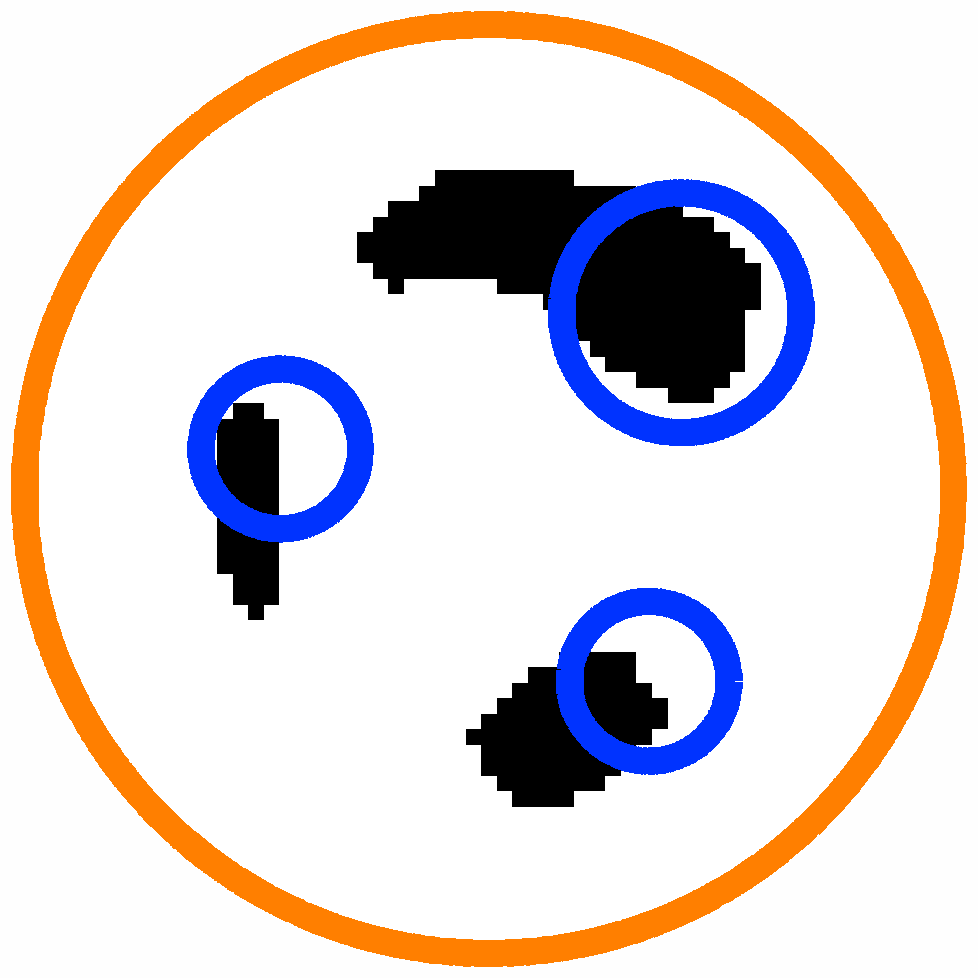} \\
ROA = 48 \%, \\ ROA$\mbox{}_{{\min}}$ = 31 \%
\end{framed}
\end{center}
\end{minipage} \end{framed} \mbox{} \vspace{-0.5cm}
\begin{framed}
Threshold 70 \% (-3 dB FS) \\ \mbox{} \vskip-0.1cm
\begin{minipage}{5.0cm}
\begin{center}
\begin{framed}
TTT 1, 16-bit, Norm.\ 100 \% (0 dB FS) \\ \mbox{} \vskip-0.1cm
\includegraphics[width=2.0cm]{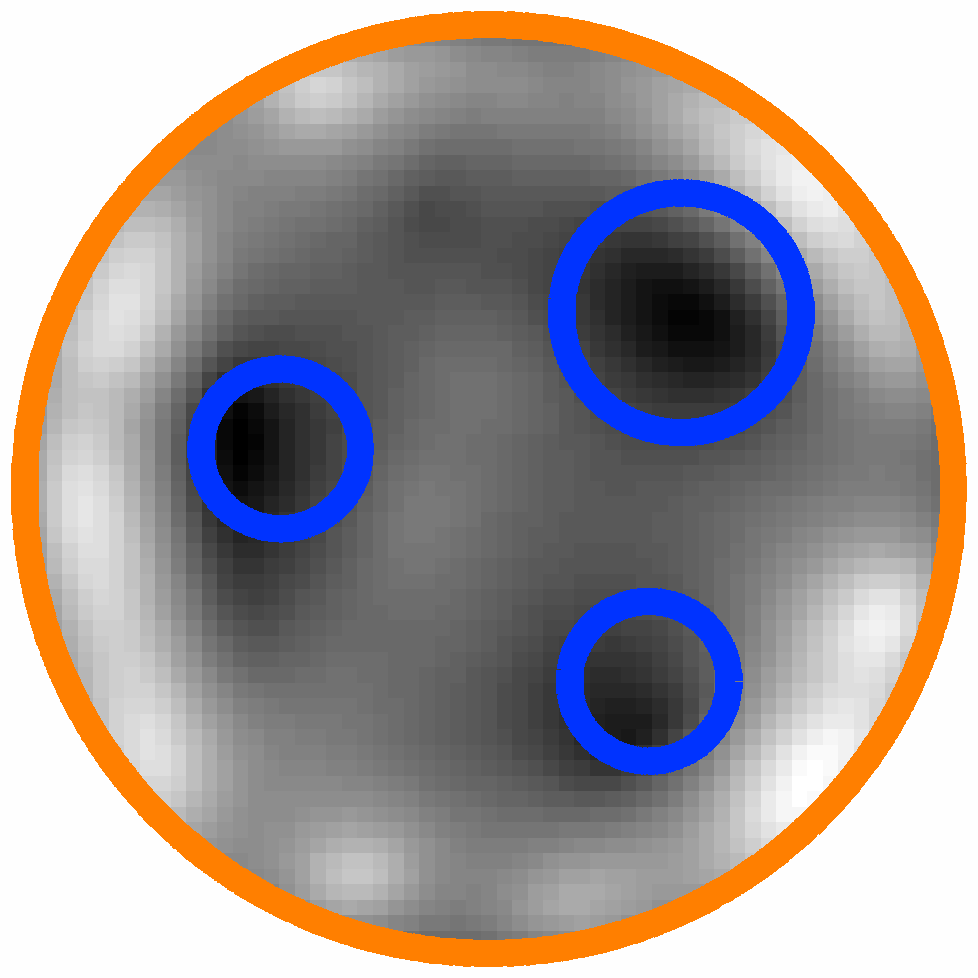} \hskip0.1cm
\includegraphics[width=2.0cm]{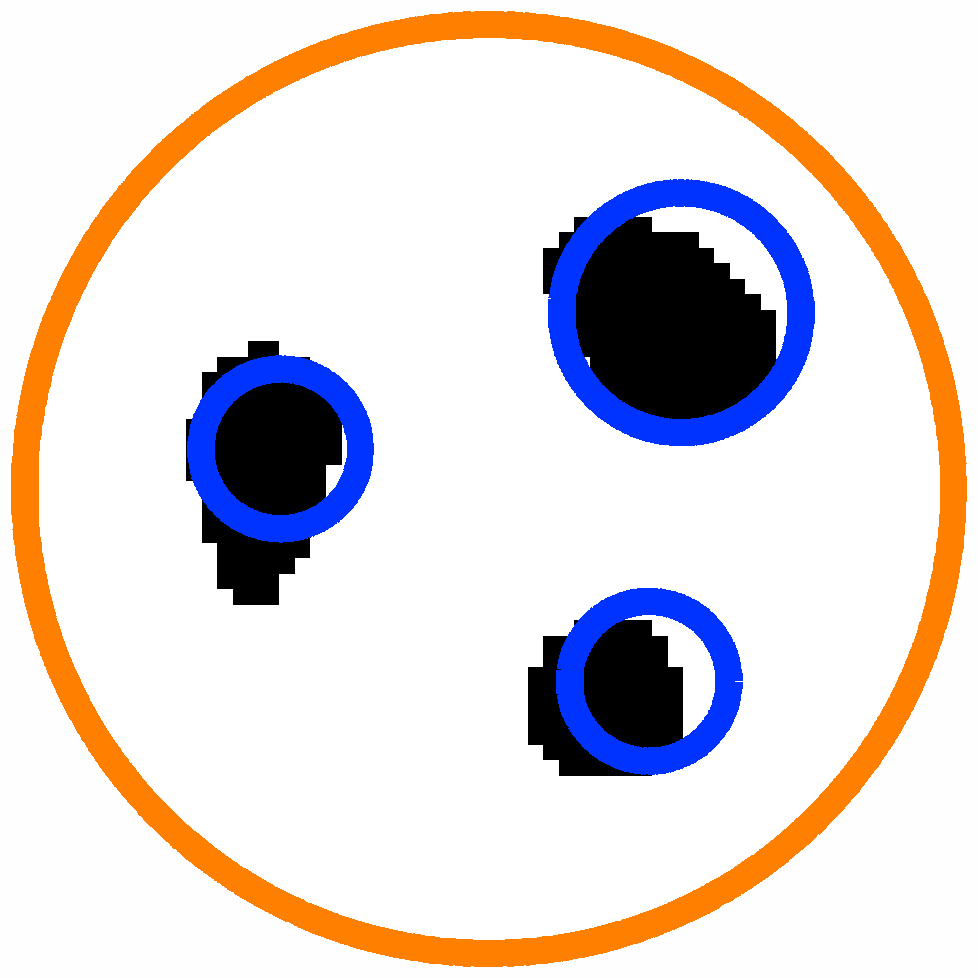} \\
ROA = 70 \%, \\ ROA$\mbox{}_{{\min}}$ =  56 \%
\end{framed}
\end{center}
\end{minipage} \hskip0.2cm
\begin{minipage}{5.0cm}
\begin{center}
\begin{framed}
TTT 1, 8-bit, Norm.\ 100 \% (0 dB FS) \\ \mbox{} \vskip-0.1cm
\includegraphics[width=2.0cm]{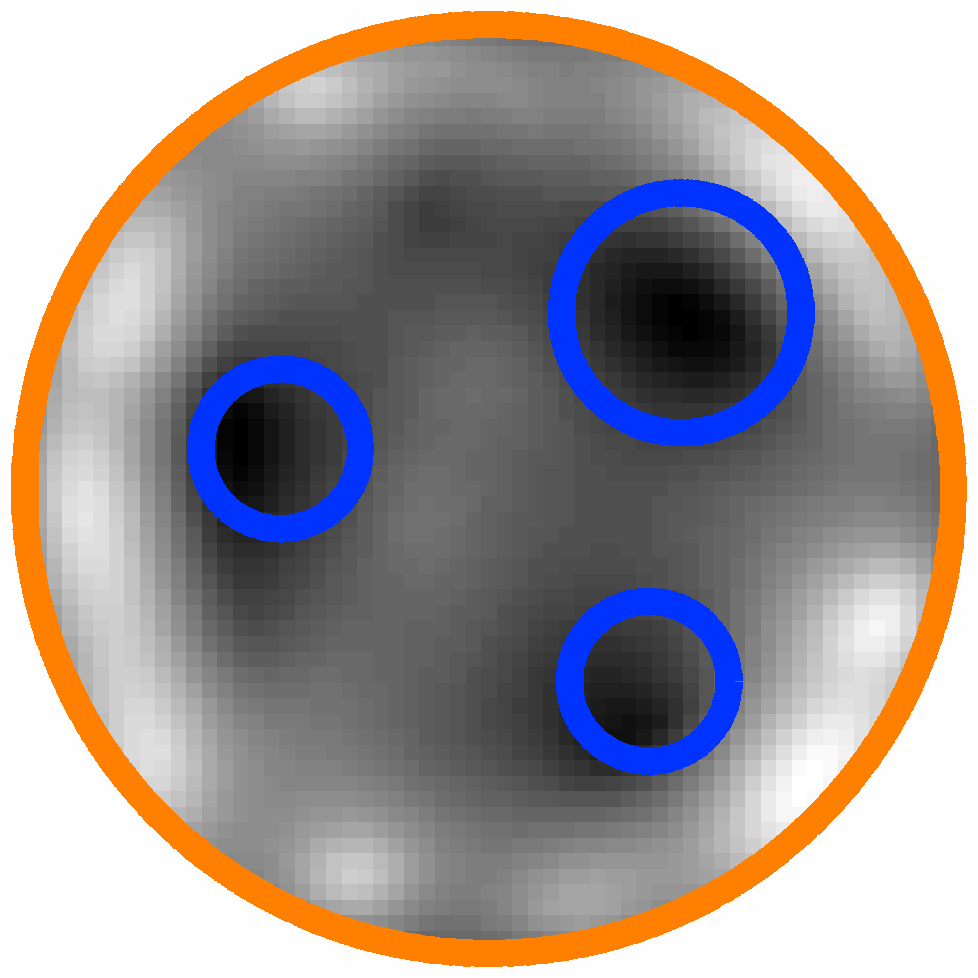} \hskip0.1cm
\includegraphics[width=2.0cm]{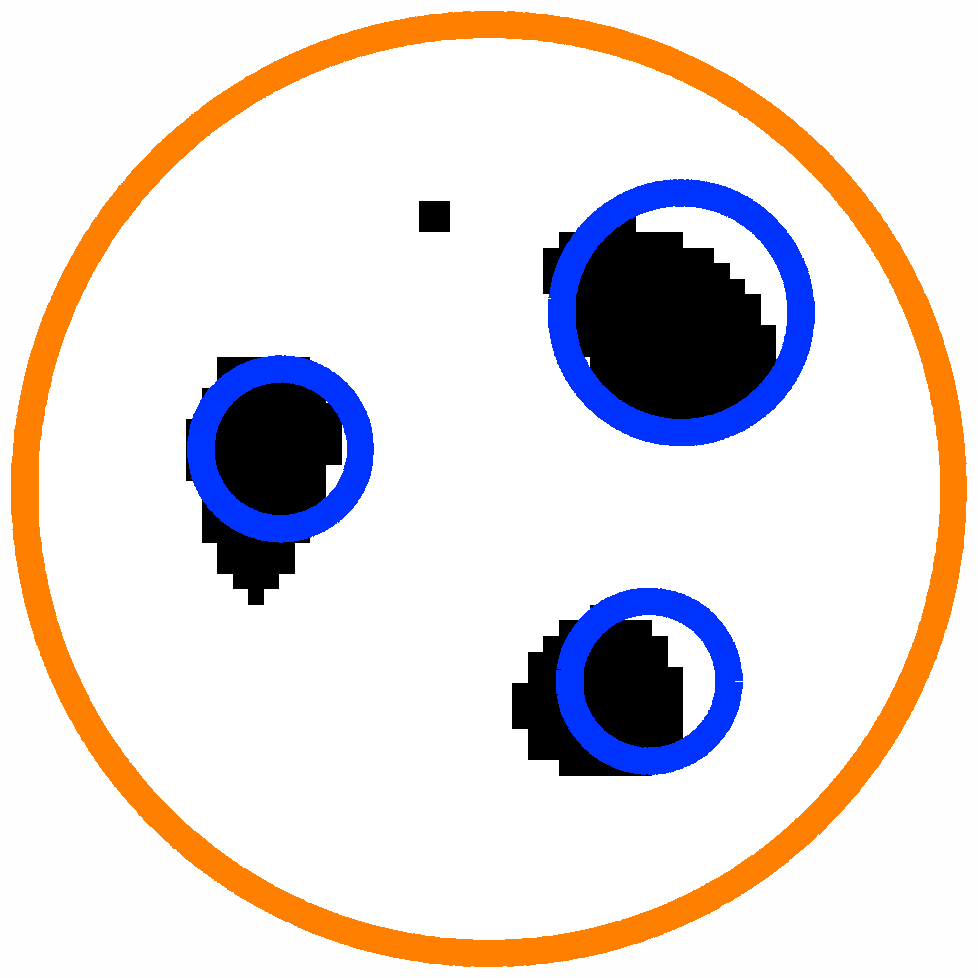} \\
ROA = 69 \%, \\ ROA$\mbox{}_{{\min}}$ = 57 \%
\end{framed}
\end{center}
\end{minipage} \hskip0.2cm
\begin{minipage}{5.0cm}
\begin{center}
\begin{framed}
TTT 1, 8-bit, Norm.\ 6 \% (-24 dB FS) \\ \mbox{} \vskip-0.1cm
\includegraphics[width=2.0cm]{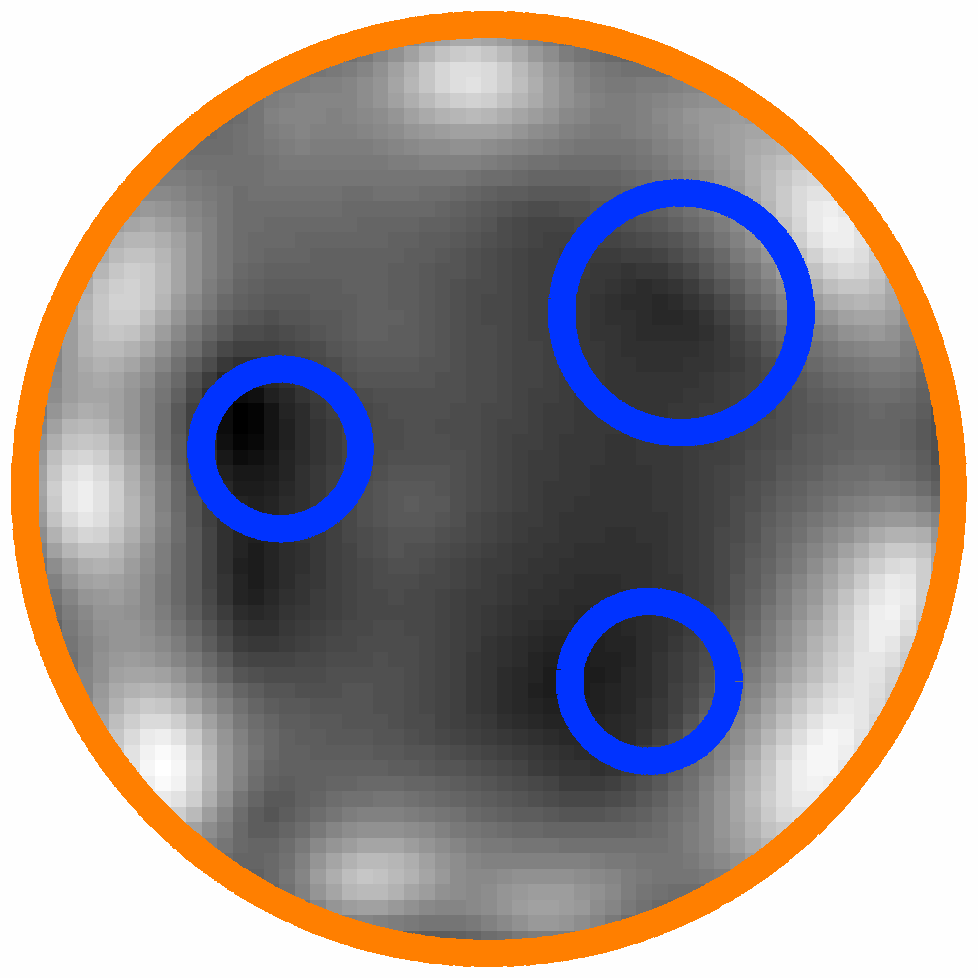} \hskip0.1cm
\includegraphics[width=2.0cm]{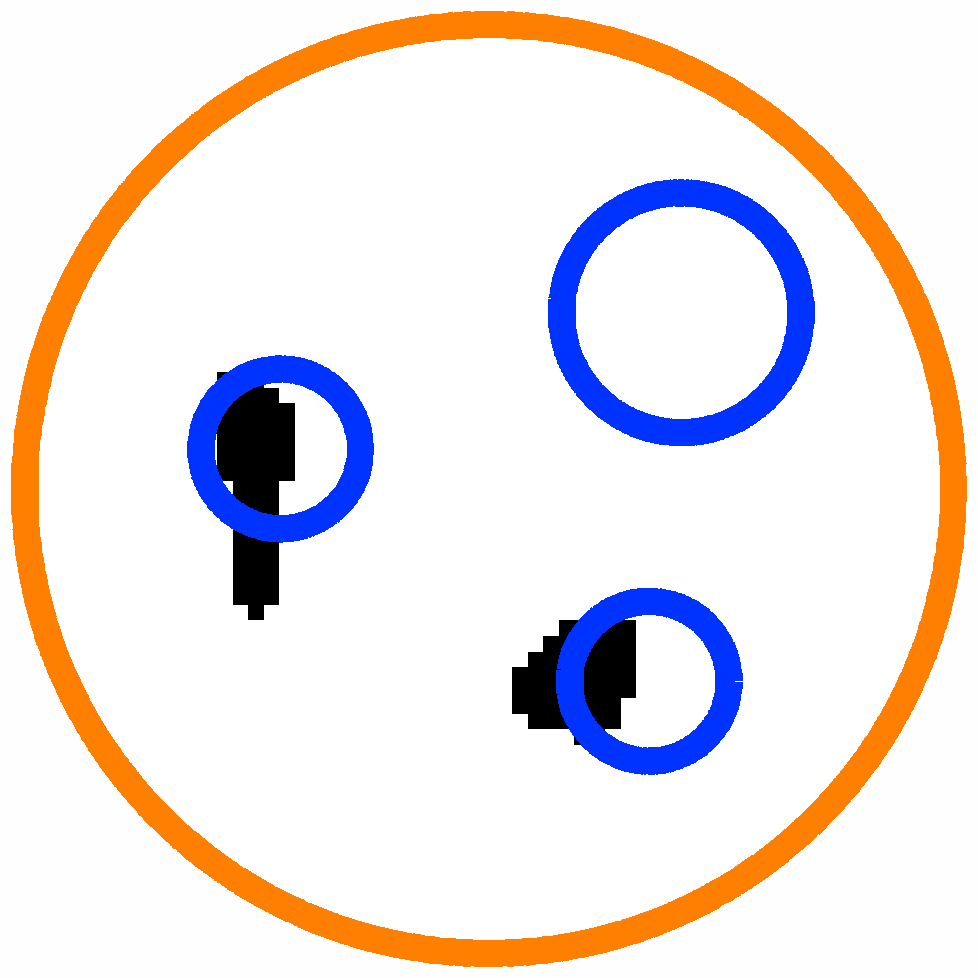} \\
ROA = 17 \%, \\ ROA$\mbox{}_{{\min}}$ = 0 \%
\end{framed}
\end{center}
\end{minipage} \end{framed} 
\end{center}
\end{scriptsize}
\caption[Thresholded/integrated travel time results]{Results with ITT used for the measured signal. Control signal was thresholded.}
\label{ttt1figure}
\end{figure*}

\begin{figure*}[t]
\begin{scriptsize}
\begin{center}
\begin{framed}
Threshold 90 \% (-1 dB FS) \\ \mbox{} \vskip-0.1cm
\begin{minipage}{5.0cm}
\begin{center}
\begin{framed}
TTT 2, 16-bit, Norm.\ 100 \% (0 dB FS) \\ \mbox{} \vskip-0.1cm
\includegraphics[width=2.0cm]{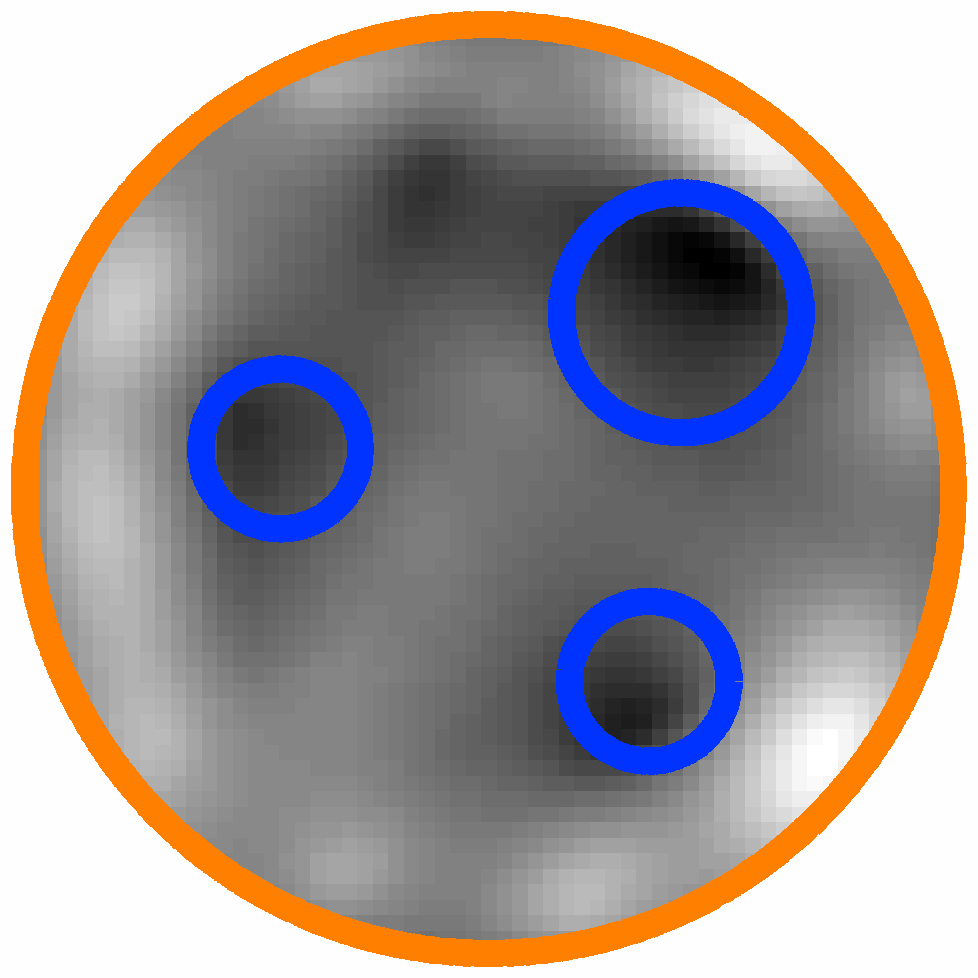} \hskip0.1cm
\includegraphics[width=2.0cm]{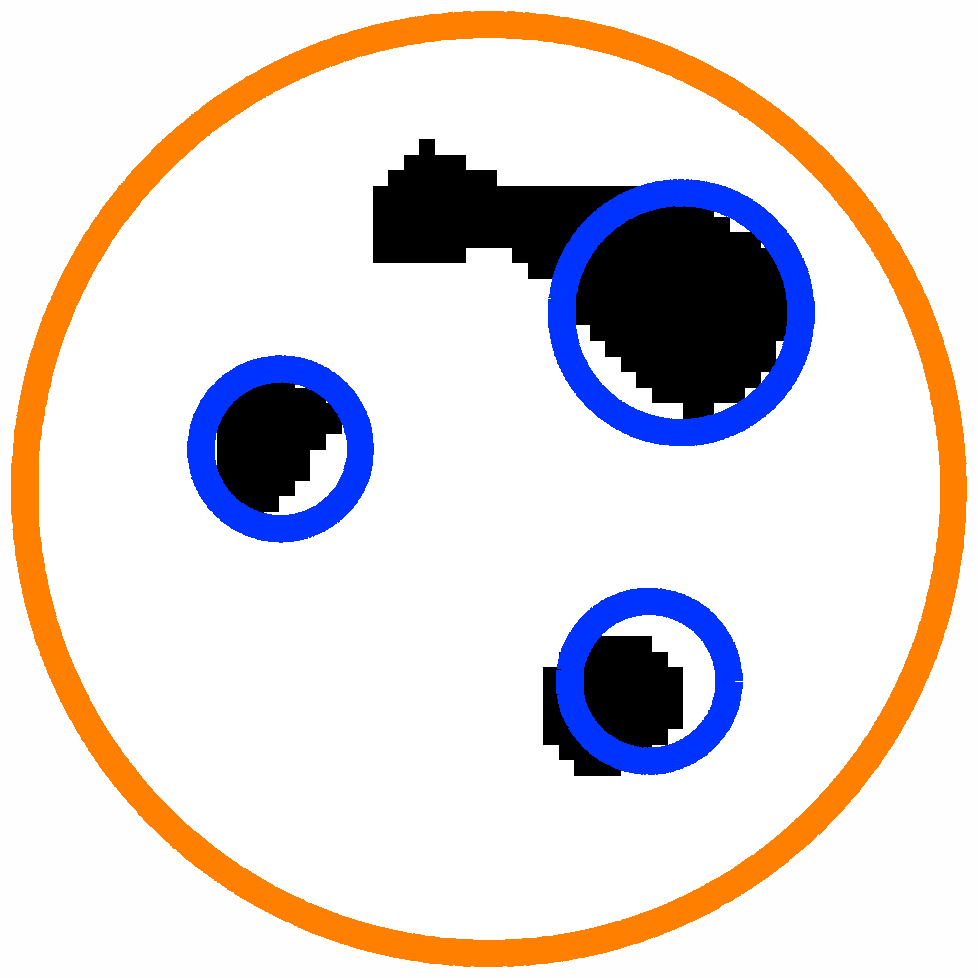} \\
ROA = 68 \%, \\ ROA$\mbox{}_{{\min}}$ = 48 \%
\end{framed} 
\end{center}
\end{minipage} \hskip0.2cm
\begin{minipage}{5.0cm}
\begin{center}
\begin{framed}
TTT 2, 8-bit, Norm.\ 100 \% (0 dB FS) \\ \mbox{} \vskip-0.1cm
\includegraphics[width=2.0cm]{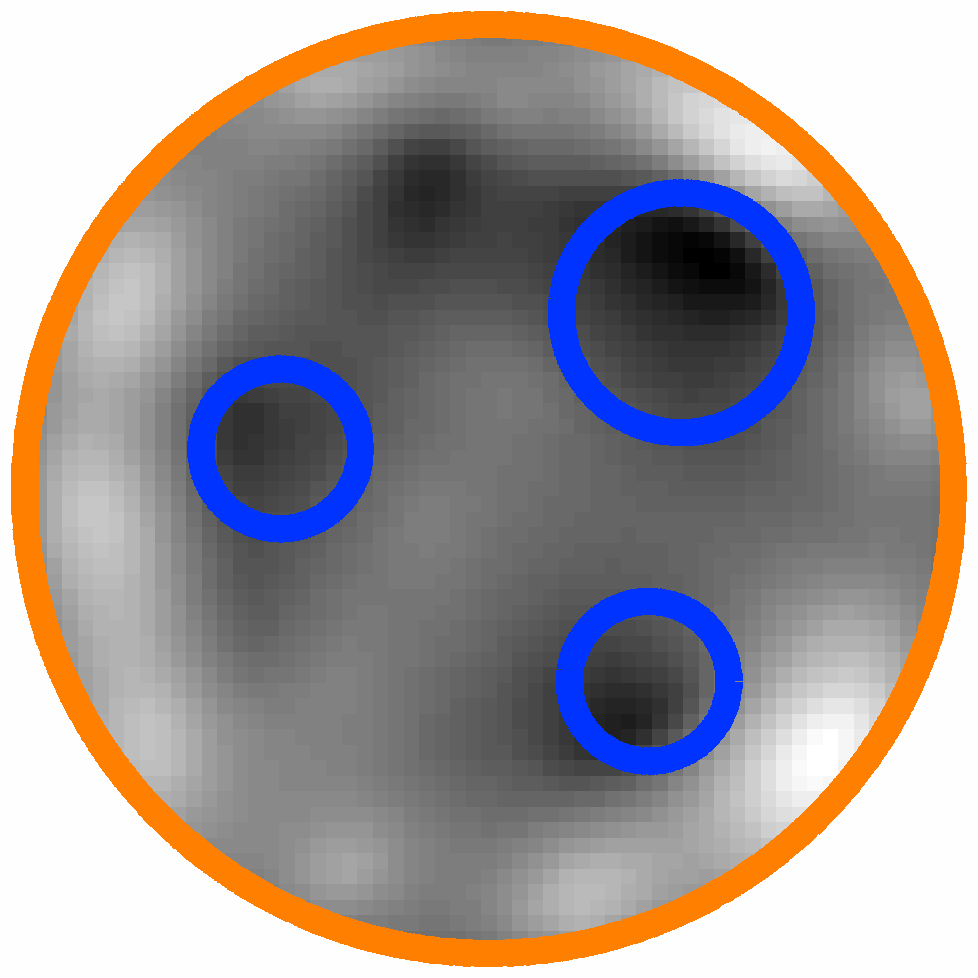} \hskip0.1cm
\includegraphics[width=2.0cm]{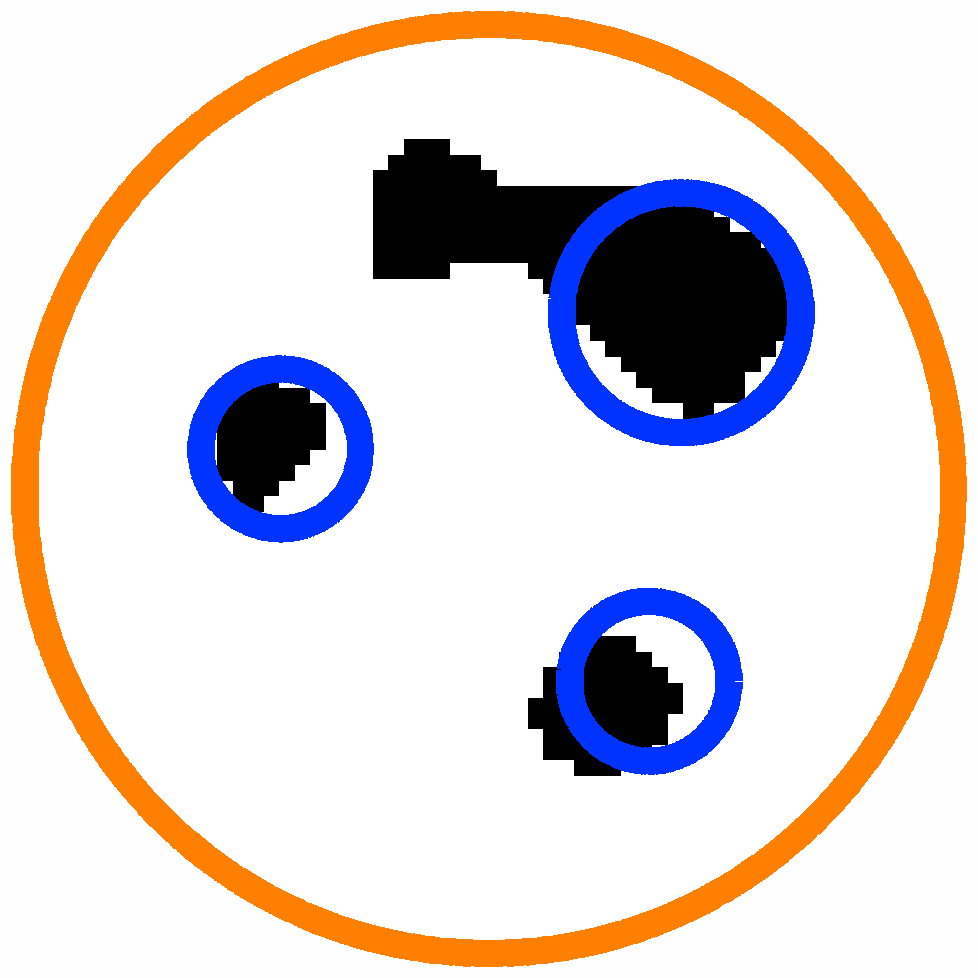} \\
ROA = 63 \%, \\ ROA$\mbox{}_{{\min}}$ = 43 \%
\end{framed}
\end{center}
\end{minipage} \hskip0.2cm
\begin{minipage}{5.0cm}
\begin{center}
\begin{framed}
TTT 2, 8-bit, Norm.\ 6 \% (-24 dB FS) \\ \mbox{} \vskip-0.1cm
\includegraphics[width=2.0cm]{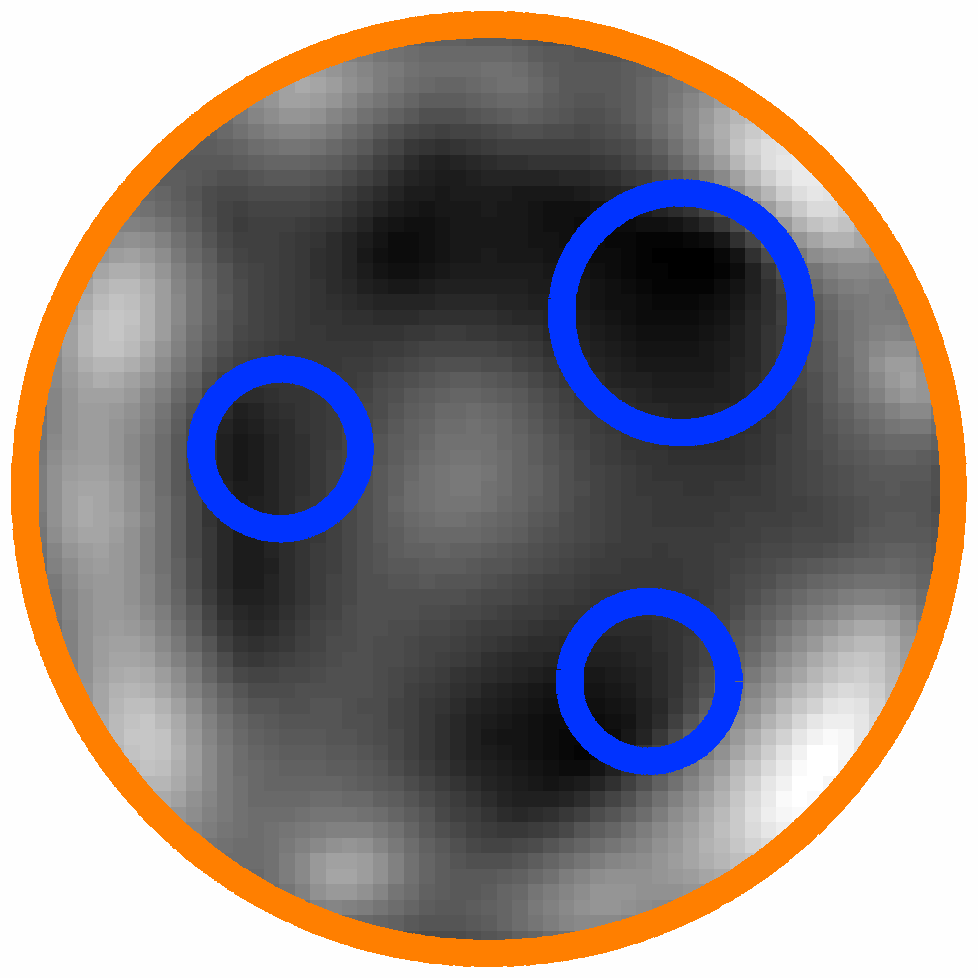} \hskip0.1cm
\includegraphics[width=2.0cm]{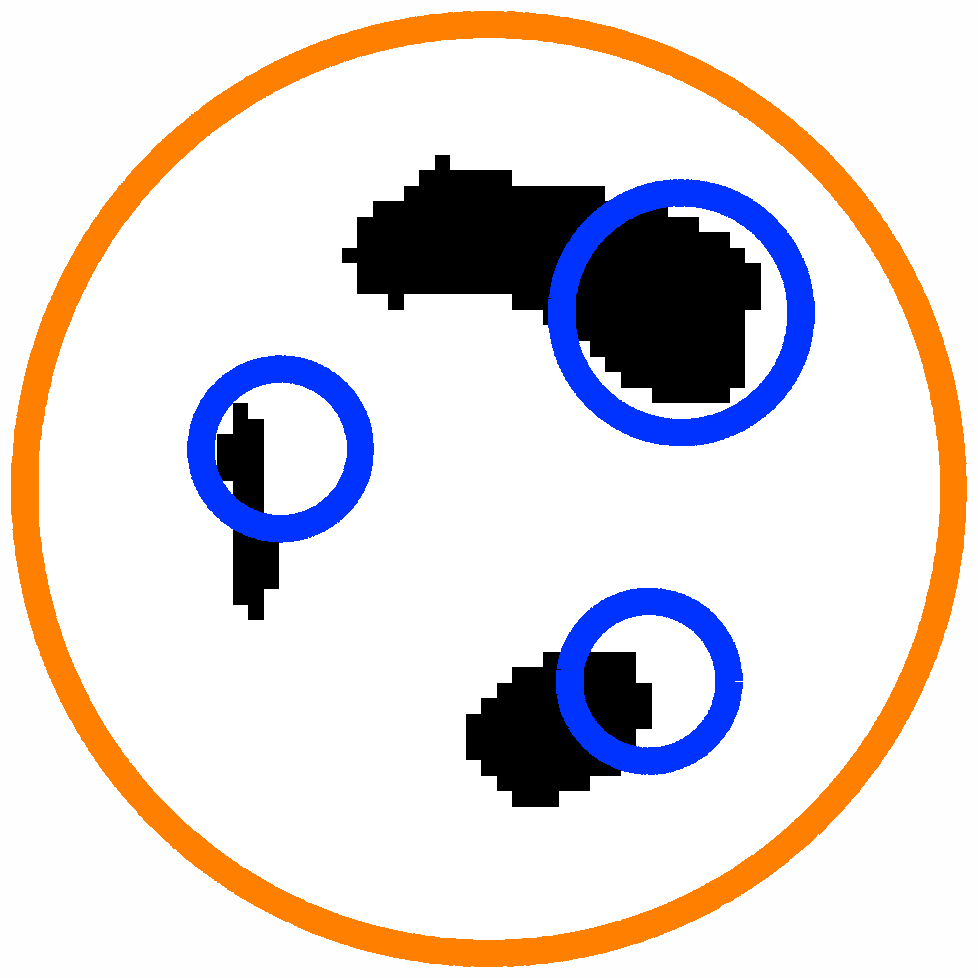} \\
ROA = 45 \%, \\ ROA$\mbox{}_{{\min}}$ = 20 \%
\end{framed}
\end{center}
\end{minipage} \end{framed} \mbox{} \vspace{-0.5cm}
\begin{framed}
Threshold 70 \% (-3 dB FS) \\ \mbox{} \vskip-0.1cm
\begin{minipage}{5.0cm}
\begin{center}
\begin{framed}
TTT 2, 16-bit, Norm.\ 100 \% (0 dB FS) \\ \mbox{} \vskip-0.1cm
\includegraphics[width=2.0cm]{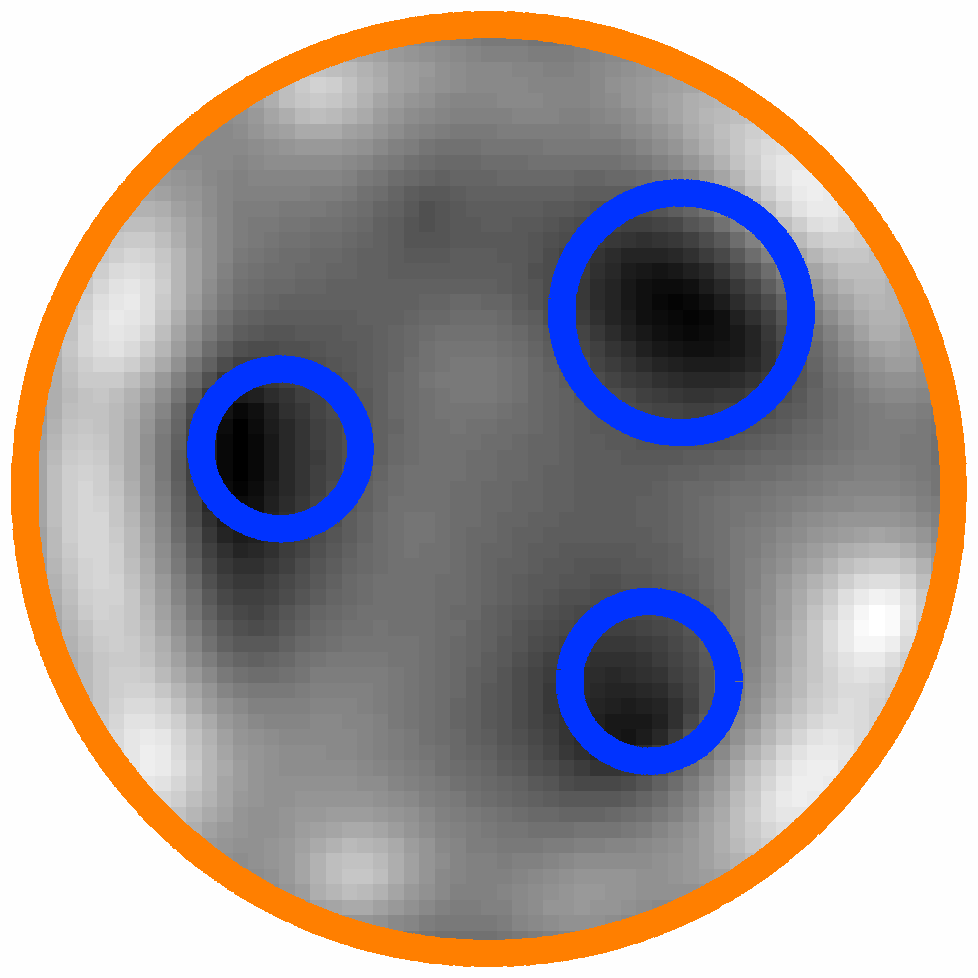} \hskip0.1cm
\includegraphics[width=2.0cm]{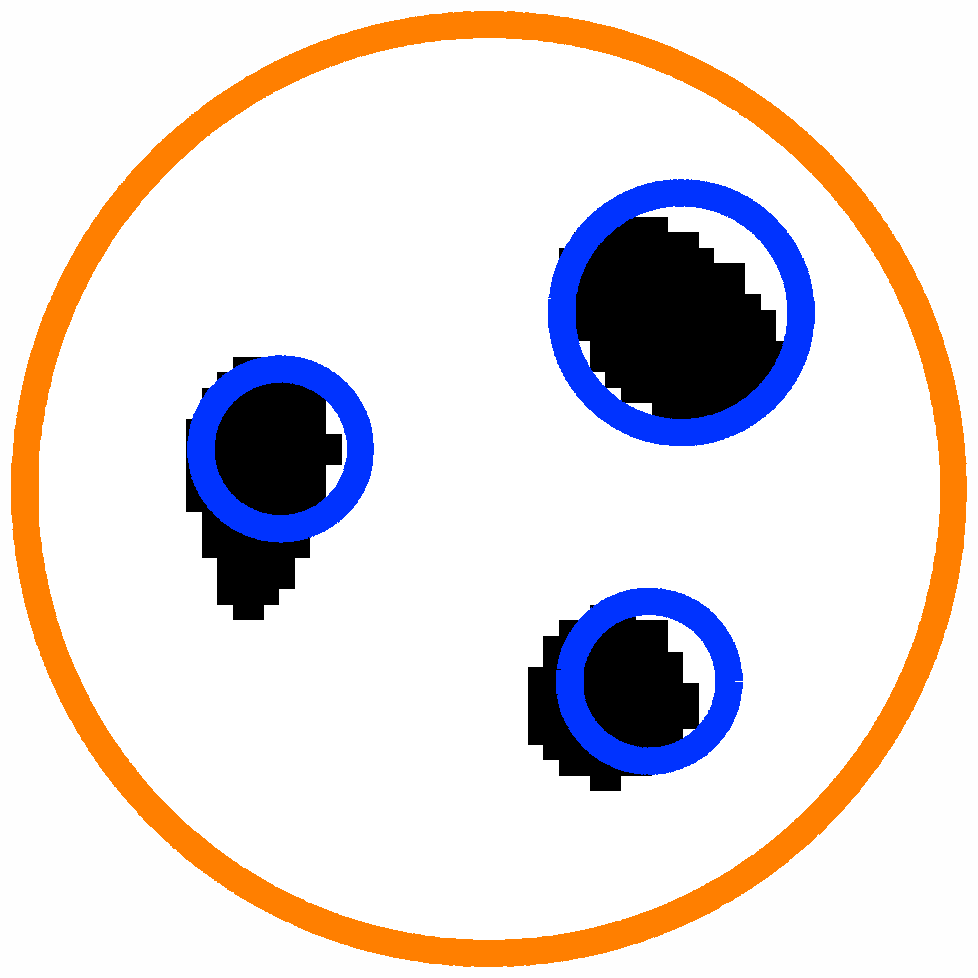} \\
ROA = 71 \%, \\ ROA$\mbox{}_{{\min}}$ =  65 \%
\end{framed}
\end{center}
\end{minipage} \hskip0.2cm
\begin{minipage}{5.0cm}
\begin{center}
\begin{framed}
TTT 2, 8-bit, Norm.\ 100 \% (0 dB FS) \\ \mbox{} \vskip-0.1cm
\includegraphics[width=2.0cm]{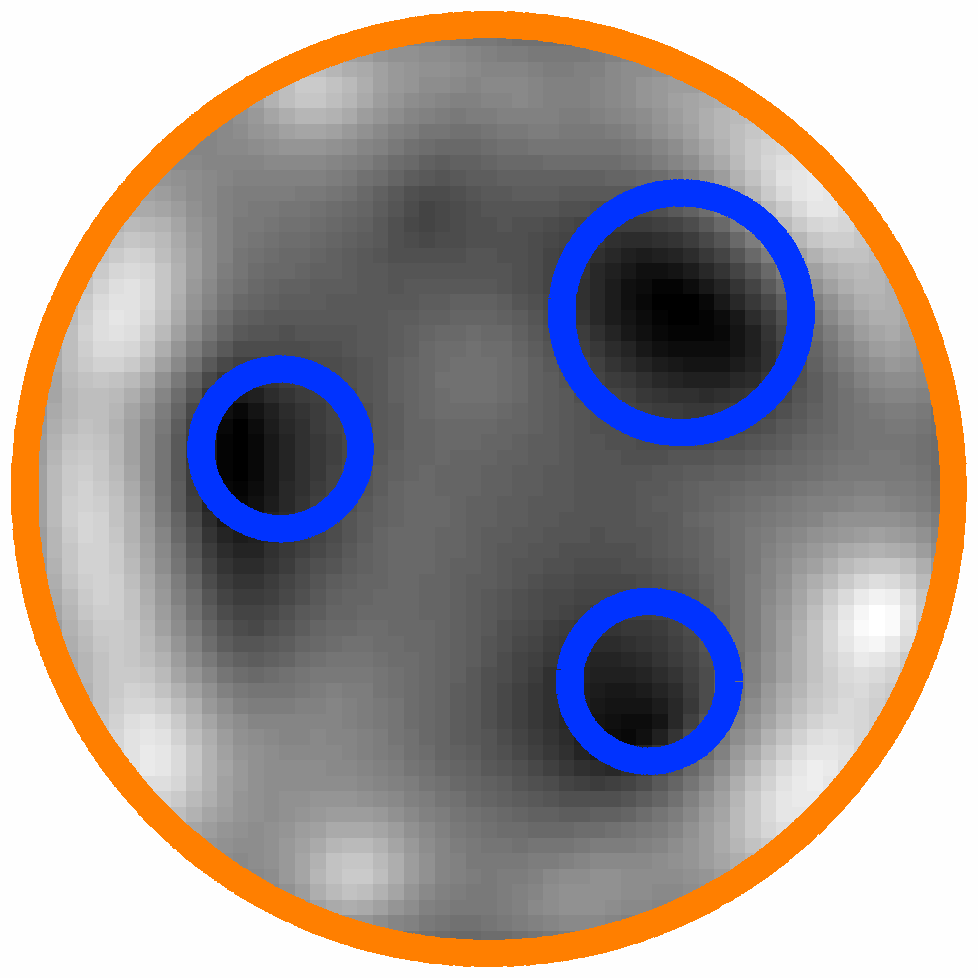} \hskip0.1cm
\includegraphics[width=2.0cm]{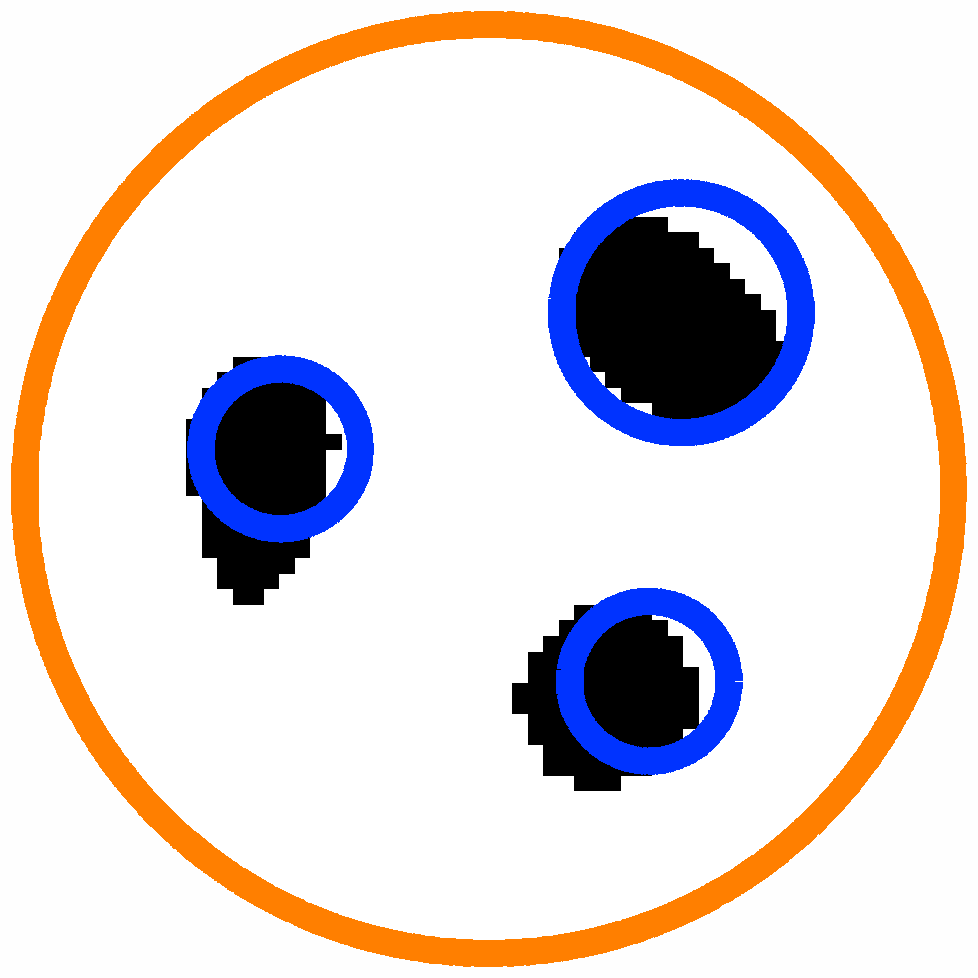} \\
ROA = 71 \%, \\ ROA$\mbox{}_{{\min}}$ = 68 \%
\end{framed}
\end{center}
\end{minipage} \hskip0.2cm
\begin{minipage}{5.0cm}
\begin{center}
\begin{framed}
TTT 2, 8-bit, Norm.\ 6 \% (-24 dB FS) \\ \mbox{} \vskip-0.1cm
\includegraphics[width=2.0cm]{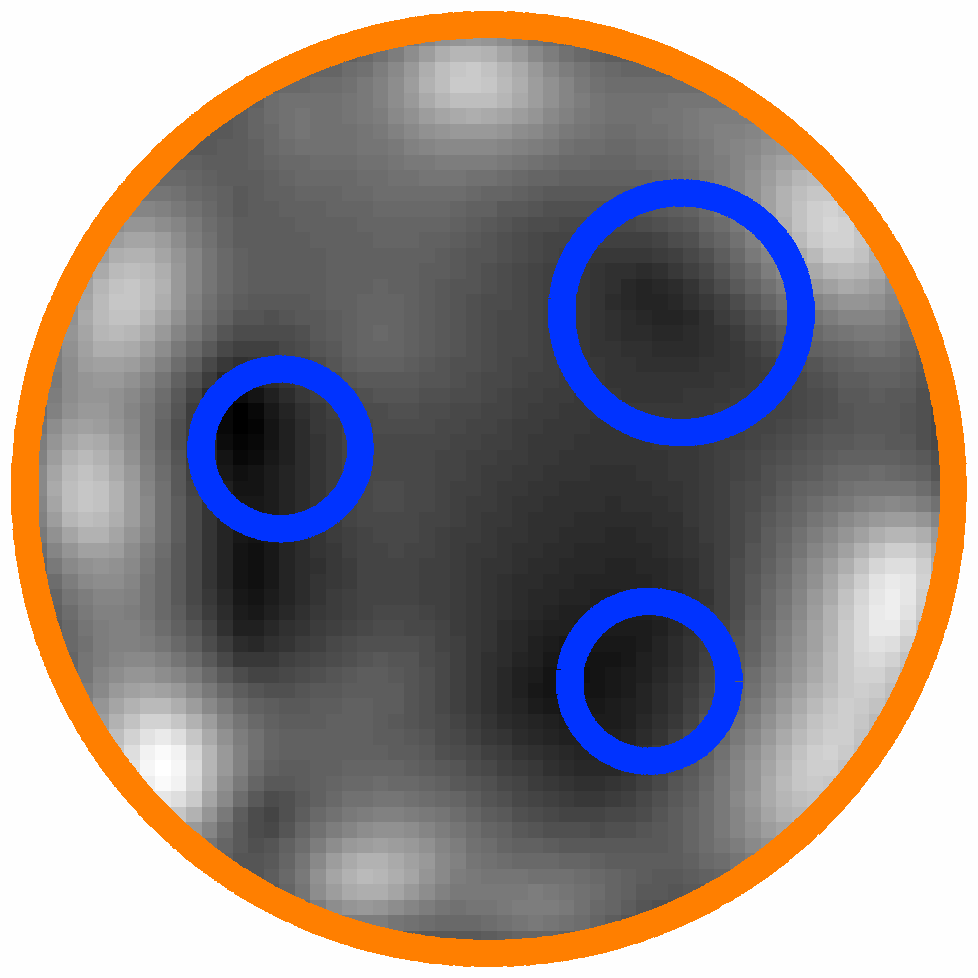} \hskip0.1cm
\includegraphics[width=2.0cm]{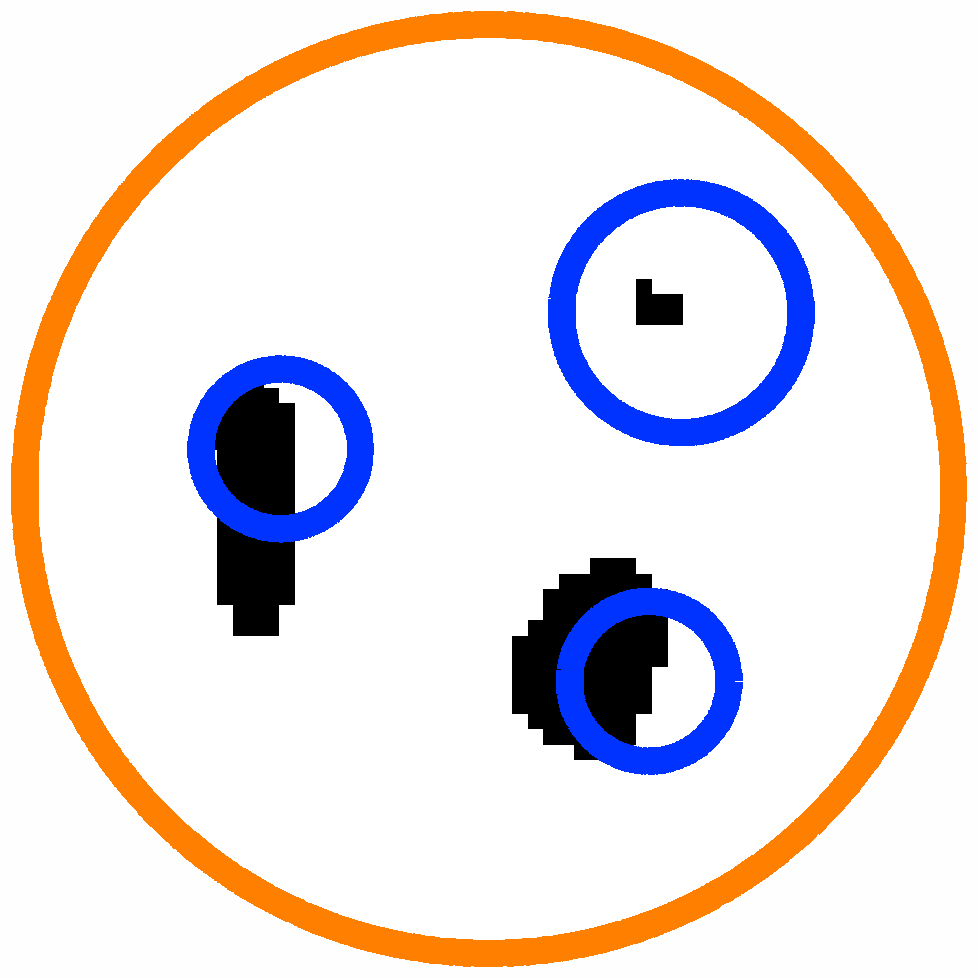} \\
ROA = 25 \%, \\ ROA$\mbox{}_{{\min}}$ = 4 \%
\end{framed}
\end{center}
\end{minipage} \end{framed} 
\end{center}
\end{scriptsize}
\caption[Thresholded/thresholded travel time results]{Results with thresholded travel time used for both the control signal and measured signal.}
\label{ttt2figure}
\end{figure*}

\begin{figure*}[t]
\begin{scriptsize}
\begin{center}
\begin{framed}
Threshold 90 \% (-1 dB FS) \\ \mbox{} \vskip-0.1cm
\begin{minipage}{5.0cm}
\begin{center}
\begin{framed}
ITT, 8-bit, Norm.\ 100 \% (0 dB FS) \\ \mbox{} \vskip-0.1cm
\includegraphics[width=2.0cm]{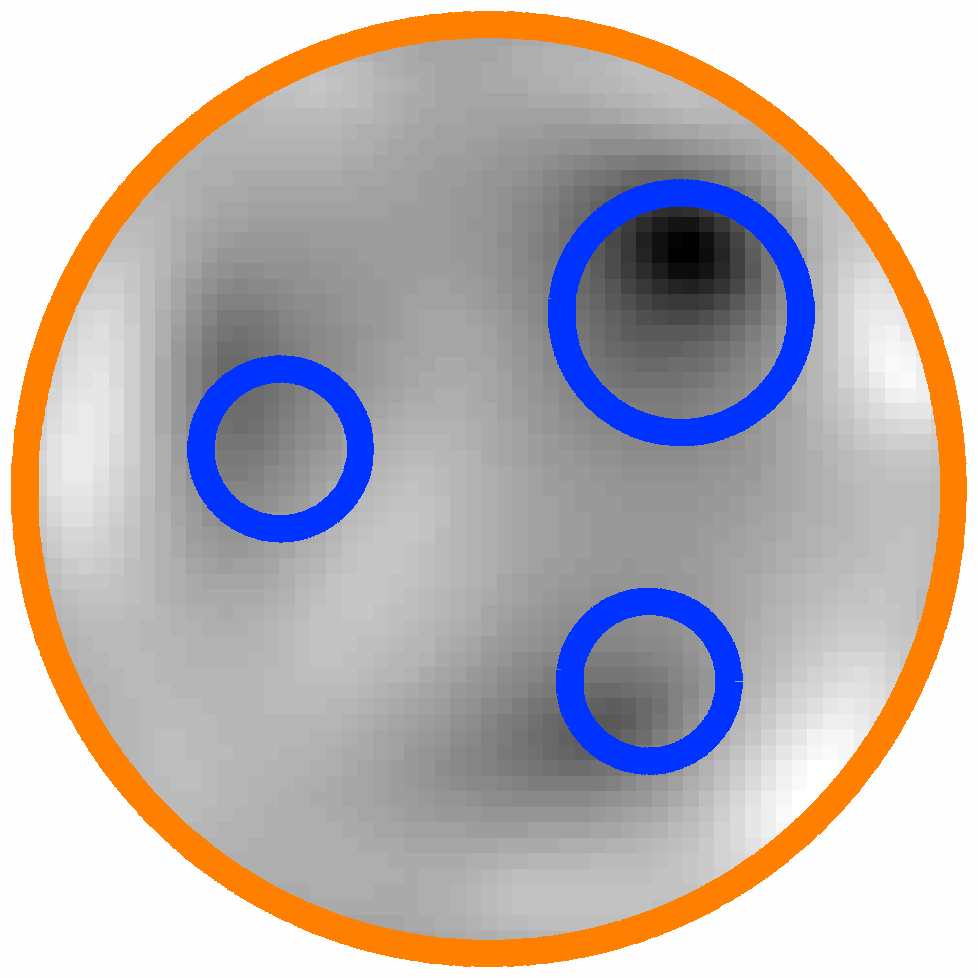} \hskip0.1cm
\includegraphics[width=2.0cm]{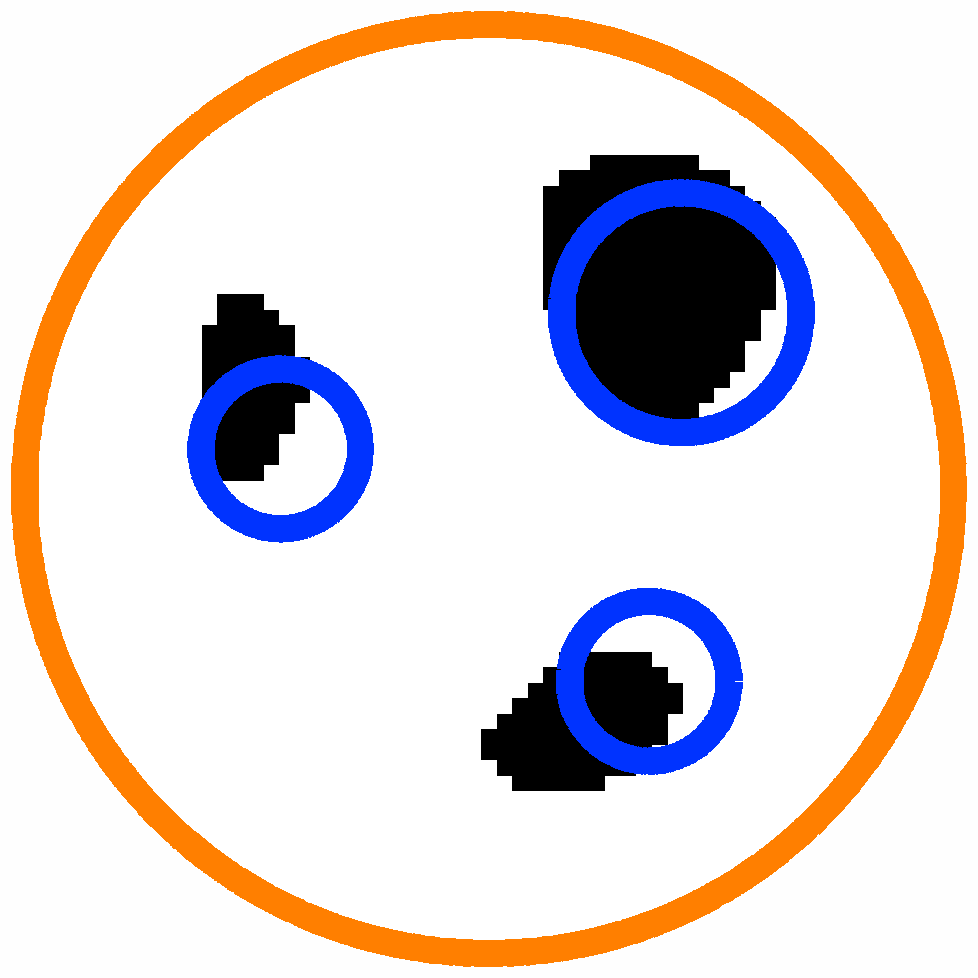} \\
ROA = 62 \%, \\ ROA$\mbox{}_{{\min}}$ = 38 \%
\end{framed} 
\end{center}
\end{minipage} \hskip0.2cm
\begin{minipage}{5.0cm}
\begin{center}
\begin{framed}
TTT 1, 8-bit, Norm.\ 100 \% (0 dB FS) \\ \mbox{} \vskip-0.1cm
\includegraphics[width=2.0cm]{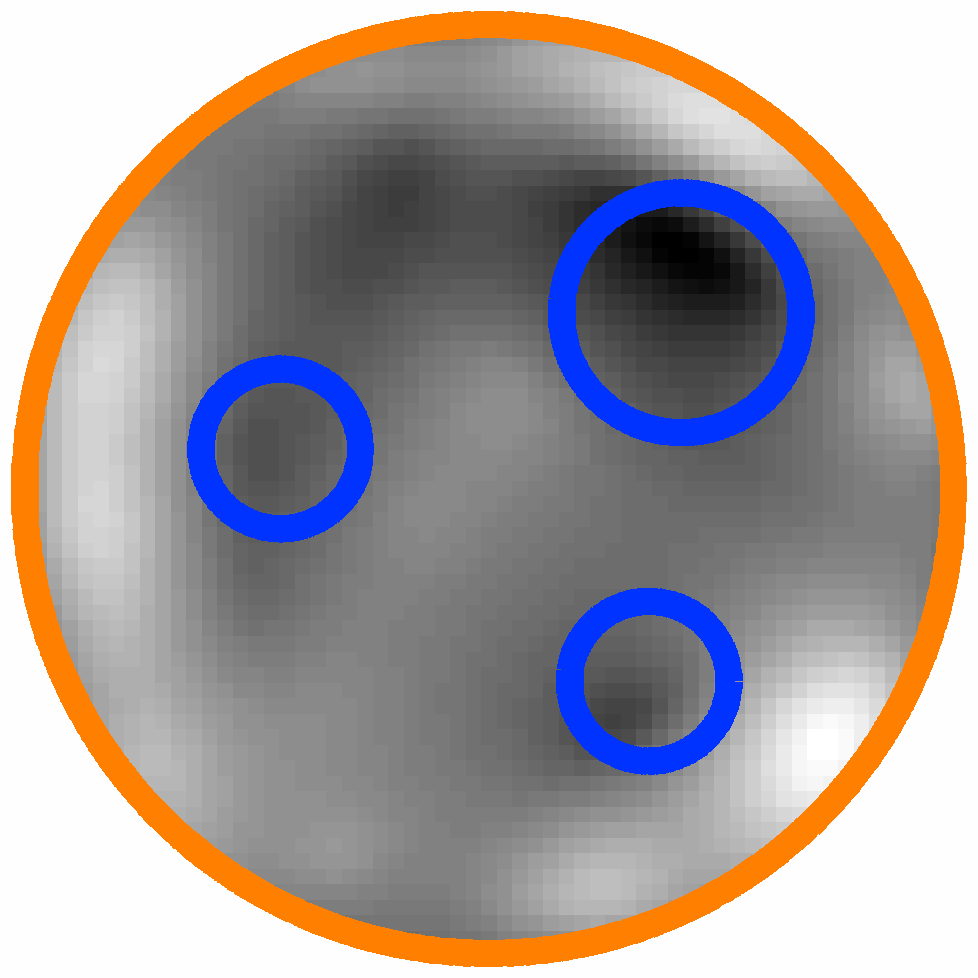} \hskip0.1cm
\includegraphics[width=2.0cm]{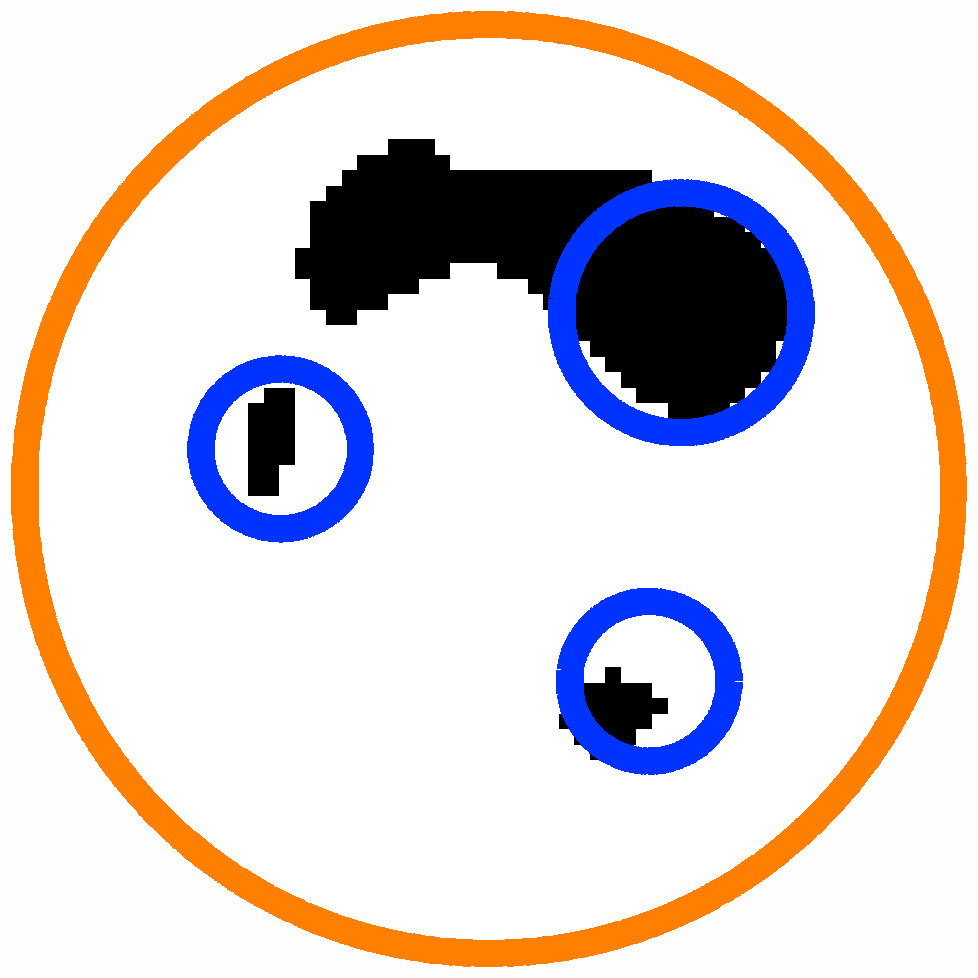} \\
ROA =53 \%, \\ ROA$\mbox{}_{{\min}}$ = 20 \%
\end{framed}
\end{center}
\end{minipage} \hskip0.2cm
\begin{minipage}{5.0cm}
\begin{center}
\begin{framed}
TTT 2, 8-bit, Norm.\ 100 \% (0 dB FS) \\ \mbox{} \vskip-0.1cm
\includegraphics[width=2.0cm]{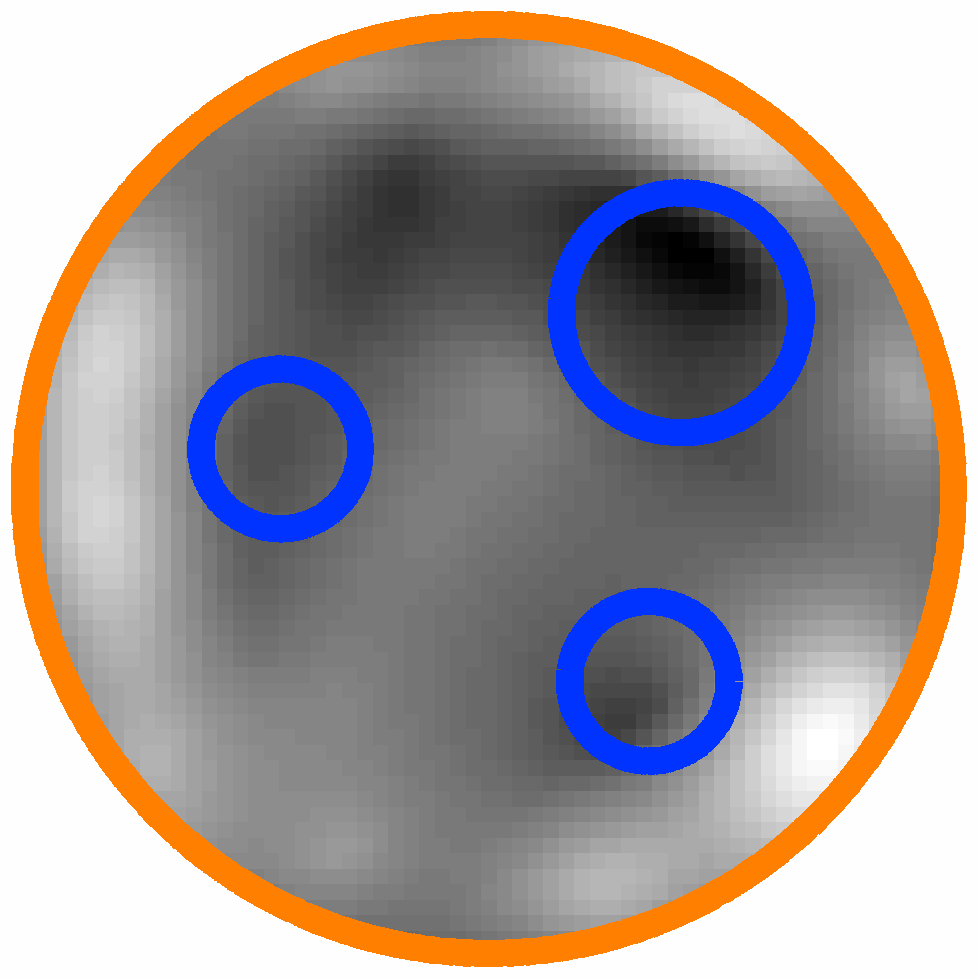} \hskip0.1cm
\includegraphics[width=2.0cm]{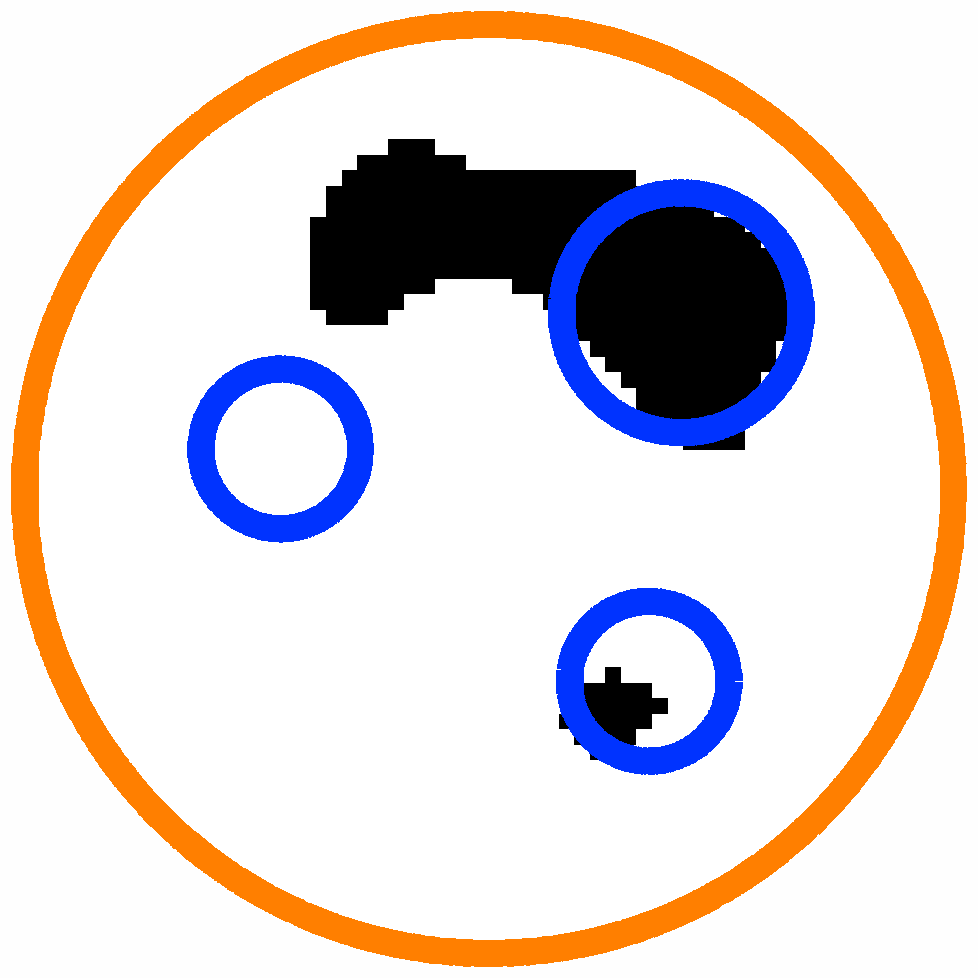} \\
ROA = 49 \%, \\ ROA$\mbox{}_{{\min}}$ = 0 \%
\end{framed}
\end{center}
\end{minipage} \end{framed} \mbox{} \vspace{-0.5cm}
\begin{framed}
Threshold 70 \% (-3 dB FS) \\ \mbox{} \vskip-0.1cm
\begin{minipage}{5.0cm}
\begin{center}
\begin{framed}
ITT, 8-bit, Norm.\ 100 \% (0 dB FS) \\ \mbox{} \vskip-0.1cm
\includegraphics[width=2.0cm]{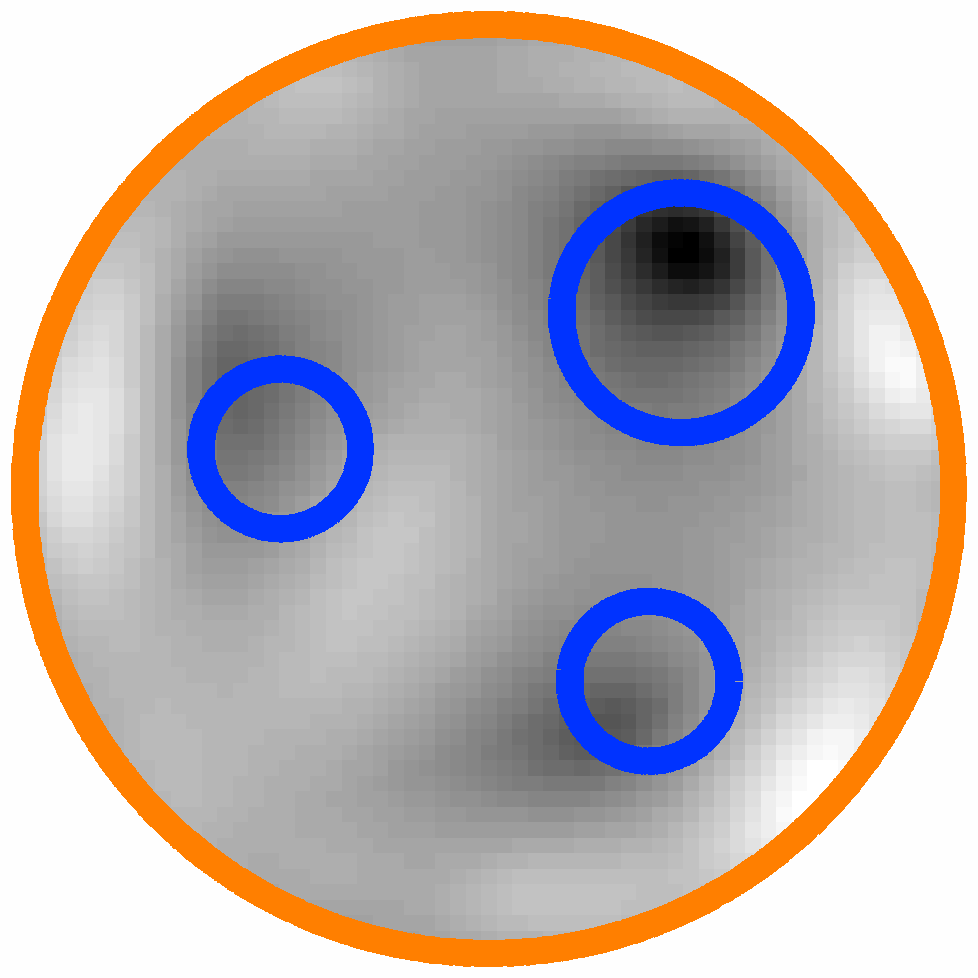} \hskip0.1cm
\includegraphics[width=2.0cm]{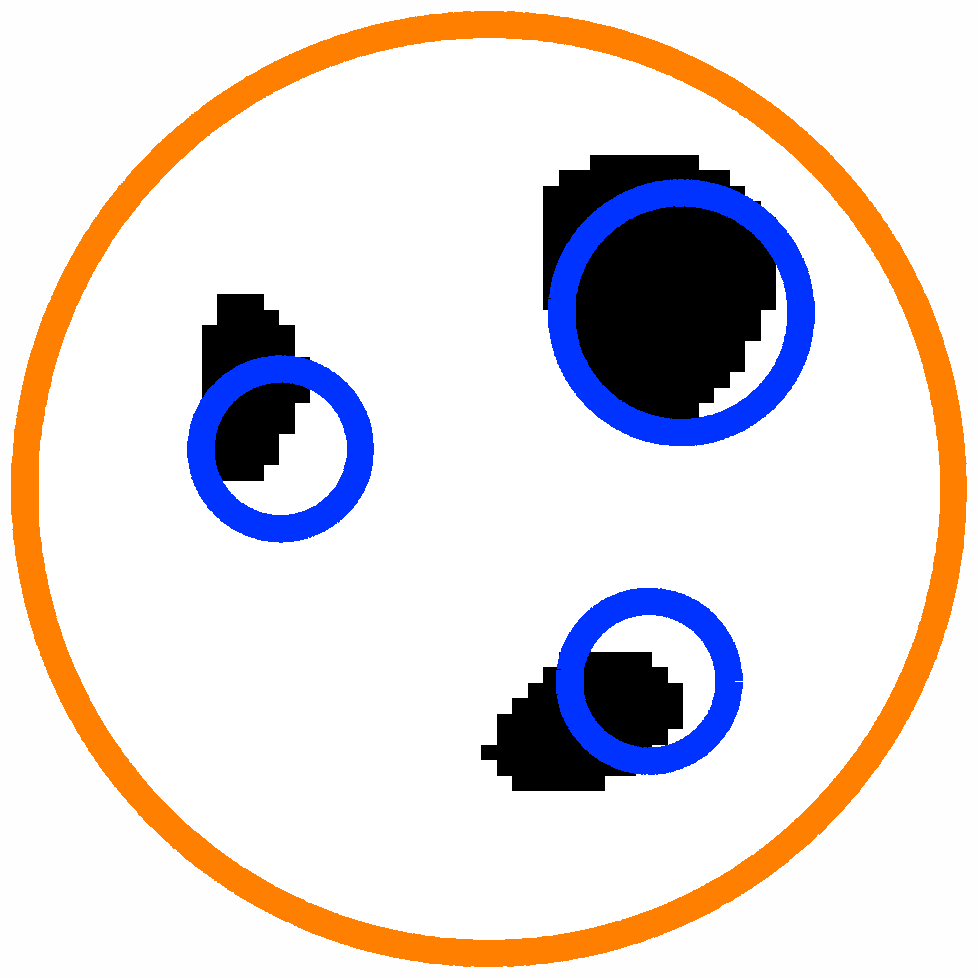} \\
ROA = 62 \%, \\ ROA$\mbox{}_{{\min}}$ =  38 \%
\end{framed}
\end{center}
\end{minipage} \hskip0.2cm
\begin{minipage}{5.0cm}
\begin{center}
\begin{framed}
TTT 1, 8-bit, Norm.\ 100 \% (0 dB FS) \\ \mbox{} \vskip-0.1cm
\includegraphics[width=2.0cm]{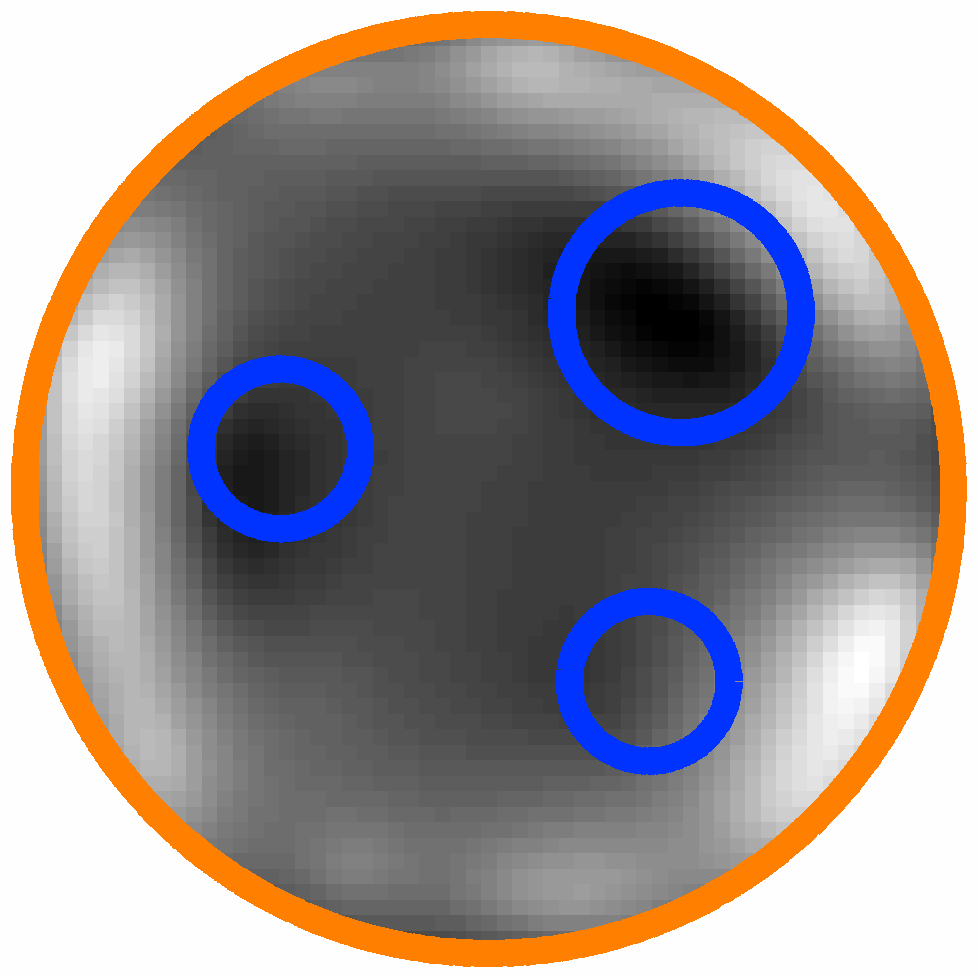} \hskip0.1cm
\includegraphics[width=2.0cm]{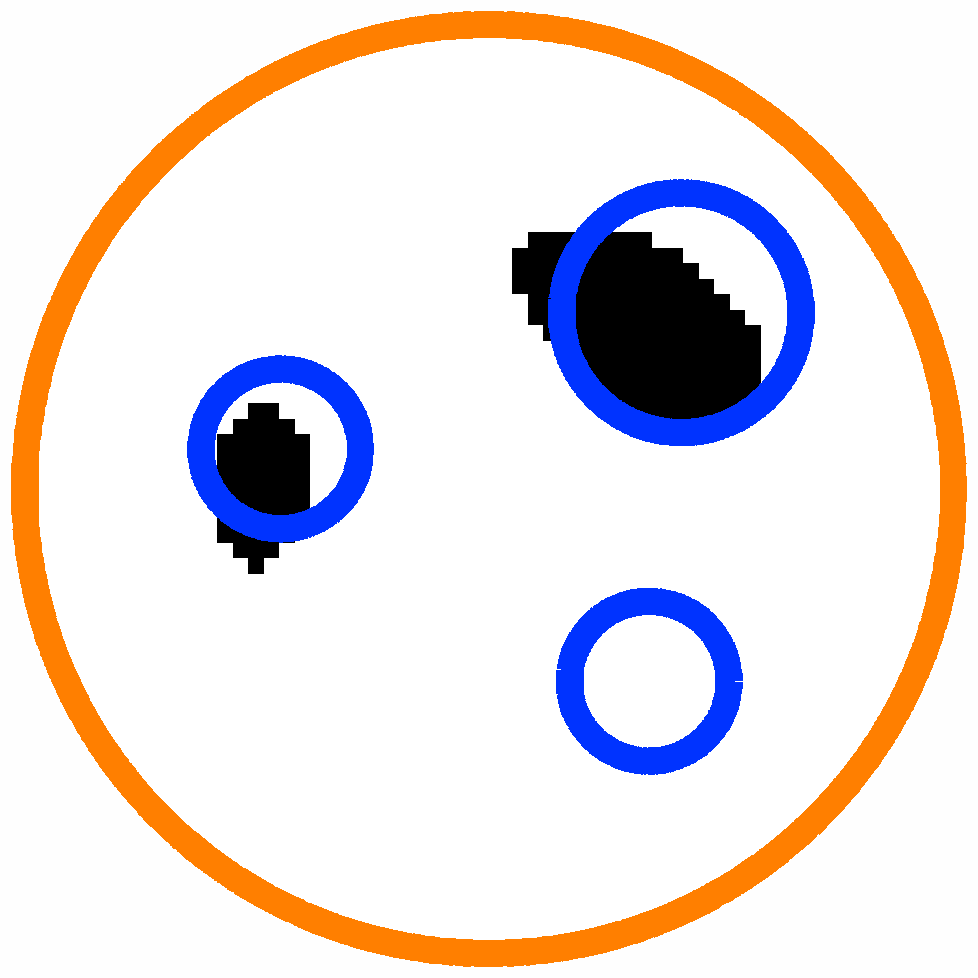} \\
ROA = 42 \%, \\ ROA$\mbox{}_{{\min}}$ = 0 \%
\end{framed}
\end{center}
\end{minipage} \hskip0.2cm
\begin{minipage}{5.0cm}
\begin{center}
\begin{framed}
TTT 2, 8-bit, Norm.\ 100 \% (0 dB FS) \\ \mbox{} \vskip-0.1cm
\includegraphics[width=2.0cm]{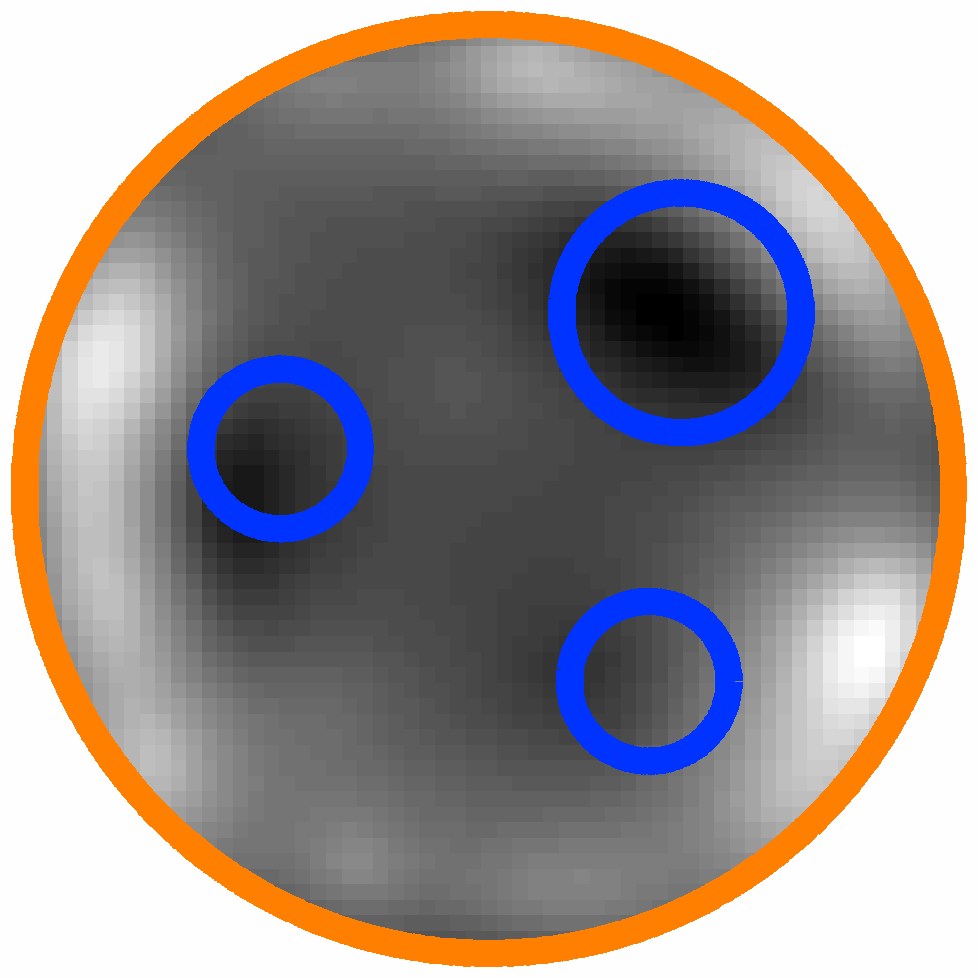} \hskip0.1cm
\includegraphics[width=2.0cm]{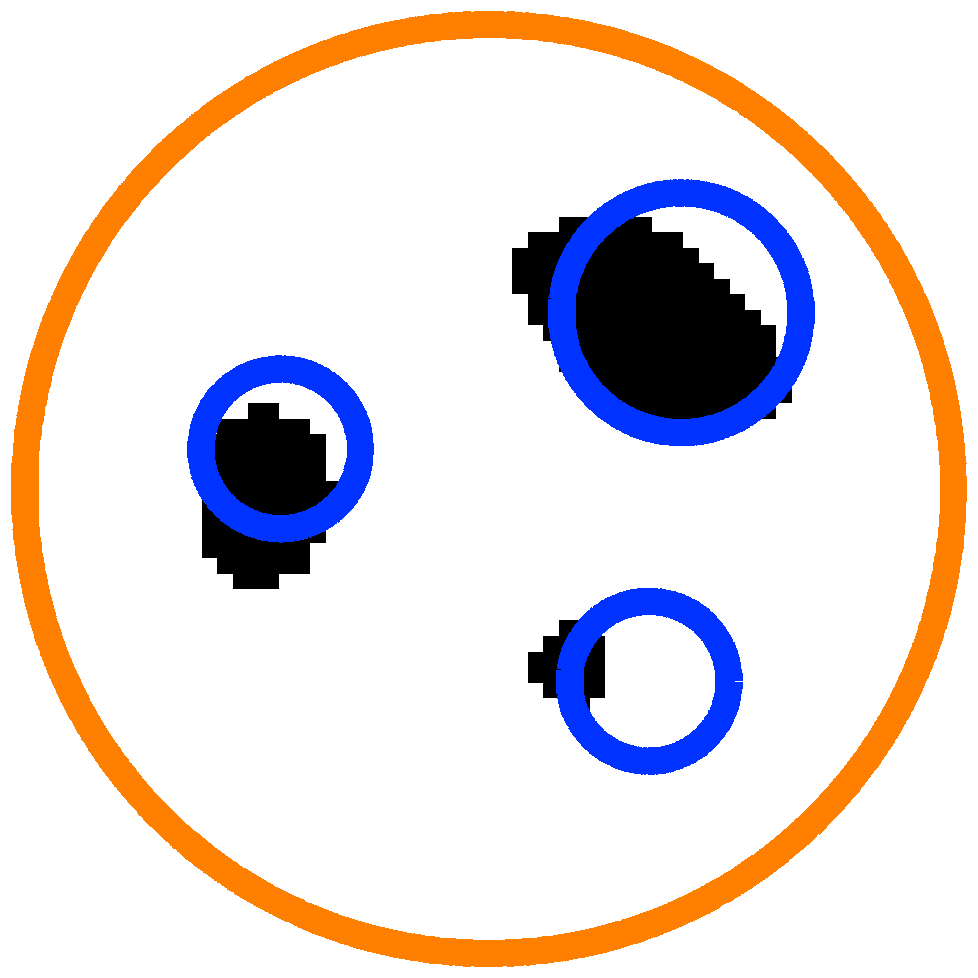} \\
ROA = 52 \%, \\ ROA$\mbox{}_{{\min}}$ = 8 \%
\end{framed}
\end{center}
\end{minipage} \end{framed} 
\end{center}
\end{scriptsize}
\caption{Result with sparse data. ITT is on the left, TTT 1 in the middle and TTT 2 on the right.}
\label{sparseresultsfigure}
\end{figure*}

Figures \ref{ittfigure}-\ref{sparseresultsfigure} show the test area recovery (inversion) results. These figures consist of image pairs. The outlines of the three foam cylinders are superimposed on the top of the images. For each image the left side image is the actual result image. From that image, an area with strongest values is collected so that it has the same area as that of the foam cylinders. The right side image shows only the solid black areas. 

ITT was found to yield robust results with respect to the A/D conversion bit resolution, signal normalization, threshold, and sparsity of the measurements. The highest ROA  ($\hbox{ROA}_{\hbox{\scriptsize min}}$) obtained with ITT was 64 \% (56 \%) and the lowest one 54 \% (38 \%).  TTT was observed to be advantageous under optimized  conditions with respect to the bit resolution, normalization, threshold, and density of the data. It was also significantly more sensitive than ITT to variations in any of these parameters. For TTT 1, the highest and lowest ROA ($\hbox{ROA}_{\hbox{\scriptsize min}}$) were 70 (57 \%) and 17 \% (0 \%), respectively.  For TTT 2, these values were 71 (68 \%) and 25 \% (0 \%), respectively. The results obtained with TTT 2 were slightly superior to those achieved using TTT 1. The 16-bit signed integer travel-time calculation module produced essentially the same outcome as  64-bit floating point arithmetics. 

The HLS method was found to work appropriately in developing the FPGA hardware. A complete prototyping cycle took a few hours, most of which time was spent implementing a new feature in C. It was found that since the algorithmic C-like code is much more maintainable than traditional designs using VHDL, further changes and reuse of the code can be done more easily via HLS.  A summary of the synthesis results for the 8-bit and 16-bit designs for the ITT and TTT is presented in Table \ref{synthesisresultstable}. The hardware for ITT was almost as fast as that of TTT, and the difference in chip area between these two methods was not significant. 

\begin{table}[h]
  \begin{center}
    \caption[Synthesis results]{Synthesis results for the travel-time calculation modules implemented with Catapult C. The percentages given for area is the percentage of available logic elements used on the FPGA chip (EP2C35F672C6N) of the DE2 board.}
    \label{synthesisresultstable}
     \begin{tabular}{lllll }
      \hline
      Mode &
	Bits & 
	 Speed  & 
	 Latency & 
Area  \\ 
&&(MHz)&(Cycles)& (Logic Elements) \\
      \hline
      Thresholding & 8 & 106.0 & 13~594 & 821 (2.5 \%)\\
      Thresholding & 16 & 99.4 & 13~558  & 1~075 (3.2 \%)\\
      Integrating & 8 & 74.8 & 13~565  & 891 (2.7 \%)\\
      Integrating & 16 & 82.6 & 13~581 & 1~253 (3.8 \%)\\
      \hline
    \end{tabular}
  \end{center}
\end{table}

\section{Discussion}

This paper focused on developing the processing and inversion of waveform tomography data  for applications in which  the signal wavelength is close to the diameter of the details that are to be recovered. In particular, FPGA-based hardware was used.  We compared the integrated and thresholded travel-time (ITT and TTT) in numerical experiments in which three foam cylinders were to be localized based on an experimental 5.8 kHz acoustic signal. We tested a 16-bit and 8-bit analog-to-digital (A/D) conversion together with two different threshold criterions and normalization levels. As reference applications of this study, we considered ({\bf 1}) microwave and ultrasonic computed tomography (MCT and UCT), ({\bf 2})  tomography of small solar system bodies (SSSBs), in particular, the CONSERT experiment and ({\bf 3}) ultrasonic/microwave detection of concrete defects.

\subsection{Experiments}

The recovery of the test object locations on the target area by using the described inversion methods was found to work appropriately.  Our results show that ITT is more stable than TTT if the signal quality decreases. This can been seen from the stability of the relative overlapping area (ROA) percentages (53~-~64~\%). These percentages are comparable to our previous research on waveform inversion within a 2D domain. In \cite{pursiainen2014} the best ROA percentage 71~\% was obtained by using the full wave data. Using TTT the best recovery result was exactly the same 71\%. ITT was 7~\% less accurate, but more stable when data preprocessing was performed using source data with lesser bit resolution. 

The results show that ITT is  invariant with respect to source data bit resolution reduction in the time-domain and the level of thresholding used to locate the signal pulse from recorded audio data.  
ITT was also found to be more reliable than TTT with respect to the normalization of the signal and the sparsity of the measurements. We normalized the 8-bit signal to two different levels, 100 \% (0 dB FS) and 6 \% (-24 dB FS) amplitude, to simulate weak receivers, i.e., to decrease signal-to-noise ratio (SNR). This is relevant in applications where the signal quality is reduced. For example, in astro-geoscientific applications, the signals can be weak in some directions. This was evident from the CONSERT experiment \cite{kofman2007, kofman1998} in which the power and quality of the received signal varied significantly depending on the direction of the measurement \cite{kofman2015}. The result concerning the sparsity of the measurements is essential for CONSERT, and other applications in which not all the data can be gathered.  

Based on the results, we suggest that ITT can be superior to TTT with respect to the robustness of the inversion. It also seems obvious that TTT might achieve a higher ROA than ITT for a high-quality signal and a well-chosen threshold parameter. Namely,  under optimal conditions TTT filters out the noisy tail part of the signal that is present in the computation of the ITT. This advantage is  utilized, e.g., in the first-arrival sound speed inversion \cite{hooi2014}. The present results suggest that the ITT method does not significantly diminish the inversion quality, and indeed, it can increase the reliability of the results with respect to the uncertainty factors and incompleteness of the data.

\subsection{Applications}

Tomography applications differ from each other by the speed and length of the electromagnetic or sound waves. The feature size $d$ that can be detected is dependent on the wave length $\lambda$ (e.g.\ $\lambda/2$). In Table \ref{relevancytable}, the significance of the test setup with respect to the present reference applications ({\bf 1})--({\bf 3}) has been summarized based on $d$, $\lambda$, the diameter of the target domain $D$ and the ratios $d/\lambda$ and  $D/\lambda$. 

The values $d$, $\lambda$, and  $D$ utilized in Table \ref{relevancytable} can be reasoned as follows.  ({\bf 1}) In MCT, signal frequencies 1--6 GHz are being used  and the diameter of the sensor ring can be, e.g., $D = 15$ cm \cite{son2015,grzegorczyk2012}. At 5 GHz,  the wavelength is  $\lambda \approx 1.9$ cm corresponding to the relative permittivity of the breast $\varepsilon_r \approx 10$  \cite{zeng2011,lazebnik2007}.   Medical UCT utilizes frequencies in the range of 1 - 20 MHz \cite{Chan2011}. Using the 1 MHz value and the speed of 1500 m/s for ultrasound propagation in human tissue \cite{cameron1991}, the wavelength is $\lambda = 1.5$ mm. In UCT and MCT, the feature to be recovered can be, e.g., a small T1 or T1a tumor which can have a maximum diameter of $d = 2$ cm and $d =0.5$ cm, respectively. ({\bf 2}) A signal frequency of 10 MHz has been suggested for the tomography of SSSBs \cite{binzel2005} matching roughly with the wavelength $\lambda \approx 15$ m ($\varepsilon_r \approx 4$, e.g., for dunite and kaolinite \cite{herique2002}) which is  the estimated resolution of the CONSERT experiment \cite{kofman2007,kofman1998}. The diameter of the SSSB can be for example $D=150$ m. ({\bf 3}) In ultrasonic material testing for concrete the velocity of the wave is 3500 m/s  \cite{engineeringcivilportal}. At a typical ultrasound frequency of 100 kHz the wavelength of the signal is $\lambda$ = $v$/$f$ = 3.5 m and the diameter of the concrete beam being tested could be, for example, $D=30$ cm. A fault or crack inside the beam could be $d = 3$ cm. 

Based on Table \ref{relevancytable}, the present experiment setup can be considered applicable to all application contexts ({\bf 1})--({\bf 3}). In MCT and UCT the best match to the test geometry is obtained with the T1 and T1a tumor size, respectively. Obviously, the difference between ITT and TTT can be less significant in applications, where the wave length is likely to be considerably smaller than the smallest detail to be detected. As an additional comparison, in the tomography of the ionosphere, the speed of the electromagnetic wave is around the speed of light and a typical frequency used in tomography is between 120 MHz and 400 MHz. Using a value of 200 MHz the wavelength is $\lambda = 1.5$ m. The size of the ionosphere is up to $D=1000$ km from the surface of the Earth and a typical vertical feature to be recovered can be $d = 100$ km, for example \cite{vandekampfmi}. Consequently, $d/\lambda$ and $D/\lambda$ can be 100 and 670, respectively. This suggests that our test scenario might be too different from  the tomography of the ionosphere to be able to draw inferences from the results.

The noise peaks in the experimental data were estimated to be mainly 10 dB below the maximum peak. This can be considered as appropriate for the applications  ({\bf 1})--({\bf 3}).  For a MCT imaging system, the relative reconstruction error has been shown to stay under 10 \% for amplitude errors down to 10 dB SNR  \cite{zeng2011}.  The total noise peak level of around 20 dB was observed in the CONSERT experiment \cite{kofman2015}. In concrete testing, structural noise \cite{maierhofer2010}, e.g., echoes from walls, can be significant resulting in noise peaks that can be comparable to the main peak. 

\begin{table*}
    \caption[Acoustic tomography compared to others]{The significance of the acoustic tomography results based on the parameters of other tomography applications. SSSB is a small solar system body, such as a comet or an asteroid. ($\mbox{}^{a}$ Signal frequency, 
$\mbox{}^{b}$ Signal wavelength, 
$\mbox{}^{c}$ Diameter of the target object, 
$\mbox{}^{d}$ Diameter of the target of reconstruction)}
    \label{relevancytable} \begin{center}
    \begin{tabular}{lllllll} \hline \\
      {Application} & 
      ${\lambda}$ {(m)}$\mbox{}^{b}$ &
      ${D}  $ {(m)}$\mbox{}^{c}$ &
      ${d}  $ {(m)}$\mbox{}^{d}$ &
      ${D}/{\lambda}$ & 
     ${d}/{\lambda}$ &
      {Significance} \\
      \hline
      {Test} &
     $59\cdot10^{-3}$& 
        {0.58} &    {0.10}
      & 10 & 1.0 &
      \\
      MCT (T1) &    
      $19\cdot 10^{-3}$ & 
      0.15 &
0.02 &
      7.9 &  
1.1 &
      Strong \\
      MCT (T1a) &    
      $19\cdot 10^{-3}$ & 
      0.15 &
$5.0 \cdot 10^{-3}$ &
      2.0 &  
0.28 &
      Medium \\
      UCT (T1)&  
       $1.5\cdot 10^{-3}$ & 
      0.15 &
0.02 &
     100 &  
13 &
      Medium\\
      UCT (T1a)&  
      $1.5\cdot 10^{-3}$ & 
      0.15 &
$5.0 \cdot 10^{-3}$ &
      25 &  
3.3 &
      Strong\\
      SSSBs& 
       $15$  & 
      $150$ &    
$15$ &
      10 & 
1.0 &
      Strong\\
      Concrete &
      $35\cdot 10^{-3}$ & 
      0.3 &
0.03& 
     10 &
0.86 &
     Strong\\
      Ionosphere&      
      1.5 & 
      $1.0\cdot10^6$ & 
     $0.1 \cdot10^6$ &  
  $670 $ & 
100 &
      Weak \\
      \hline
    \end{tabular} \\
\end{center}
\end{table*}

\subsection{Hardware}

The use of high-level synthesis (HLS) to develop the test hardware was found to be essential, as the specifications for the travel time calculation changed during the implementation phase and so changes to the hardware had to be made quickly. We aimed at a very direct workflow in implementing the hardware, and thus some optimization methods in the design partitioning and in the HLS tool were not utilized. In addition, having the different bit resolution implementations in separate files increased the work when changes had to be made. The resulting digital hardware was synthesized on an FPGA platform for demonstration and to facilitate further development of the data gathering system towards a laboratory instrument.

Many operations in the computation scripts, such as the normalization of values, require divisions. In the ITT calculation formula  (\ref{traveltimeintegration}) there is a large-valued divisor that first sums the squares of the values and then divides the sum with that value. Catapult generates divisions as combinatorial logic if written directly as it is in typical C code \cite{hlsbluebook}. Combinatorial logic \cite{grout2011digital}  is not synchronized with the hardware clock, which makes it unreliable in use. This gives a false sense of flexibility in the HLS tool and the user has to know what is being generated. Catapult has a math library which has synthesizable, basic algorithmic division operators for integers. Changing all divisions to use division operators offered by this library improved the results, and resulted in a  working design for most division operations whereas combinatorial dividers did not. 

Travel-time calculation can be seen as an extreme form of compression in time domain. There are more compression methods such as filtering in the frequency domain. Our work used reduced the bit resolution for input audio data. Calculation of the travel times was also performed with integer arithmetics. These reduce the accuracy of the results. bit resolution reduction is relevant when using FPGAs in general. This is because vendor-provided hardware multipliers available for use on FPGA chips such as Altera's Cyclone II have limited bit resolutions \cite{alteracycloneIIhandbook}. For example, a FIR filter can use these hardware multipliers on FPGAs as in \cite{Birk2011}. Because the operator is a multiplication that can increase the required bit resolution to store the intermediate results, the original bit resolution of the data has to be reduced. One system \cite{Birk2011} has implemented Fast Fourier Transform (FFT) filtering in the frequency domain and inverse FFT (IFFT) signal reconstruction back to the time domain. In currently available FPGA DSP chips, the available bit resolutions for multipliers are much larger \cite{alterastratixdsp}, but reduction in bit resolution can still be required. This makes the invariance of the source data bit resolution of our ITT method an important point of interest. An example of decreased signal quality is found in Ground Penetrating Radar (GPR) applications. In \cite{leucci2008} an instrumentation system with either 8-bit or 16-bit Analog-to-Digital conversion was employed. This shows that bit resolution limitations are also found in instrument hardware.

\subsection{Outlook}

Finally, the present results indicate various directions for future work. We will study hardware-level solutions regarding (i) processing and (ii) inverting waveform data.  (i) Investigating harware constraints utilizing a more sophisticated statistical travel-time detection model, such as the akaike information criterion \cite{li2009},  would be an interesting goal. The calculation of travel-time values is only one data preprocessing method that can be used, and mathematical methods to accomodate other types of filtering, such as compressing in the frequency domain could be developed. The goal in preprocessing depends on the application: in the space environment minimal data transfer between the sensors and the computation unit is important for the limited communication capacity available, whereas in biomedical and civil engineering the main objective can be to optimize the speed of the procedure in order to allow recording  as much data as possible. (ii) We aim to develop inversion approaches utilizing the FPGA environment, so extending the current study of hardware-level constraints to include inversion algorithms is essential. 

\section*{Acknowledgements}
  
M.T. and S.P. were supported by the Academy of Finland Key Project 305055 and the Academy of Finland Centre of Excellence in Inverse Problems Research. 

\subsection{Appendix}
\label{a_inversion}

In the inversion procedure (\ref{tv_iteration}), the matrix ${\bf D}$ is symmetric and can thus be diagonalized. When $\beta>0$, ${\bf D}$ is also positive definite and also invertible.   Hence,  one can define $\hat{\bf x } = {{\bf D}} {\bf x}$ and $\hat{\bf L}= {\bf L} {{\bf D}^{-1}}$. Substituting  $\hat{\bf x}$ and $\hat{\bf L}$ into (\ref{tv_iteration}) leads to the following form \cite{pursiainen2014, pursiainen2014b, kaipio2004} \begin{equation} \label{iter} \hat{\bf x }_{\ell+1}  = (\hat{\bf L}^T \hat{\bf L}  + \alpha \hat{\bf \Gamma}_{\ell} )^{-1} \hat{\bf L}^T {\bf y}, \quad \! \! \hat{\bf \Gamma}_{\ell} = \hbox{diag} ( |\hat{\bf {x}}_{\ell}|)^{-1}, \quad \! \! \hat{\bf \Gamma}_{0} = {\bf I} . \end{equation} which can be associated with alternating conditional  minimization of the function 
$H(\hat{\bf x}, \hat{\bf z})  =    \| \hat{\bf L} \hat{\bf x} - {\bf y} \|^2_2 + {\alpha} \sum_{j = 1}^M  \frac{\hat{x}^2_j}{\hat{z}_j} + {\alpha} \sum_{j = 1}^M \hat{z}_j$ 
in which $z_j>0$, for $i = 1, 2, \ldots, M$. As $H(\hat{\bf x}, \hat{\bf z})$ is quadratic  with respect to $\hat{\bf x}$,  the conditional minimizer $\hat{\bf x}'  =\arg \min_{\hat{\bf x}} H(\hat{\bf x} \mid \hat{\bf z})$ is given by the least-squares solution of the form
$\hat{\bf x}' =  (\hat{\bf L}^T \hat{\bf L}  + \alpha \hat{\bf \Gamma}_{\hat{\bf z}} )^{-1} \hat{\bf L}^T {\bf y}$, $\hat{\bf \Gamma}_{\hat{\bf z}} = \hbox{diag} ( \hat{\bf z})^{-1}$ and $\hat{\bf \Gamma}_0 = {\bf I}$. At the point of the conditional minimizer $\hat{\bf z}'  =\arg \min_{\hat{\bf z}} H(\hat{\bf z} \mid \hat{\bf x})$, the gradient of $H( \hat{\bf z} \mid \hat{\bf x} )$ vanishes with respect to $\hat{\bf z}$, i.e., 
\begin{equation}
\frac{\partial H( \hat{\bf x}, \hat{\bf z})}{\partial \hat{z}_j} \Big|_{\hat{\bf z}'} =  - {\alpha} \frac{\hat{x}^2_j}{(\hat{z}'_j)^2} + 1 = 0, \quad \hbox{i.e.} \quad \hat{z}'_j = |\hat{x}_j|  \sqrt{\alpha}.
\end{equation} 
Hence, the global minimizer can be estimated via the following alternating iterative algorithm. 
\begin{enumerate}
\item Set $\hat{\bf z}_{0} = (1, 1, \ldots, 1)$ and $\ell = 1$. For a desired number of iterations repeat the following two iteration steps.
\item  Find the conditional minimizer $\hat{\bf x}_\ell = \arg \min_{\hat{\bf x}} H(\hat{\bf x}, \hat{\bf z}_{\ell-1})$. 
\item Find the conditional minimizer  $\hat{\bf z}_\ell = \arg \min_{\hat{\bf z}} H(\hat{\bf x}_\ell, \hat{\bf z})$.
\end{enumerate}
The sequence $\hat{\bf x}_1, \hat{\bf x}_2, \ldots$ produced by this algorithm is identical to that of (\ref{iter}) and ${\bf x}_\ell = {\bf D}^{-1} \hat{\bf x}_\ell$ equals to the $\ell$-th iterate of (\ref{tv_iteration}). If for some $\ell < \infty$ the pair $({\bf \hat{x}}_\ell, {\bf \hat{z}}_\ell)$ is a global minimizer of  $H({\bf x}, {\bf z})$, then, since $(\hat{z}_\ell)_j =  |(\hat{x}_\ell)_j|$, $j = 1, 2, \ldots, M$,  then it is also the minimizer of   \begin{eqnarray} \hat{\Psi}(\hat{\bf x}) & = & H(\hat{x}_1, \hat{x}_2, \ldots , \hat{x}_M, |\hat{x}_j|, |\hat{x}_j|, \ldots, |\hat{x}_M|) \nonumber \\ & =&      \| \hat{\bf L} \hat{\bf x} - {\bf y} \|^2_2 + {\alpha} \sum_{j = 1}^M  \frac{\hat{x}_j^2}{\hat{z}_j} +\sum_{j = 1}^M \hat{z}_j \nonumber \\ 
& = &  \| \hat{\bf L} \hat{\bf x} - {\bf y} \|^2_2 + {\alpha} \sum_{j = 1}^M  \frac{\hat{x}_j^2}{|\hat{x}_j| \sqrt{\alpha}} +\sum_{j = 1}^M |\hat{x}_j| \sqrt{\alpha} \nonumber \\
& = &  \| \hat{\bf L} \hat{\bf x} - {\bf y} \|^2_2 + 2\sqrt{\alpha} \|\hat{\bf x} \|_1.
\end{eqnarray}
Consequently, ${\bf x}_\ell = {\bf D}^{-1} \hat{\bf x}_\ell$ is the minimizer of $\Psi({\bf x}) $, as  $\Psi({\bf x}) = \hat{\Psi}(\hat{\bf x})$. Furthermore, the minimizer of $\hat{\Psi}(\hat{\bf x})$ is also the 1-norm regularized solution of the linearized inverse problem.

\bibliographystyle{IEEEtran}
\bibliography{references,references1,references2,references_uusi,mt_mscthesis_refs,takala_pursiainen_artikkeli_references}


\begin{IEEEbiography}[{\includegraphics[width=1in,height=1.25in,clip,keepaspectratio]{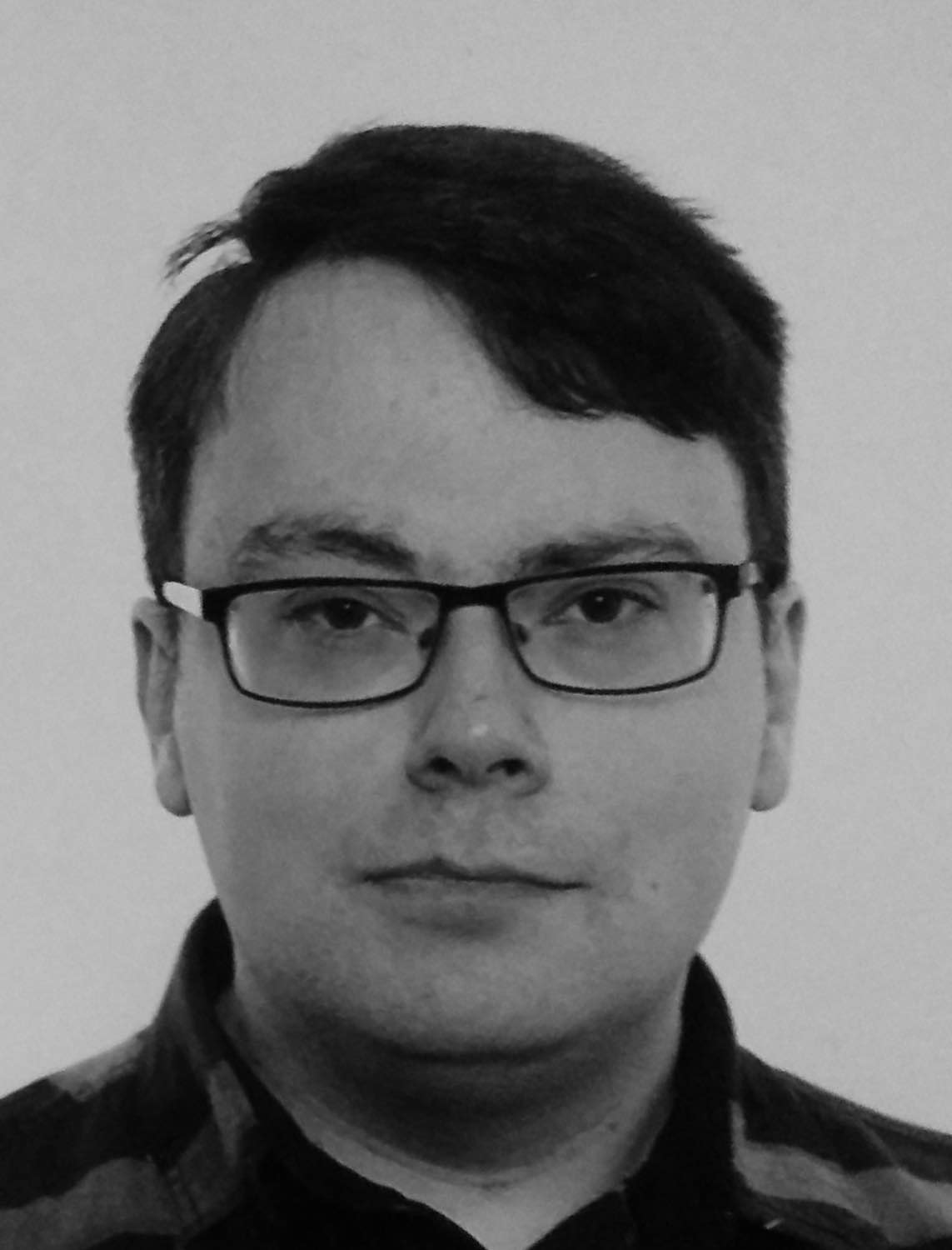}}]{Mika Takala} (M’16) was born in Nurmo, Finland, in 1982. He received the B.Sc. degree in electrical engineering from the Tampere University of Technology (TUT), in 2015, and the M.Sc.(tech.) degree from TUT in 2016. His master’s thesis in the field of embedded systems concentrated on implementation of signal preprocessing modules with High-Level Synthesis for waveform inversion applications. 
In 2016 Mr. Takala started working at the Department of Mathematics, TUT, as a PhD student. He currently works on his PhD research related geophysical inversion strategies and embedded systems. He also works as a software architect in Granite Devices, Inc., Tampere, Finland. \end{IEEEbiography}

\begin{IEEEbiography}[{\includegraphics[width=1in,height=1.25in,clip,keepaspectratio]{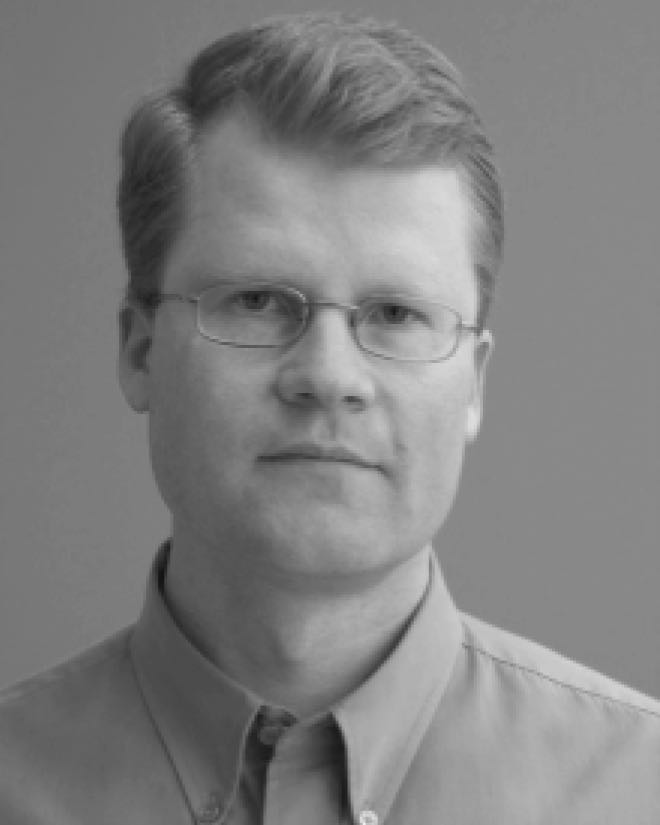}}]{Timo D. H\"{a}m\"{a}l\"{a}inen} (M’95) received the M.Sc.
and Ph.D. degrees from Tampere University of Technology (TUT), Tampere, Finland, in 1993 and 1997, respectively. He has been a Full Professor with TUT since 2001 and is currently the 
Head of the Laboratory of Pervasive Computing. 
He has authored over 70 journals and 210 conference publications. He holds several patents. His research interests include design methods and tools for 
multiprocessor systems-on-a-chip and parallel video codec implementations. 
\end{IEEEbiography}

\begin{IEEEbiography}[{\includegraphics[width=1in,height=1.25in,clip,keepaspectratio]{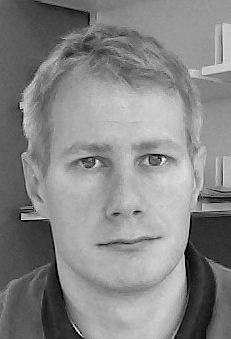}}]{Sampsa Pursiainen} received his MSc(Eng) 
and PhD(Eng) degrees (Mathematics)  in the Helsinki University of Technology  (Aalto University since 2010), Espoo, Finland, in 2003 and 2009. He focuses on various forward and inversion techniques of applied mathematics. In 2010--11, he stayed at the Department of Mathematics, University of Genova, Italy collaborating also with the Institute  for Biomagnetism and Biosignalanalysis (IBB), University of M\"{u}nster, Germany. In 2012--15, he worked at the  Department of Mathematics and  Systems Analysis, Aalto University,  Finland and also at the Department of Mathematics, Tampere University of Technology, Finland, where he currently holds an Assistant Professor position. 
\end{IEEEbiography}

\medskip
Received xxxx 20xx; revised xxxx 20xx.
\medskip

\end{document}